\documentclass[prx,aps,floatfix,twocolumn,superscriptaddress,amsmath,amssymb]{revtex4-2}
\usepackage{graphicx}
\usepackage{xcolor}
\usepackage{bm}
\usepackage{hyperref}
\hypersetup{colorlinks=true,citecolor=blue,linkcolor=blue,urlcolor=blue}
\usepackage[T1]{fontenc}
\usepackage{newtxtext,newtxmath}
\usepackage[normalem]{ulem}

\allowdisplaybreaks

\graphicspath{%
{fig/},
}

\begin{document}

\title{%
Microscopic Mechanism of High-Temperature Superconductivity Revealed by \textit{Ab Initio} Studies\\
on Hole-Doped Multilayer Cuprates
{\boldmath $\mathrm{HgBa_2Ca_2Cu_3O_8}$}
under Pressure
}

\author{Ryui Kaneko}
\email{ryuikaneko@sophia.ac.jp}
\affiliation{%
Physics Division, Sophia University, Chiyoda, Tokyo 102-8554, Japan}

\author{Masatoshi Imada}
\email{imada@g.ecc.u-tokyo.ac.jp}
\affiliation{%
Faculty of Engineering, The University of Tokyo, 7-3-1 Hongo, Bunkyo-ku, Tokyo 113-8656, Japan}
\affiliation{%
Physics Division, Sophia University, Chiyoda, Tokyo 102-8554, Japan}

\date{\today}

\begin{abstract}
Triple-layer cuprate superconductor $\mathrm{HgBa_2Ca_2Cu_3O_8}$ (Hg1223)
keeps the record of the highest superconducting (SC) critical temperature $T_{\mathrm{c}}\sim 134$\,K among all the existing materials at ambient pressure. $T_{\mathrm{c}}$ further increases under pressure up to $T_{\mathrm{c}}\sim 160$\,K, thereby holding the key to the realistic and comprehensive understanding of the cuprate superconductors, which is one of the grand challenges in physics. However, the microscopic mechanism of high $T_{\mathrm{c}}$ of Hg1223 remains to be elucidated. Here, we perform first-principles calculations of Hg1223 both at ambient and applied pressures by solving \textit{ab initio} Hamiltonians with an accurate variational solver supplemented by a neural network. The pressure dependence of the calculated $d$-wave SC order parameter and estimated $T_{\mathrm{c}}$ shows a
structure supporting a $T_{\mathrm{c}}$ peak around $30$\,GPa in quantitative agreement with the reported experimental indications.
The origin of the strong SC amplitude of Hg1223 at ambient pressure compared to other cuprates is identified as strong local Coulomb repulsion $U$ attributed to poorer screening of the Coulomb interaction in multilayer compounds. 
The origin of the further increase in $T_{\mathrm{c}}$ under pressure is ascribed to a unique interplay of three elements, namely increased electron hopping $t$, decreased $U$, and more importantly, strongly reduced non-local Coulomb repulsion $V$ with increasing pressure. 
The origin of Cooper pairing is identified as the emergent local attraction counterintuitively arising from the strong bare local repulsion $U$ for Hg1223. It explains why Hg1223 shows the highest $T_{\mathrm{c}}$ among other cuprates, while the pairing mechanism is common in other cuprates as well. 
The emergent attraction is interpreted from ``attraction from reduced repulsion'', 
originating from the release of the fluctuating doubly-occupied sites characterized from the ``false vacuum'' in the Mott insulator to the double-occupation-free $d$-wave SC states caused by the carrier doping. This instantaneous and local attraction is in marked contrast with the case of the conventional BCS SC mediated by bosonic glues. The local attraction is consistent with the electron fractionalization supported in experimental analyses. The coexistence of SC and antiferromagnetic (AF) orders is also demonstrated as a characteristic feature of the multilayer system with self-doping in the underdoped region. 
The quantitative and microscopic understanding of the high $T_{\mathrm{c}}$ in
Hg1223 provides a new perspective on the origin of the cuprate
superconductivity
and offers a new route explicitly using this channel of the emergent attraction for future superconductivity research aiming at designing and optimizing SC materials.
\end{abstract}

\maketitle

\section{Introduction}
\label{sec:intro}

Since the discovery of high-temperature superconductors (SCs)
in copper oxides (cuprates) in 1986~\cite{bednorz1986},
cuprate SCs continue to hold the record of
the highest
SC critical temperature ($T_{\mathrm{c}}$) at ambient pressure.
Although a tremendous amount of research has been conducted with significant advances, understanding the microscopic mechanisms behind materials dependencies
of $T_{\mathrm{c}}$ and high-$T_{\mathrm{c}}$ superconductivity itself continues to be a central challenge
in condensed matter physics.
Strong electron correlations that 
characterize the cuprate SCs accompanied by severe competition among intertwined different states and fluctuations
require accurate many-body solutions.

Recently, by using state-of-the-art quantum many-body solvers, \textit{ab initio}  low-energy effective Hamiltonians for single-, double-, and infinite-layer cuprates have been solved, which successfully reproduced the rich
material-dependence of $T_{\mathrm{c}}$~\cite{schmid2023}.
More specifically, based on the effective Hamiltonians for
$\mathrm{CaCuO_2}$ with an infinite-layer structure, $\mathrm{HgBa_2CuO_4}$ (Hg1201), $\mathrm{Bi_2Sr_2CuO_6}$ (Bi2201) with a single-layer structure and $\mathrm{Bi_2Sr_2CaCu_2O_8}$ (Bi2212)
with a double-layer structure by using an accurate variational Monte Carlo (VMC)
method~\cite{tahara2008,misawa2019} combined with a neural network algorithm~\cite{nomura2017} using only experimental crystal structures as inputs,
the material and doping concentration dependencies
of the SC order parameter and $T_{\mathrm{c}}$ have been quantitatively reproduced.
The materials dependence has been understood primarily from the difference in an \textit{ab initio} Hamiltonian parameter for the dimensionless ratio of the local interaction $U$ to the nearest-neighbor hopping amplitude $|t_1|$.
The severe competition among the SC, antiferromagnetic (AF), and stripe phases is in agreement with the observations in the experiments as well.

\begin{figure}[!t]
\centering
\includegraphics[width=0.6\columnwidth]{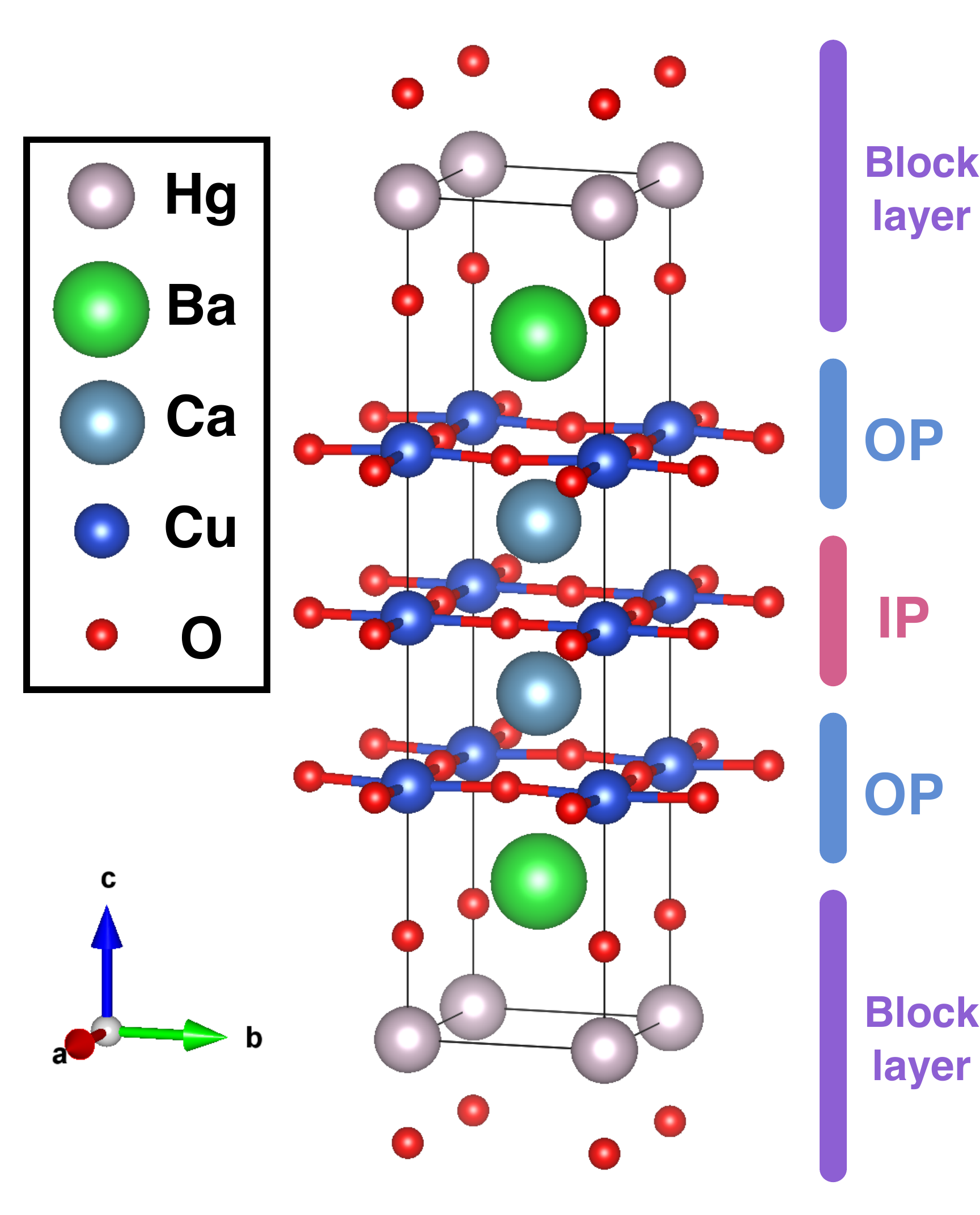}
\caption{%
Crystal structure of Hg1223
and directions of primitive vectors
$(\bm{a},\bm{b},\bm{c})$.
The inner $\mathrm{CuO_2}$ plane (IP)
sandwiched by the two outer $\mathrm{CuO_2}$ planes (OPs)
is separated from the equivalent next three $\mathrm{CuO_2}$ planes by
the block layers, and the neighboring IP and OP are separated by interstitial Ca atoms.
The crystal structure~\cite{finger1994} is visualized from data in
the Crystallography Open Database~\cite{downs2003,grazulis2009,vaitkus2023}
using VESTA~\cite{momma2011}.
}
\label{fig:hg1223}
\end{figure}

However, significantly higher $T_{\mathrm{c}}$ was observed in multilayer, especially triple-layer cuprates,  while the numerical analysis of the multilayer cuprates on the same level as Ref.~[\onlinecite{schmid2023}] is so far missing because of the demanding computational cost ascribed to the large unit cell.

Among all the available materials at ambient pressure, the highest $T_{\mathrm{c}}$  was reported at $T_{\mathrm{c}}\sim 134$\,K in a triple-layer cuprate compound $\mathrm{HgBa_2Ca_2Cu_3O_8}$ (Hg1223)~\cite{putilin1993,schilling1993,gao1994,dai1995,yamamoto2015}
(see Fig.~\ref{fig:hg1223} for the crystal structure).
When the pressure is applied, Hg1223 shows a further increased $T_{\mathrm{c}}$ forming dome-like pressure dependence and reaches the peak at around $30$\,GPa, as high as $T_{\mathrm{c}}\sim 160$\,K, which is the highest record among the cuprates~\cite{gao1994,yamamoto2015}.
Therefore, without identifying the microscopic origin of such high $T_{\mathrm{c}}$ values based on the \textit{ab initio} parameter-free framework, the mechanism of the cuprate SC would not be ultimately understood.

In this study, we
quantitatively elucidate why multilayer cuprates including triple-layer Hg1223 show the substantially higher $T_{\mathrm{c}}$ than the single-, double-, and infinite-layer compounds. We also
discuss the origin of the observed dome-like pressure dependence of $T_{\mathrm{c}}$, which peaks around $30$\,GPa.
The materials dependence of the single-, double-, and infinite-layer compounds was captured without paying serious attention to the difference in the off-site interactions because they are similar at ambient pressure~\cite{schmid2023}. Here, by contrast, we show that the pressure-induced reduction of off-site interactions is substantial and plays a major role in enhancing SC. We also clarify that, in the multilayer compounds, the induced self-doping brings about the difference in carrier concentration between the IP and OPs, which allows for the emergence of coexisting SC and AF
order as was experimentally observed~\cite{mukuda2012,kurokawa2023}.

Based upon the successful and quantitative reproduction of the materials dependence of the SC for the cuprates including the triple-layer Hg1223, we identify the realistic mechanism of SC: The Cooper pair is induced by the instantaneous and local attraction caused by the reduction of repulsion in the background of strong ``vacuum fluctuation'', which is a unique property of the strongly correlated doped Mott insulator. This mechanism is radically different from the conventional BCS SC~\cite{BCS1957}, where the retardation is essential to induce the boson-mediated weak-coupling SC to circumvent the unavoidable electron-electron Coulomb repulsion.

Our calculations
are based on the \textit{ab initio} low-energy effective Hamiltonians of Hg1223 derived in Ref.~[\onlinecite{moree2024}].
The pressure-dependent lattice constants and atomic positions used in deriving these Hamiltonians were obtained from the first-principles structural optimization performed in Ref.~[\onlinecite{moree2024}].
The crystal structure was parameterized by the lattice constants $a$ and $c$ and five internal structure parameters, from which the atomic coordinates at each pressure were determined.
The relevant hopping parameters were obtained from the matrix elements between maximally localized Wannier orbitals.
In the procedure to derive the effective Hamiltonians, Refs.~\cite{moree2022,moree2024} employed a state-of-the-art methodology of constrained GW (cGW) scheme~\cite{aryasetiawan2009,hirayama2013,hirayama2017}, which completely removes the double counting problem known in the constrained RPA~\cite{aryasetiawan2004,imada2010}. In addition, the self-interaction correction (SIC) together with the level renormalization feedback (LRFB)~\cite{hirayama2019} are taken into account, in which the charge distribution among the atomic O$2p$ and Cu$3d_{x^2-y^2}$ orbitals, which strongly hybridize to form the antibonding and bonding orbitals, is appropriately treated in a self-consistent fashion.
This procedure in total is called cGW-SIC+LRFB scheme. The antibonding orbital constituting the low-energy effective Hamiltonian contains a small portion of other atomic orbitals, such as hybridizing Cu$3d_{3z^2-r^2}$ and O$2p_z$ orbitals, which contribute to the interlayer hopping.
The ground state of the derived effective Hamiltonian for the antibonding orbital on lattices is obtained in this paper by applying the VMC method~\cite{tahara2008,misawa2019} refined from the seminal work~\cite{yokoyama1987,gros1987,gros1988,capriotti2001}.
We further enhance the accuracy by combining the VMC method with neural network techniques~\cite{carleo2017,nomura2017,nomura2021}.

This paper is organized as follows:
In Sec.~\ref{sec:methods},
we briefly describe the \textit{ab initio} low-energy effective Hamiltonian
and the VMC method.
In Sec.~\ref{sec:ambient},
we present the ground states of Hg1223 at ambient pressure obtained by the VMC method.
In Sec.~\ref{sec:pressure},
we present the pressure dependence of the SC order parameter and $T_{\mathrm{c}}$ in Hg1223.
In Sec.~\ref{sec:effective_attraction},
we discuss the effective attraction emergently arising from the strong local repulsion as the origin of Cooper pairing of Hg1223 and compare with the cases of other cuprates with lower $T_{\mathrm{c}}$ on the basis of the successful and realistic reproductions of the experimental trend.
Section~\ref{sec:discussion} is devoted to discussions.
Finally, in Sec.~\ref{sec:conclusions},
we summarize our findings and discuss their consequences. We also provide perspectives for future studies.

\section{%
Methods
}
\label{sec:methods}

\subsection{%
\textit{Ab initio} effective Hamiltonians
}
\label{subsec:abinitio}

We solve the \textit{ab initio} triple-layer effective Hamiltonian
of Hg1223 for the antibonding orbital consisting of strongly hybridized Cu$3d_{x^2-y^2}$ and O$2p_{\sigma}$ orbitals as derived in Ref.~[\onlinecite{moree2024}], which has the form
\begin{align}
\label{eq:ham}
 \mathcal{H}
 &=
 \mathcal{H}_{t} + \mathcal{H}_{U} + \mathcal{H}_{V} + \mathcal{H}_{\mu}
\end{align}
with
\begin{align}
\label{eq:ham_ht}
 \mathcal{H}_{t}
 &=
 \sum_{I, J, \sigma} t_{IJ} c^{\dagger}_{I,\sigma} c_{J,\sigma},
\\
\label{eq:ham_hu}
 \mathcal{H}_{U}
 &=
 \sum_{I} U_{I} n_{I,\uparrow} n_{I,\downarrow},
\\
\label{eq:ham_hv}
 \mathcal{H}_{V}
 &=
 \frac{1}{2}\sum_{I \neq J} V_{IJ} n_{I} n_{J},
\\
\label{eq:ham_hmu}
 \mathcal{H}_{\mu}
 &=
 \sum_{I} \mu_{I} n_{I}.
\end{align}
Here, $I$ and $J$ are multiindices of site and layer,
e.g., $(i,\alpha)$ and $(j,\beta)$, respectively;
sites $i$ and $j$ correspond to
positions $\bm{r_i}=(x_i,y_i)$ and $\bm{r_j}=(x_j,y_j)$
on an $L\times L$ square lattice,
whereas $\alpha$ and $\beta$ indicate the layer, namely, IP or OP.
In Fig.~\ref{fig:hg1223}, here and later, two OPs should be distinguished as OP1 and OP2, where the summations over $\alpha$ and $\beta$ run over IP, OP1, and OP2.
The operator $c^{\dagger}_{I,\sigma}$ ($c_{I,\sigma}$)
creates (annihilates) an electron with spin $\sigma$
and index $I$.
The operator $n_{I,\sigma} = c^{\dagger}_{I,\sigma} c_{I,\sigma}$
counts the number of electrons with spin $\sigma$
and index $I$,
and the total number operator is given by
$n_{I} = n_{I,\uparrow} + n_{I,\downarrow}$.
The system contains one IP and two OPs as shown in Fig.~\ref{fig:hg1223},
and the total number of sites is $N_{\mathrm{s}}=N_{l} L^2$ with $N_{l}$ being the number of layers ($N_l=3$ for the triple-layer Hg1223).
The hopping parameter is given by $t_{IJ}$
and the strengths of the onsite and off-site Coulomb interactions
are represented as $U_I$ and $V_{IJ}$, respectively.
Both hopping and off-site interaction parameters are functions of
the relative coordinate vector $\bm{r}_i - \bm{r}_j$
between sites $i$ and $j$
because of the translational invariance of the crystal structure.
The chemical potential of the IP $\mu_{I}$ relative to that of the OP controls the relative carrier densities
under the condition of a fixed total number of electrons for a fixed size in the canonical ensemble and is a function of the total number of
electrons $N_{\mathrm{e}}[=(1-\delta)N_{\mathrm{s}}]$.

\begin{table}[!t]
\centering
\caption{%
Selected parameters for the \textit{ab initio} low-energy effective Hamiltonian
of Hg1223 at $0$\,GPa, $30$\,GPa, and $60$\,GPa.
We show the nearest-neighbor hopping parameter $t_1$,
the onsite Coulomb interaction $U$,
and the nearest-neighbor off-site Coulomb interaction $V_1$
for the IP and the OP.
All values are given in eV.
Complete lists including other parameters are given in Tables~\ref{tab:parameters_0GPa}, \ref{tab:parameters_30GPa}, and \ref{tab:parameters_60GPa}.
}
\label{tab:selected_parameters_0,30,60GPa}
\begin{tabular}{cccc}
\hline
\hline
 IP & $t_1$ & $U$ & $V_1$ \\
\hline
  $0$\,GPa & $-0.4857$ & $4.5184$ & $0.9878$ \\
 $30$\,GPa & $-0.5833$ & $4.3615$ & $0.8752$ \\
 $60$\,GPa & $-0.6492$ & $4.1751$ & $0.6842$ \\
\hline
 OP & $t_1$ & $U$ & $V_1$ \\
\hline
  $0$\,GPa & $-0.4849$ & $4.5443$ & $0.9785$ \\
 $30$\,GPa & $-0.5554$ & $4.4986$ & $0.8056$ \\
 $60$\,GPa & $-0.5812$ & $4.2381$ & $0.5832$ \\
\hline
\hline
\end{tabular}
\end{table}

We consider hopping and interaction parameters
that extend over long distances
and assume that their values are nonzero
within the range satisfying
$r=|\bm{r}_i-\bm{r}_j| \le 3\sqrt{2}$
between sites $i$ and $j$. 
Here, the distance is counted in the length unit of the nearest Cu-Cu distance both within the $xy$ plane and along the interlayer $z$ direction.
Beyond this range, the parameters are negligibly small.
The dominant hopping and interaction parameters 
employed in the present calculations
are summarized in Table~\ref{tab:selected_parameters_0,30,60GPa},
for which we employ the values from Ref.~[\onlinecite{moree2022}].
For other parameters,
see complete lists in Tables~\ref{tab:parameters_0GPa}, \ref{tab:parameters_30GPa} and \ref{tab:parameters_60GPa}
in Appendix~\ref{App:parameters}.
They include both intralayer and interlayer parameters.
In most cases, the parameters for hopping and interaction are smaller than $10$\%
of the largest nearest-neighbor hopping and onsite interaction, respectively, when $r=|\bm{r}_i - \bm{r}_j|>2$.
Although parameters with such small values
contribute little to qualitative physical properties,
we keep values at longer distances with $r>2$
up to the $11$th neighbor to perform more realistic simulations of actual systems and for quantitative comparisons with real compounds.
The truncated hopping parameters employed in the VMC calculations accurately reproduce the cGW-SIC+LRFB band dispersion near the Fermi level.
The resulting Fermi surfaces of the individual bands agree well with those obtained by the full parameter set of the cGW-SIC+LRFB calculations.

The chemical potential difference
$\mu = \mu_{\mathrm{OP}} - \mu_{\mathrm{IP}}$
between the IP and the OP
is determined in a way such that
$\mu$ reproduces the carrier density in each layer
obtained by GW calculations.
For the details of the determination of $\mu$,
see Appendix~\ref{App:estimate_mu}.
The final hole doping $\delta$ dependencies of $\mu$
at $0$\,GPa, $30$\,GPa, and $60$\,GPa
are given by
\begin{align}
\label{eq:mu_vs_delta_0GPa}
 \mu_{\mathrm{0\,GPa}} &= 5.192 - 9.78 \delta,
\\
\label{eq:mu_vs_delta_30GPa}
 \mu_{\mathrm{30\,GPa}} &= 5.087 - 11.2 \delta,
\\
\label{eq:mu_vs_delta_60GPa}
 \mu_{\mathrm{60\,GPa}} &= 2.990 - 6.36 \delta,
\end{align}
respectively.
At ambient pressure, $\delta_{\rm IP}$ is always smaller than $\delta_{\rm OP}$.
With increasing pressure and doping $\delta$,
the chemical potential difference $\mu$ becomes smaller,
and the carrier densities of the IP and the OP
tend to be closer to each other as one can see later in Figs.~\ref{fig:self-doping} and \ref{fig:self-doping_3060GPa}.

\subsection{%
Numerical methods
}
\label{subsec:vmc}

We employ an elaborated VMC method, which is available as an open source software~\cite{tahara2008,misawa2019,mVMC}
for calculating the ground state properties of Hg1223.
The variational wave function is given by
\begin{align}
\label{eq:wf}
 |\psi\rangle
 =
 \mathcal{P}_{\mathrm{G}}
 \mathcal{P}_{\mathrm{J}}
 \mathcal{P}_{\mathrm{dh}}
 |\phi_{\mathrm{pair}}\rangle,
\end{align}
where the Gutzwiller factor $\mathcal{P}_{\mathrm{G}}$~\cite{gutzwiller1963},
the charge Jastrow correlation factor $\mathcal{P}_{\mathrm{J}}$~\cite{jastrow1955,capello2005},
and the doublon-holon correlation factor $\mathcal{P}_{\mathrm{dh}}$~\cite{yokoyama1990}
are defined by
\begin{align}
\label{eq:wf_projection}
 \mathcal{P}_{\mathrm{G}}
 &=
 \exp\left( \sum_{I} g_I n_{I,\uparrow} n_{I,\downarrow} \right),
\\
 \mathcal{P}_{\mathrm{J}}
 &=
 \exp\left( \sum_{I<J} v_{IJ} n_I n_J \right),
\\
 \mathcal{P}_{\mathrm{dh}}
 &=
 \exp\left[
 \sum_{t=1,2} \sum_{m=0}^{4} \sum_{l=\mathrm{d,h}} \sum_{I}
 \gamma^{(t,m,l)}_{I} \xi^{(m,l)}_{I}
 \right],
\end{align}
respectively. Here, the doublon is a site occupied by both up- and down-spin electrons and the holon represents an empty site.
Variational parameters are $g_I$, $v_{IJ}$, and $\gamma^{(t,m,l)}_{I}$.
We assume that $g_I=g_{(i,\alpha)}$ does not depend on the position $\bm{r}_i$,
but can take different values depending on the layer $\alpha$.
On the other hand, $v_{IJ}=v_{(i,\alpha)(j,\beta)}$ is chosen to be a function
of the relative position $|\bm{r}_i-\bm{r}_j|$
and can vary for different layer pairs $(\alpha,\beta)$.
The many-body operator $\xi^{(m,l)}_{I}$ is diagonal in the real space representation
and takes $1$ if a doublon ($l=\mathrm{d}$)
[holon ($l=\mathrm{h}$)] exists at index $I$
and $m$ holons (doublons) at $t$th nearest neighbors surround it;
otherwise, $\xi^{(m,l)}_{I}$ is $0$.
The parameter $\gamma^{(t,m,l)}_{I}=\gamma^{(t,m,l)}_{(i,\alpha)}$
does not depend on the position $\bm{r}_i$, but can take different values
depending on the layer $\alpha$ and $t$th nearest neighbors.
The pair-product wave function~\cite{bouchaud1988,bajdich2008,tahara2008}
is given by
\begin{align}
\label{eq:wf_pair}
 |\phi_{\mathrm{pair}}\rangle
 =
 \left(
 \sum_{I,J=1}^{N_{\mathrm{s}}} \sum_{\sigma,\sigma'}
 f^{\sigma\sigma'}_{IJ}
 c^{\dagger}_{I,\sigma}
 c^{\dagger}_{J,\sigma'}
 \right)^{N_{\mathrm{e}}/2}
 |0\rangle,
\end{align}
where $f^{\sigma\sigma'}_{IJ}$ are variational parameters
and are assumed to have a sublattice structure with
translational invariance.

The variational wave function can further be improved by
applying additional correlation factors to the aforementioned wave function,
and the corresponding energy becomes closer to the exact values by
performing a zero-variance extrapolation.
We typically consider the wave function with the restricted Boltzmann machine
(RBM) correlation factor~\cite{carleo2017,nomura2017,nomura2021},
which has the form
\begin{align}
\label{eq:wf_rbm}
 |\psi_{\mathrm{RBM}}\rangle
 =
 \mathcal{P}_{\mathrm{G}}
 \mathcal{P}_{\mathrm{J}}
 \mathcal{P}_{\mathrm{dh}}
 \mathcal{P}_{\mathrm{RBM}}
 |\phi_{\mathrm{pair}}\rangle.
\end{align}
Here, the RBM correlation factor is chosen as
\begin{align}
\label{eq:wf_rbm_2}
 \mathcal{P}_{\mathrm{RBM}}
 =
 \exp\left( \sum_{I,\sigma} a_{I\sigma} n_{I,\sigma} \right)
 \prod_{J}^{N_{\mathrm{h}}} \cosh 
 \left( b_J + \sum_{I,\sigma} W_{IJ\sigma} n_{I,\sigma} \right),
\end{align}
with $a_{I\sigma}$, $b_J$, and $W_{IJ\sigma}$ being variational parameters.
Increasing the ratio of the number of hidden parameters $N_{\mathrm{h}}$
to the number of sites, namely,
$\alpha_{\mathrm{RBM}}=N_{\mathrm{h}}/N_{\mathrm{s}}$,
improves the accuracy of the variational wave function.
In practice, we apply the RBM correlation factor
on top of the optimized wave function
$ |\psi\rangle =
 \mathcal{P}_{\mathrm{G}}
 \mathcal{P}_{\mathrm{J}}
 \mathcal{P}_{\mathrm{dh}}
 |\phi_{\mathrm{pair}}\rangle$,
although the RBM correlation factor itself can simulate the effects
coming from $\mathcal{P}_{\mathrm{G}}$ and $\mathcal{P}_{\mathrm{J}}$.
Another way of improving the wave function is to consider
the Lanczos method~\cite{sorella2001,heeb1993}.
We choose the appropriate factors $\alpha_n$ in the wave function
\begin{align}
\label{eq:wf_lanczos}
 |\psi_M\rangle
 =
 \left(
 1 + \sum_{n=1}^{M} \alpha_n \mathcal{H}^{n}
 \right)
 |\psi\rangle
\end{align}
so that its energy is minimized.
We apply the Lanczos method up to the first step
($M=1$)~\cite{misawa2014a,ido2022} in this study
because of the high computational cost,
which increases exponentially in $M$.

We assume antiperiodic-periodic boundary conditions
and calculate ground-state candidate states
of lattices with $N_{\mathrm{s}}=3L^2$ sites.
The system sizes that we have studied range from $L=16$
to $L=28$, where the system contains $2352$ sites at $L=28$.
We first calculate ground-state candidate states
and then compare the energy per site
$E/N_{\mathrm{s}} = \langle H \rangle/N_{\mathrm{s}}$
of each state after
the variance extrapolation~\cite{imada2000,kashima2001,wu2024}
to the zero limit with the help of results obtained after Lanczos and RBM operations if necessary.
The variance is evaluated by
$(\delta E)^2 = (\langle H^2 \rangle - \langle H \rangle^2)/\langle H \rangle^2$
and satisfies $\delta E=0$ for true eigenstates.
We determine the ground state by choosing the one with the lowest energy
among the candidate states.
Such ground-state candidate states include
the SC, metallic, staggered AF,
and C4S8 stripe~\cite{tranquada1995,tranquada1997} states as well as the coexistence of the SC and AF.
See the detailed procedure in Appendix~\ref{App:variance_extrapolate_sc}.
Since the Lanczos procedure is generally considered to have negligible effects on analyzed physical quantities such as correlations and order parameters~\cite{iqbal2013,misawa2014a,zhao2017,ido2018},
we present order parameters without applying the Lanczos method hereafter.

For important quantities such as SC and AF order parameters, we take the size extrapolation to the thermodynamic limit. In some cases below, however, we only show the results for $24\times 24 \times 3$ lattice for Hg1223 with an explicit remark to save the demanding computational time for the size extrapolation, when the size dependence is small.

\subsection{%
Physical quantities
}
\label{subsec:physquant}

We calculate the spin and charge structure factors
to distinguish whether the state exhibits magnetic and charge orders.
The spin structure factor in each layer
$\alpha(=\mathrm{IP},\mathrm{OP})$ is defined by
\begin{align}
\label{eq:sf}
 S^{\alpha}_{\mathrm{s}}(\bm{q})
 =
 \frac{1}{L^2} \sum_{i,j}
 \langle \bm{S}_{i,\alpha} \cdot \bm{S}_{j,\alpha} \rangle
 e^{i\bm{q}\cdot(\bm{r}_i-\bm{r}_j)},
\end{align}
where $\bm{S}_i=(S^x_i,S^y_i,S^z_i)
=\sum_{\tau,\tau'} c^{\dagger}_{i,\tau} \bm{\sigma}_{\tau,\tau'} c_{i,\tau'}/2$
is the spin-$1/2$ operator and $\bm{\sigma}_{\tau,\tau'}$ is the Pauli matrix.
Similarly, the charge structure factor in each layer is defined by
\begin{align}
\label{eq:cf}
 S^{\alpha}_{\mathrm{c}}(\bm{q})
 =
 \frac{1}{L^2} \sum_{i,j}
 \langle n_{i,\alpha} n_{j,\alpha} \rangle
 e^{i\bm{q}\cdot(\bm{r}_i-\bm{r}_j)}.
\end{align}
The system exhibits long-range order
when $S^{\alpha}_{\mathrm{s}}(\bm{Q})/L^2$
or $S^{\alpha}_{\mathrm{c}}(\bm{Q})/L^2$ at the Bragg peak $\bm{Q}$
is extrapolated to a nonzero value in the thermodynamic
limit ($L\to\infty$).
The order parameter in the thermodynamic limit is evaluated by
\begin{align}
\label{eq:msc}
 m^{\alpha}_{\mathrm{s,c}}=\sqrt{\lim_{L\to\infty} \frac{S^{\alpha}_{\mathrm{s,c}}(\bm{q})}{L^2} }.
\end{align}

The SC states in the present study are of $d$-wave type,
and the corresponding SC correlation function in each layer $\alpha$
is given by
\begin{align}
\label{eq:scf}
 P^{\alpha}_{d}(\bm{r})
 =
 \frac{1}{L^2} \sum_{\bm{r}_i}
 \langle
 {\Delta^{\alpha}_d}^{\dagger}(\bm{r}_i)
 \Delta^{\alpha}_d(\bm{r}_i+\bm{r})
+\mathrm{H.c.}
 \rangle,
\end{align}
where $\Delta^{\alpha}_d(\bm{r}_i)$ is the order parameter
with the form
\begin{align}
 \Delta^{\alpha}_d(\bm{r}_i)
 =
 \frac{1}{\sqrt{2}} \sum_{\bm{r}}
 f_d(\bm{r}) (
 c_{\bm{r}_i,\alpha,\uparrow}
 c_{\bm{r}_i+\bm{r},\alpha,\downarrow}
 -
 c_{\bm{r}_i,\alpha,\downarrow}
 c_{\bm{r}_i+\bm{r},\alpha,\uparrow}
 ).
\end{align}
The form factor at $\bm{r}=(x,y)$
that reflects the $d_{x^2-y^2}$ symmetry is given by
\begin{align}
 f_d(\bm{r})
 =
 \delta_{y,0} (\delta_{x,1} + \delta_{x,-1})
 -
 \delta_{x,0} (\delta_{y,1} + \delta_{y,-1}).
\end{align}
For SC states in large systems, the SC correlation function
nearly saturates for sufficiently long distances.
This long-distance correlation is proportional to the square of
the SC order parameter $F^{\alpha}_{\mathrm{SC}}(L)$ for a finite system in layer $\alpha$.
To practically estimate this value, we average
the long-distance part of $P^{\alpha}_{d}(\bm{r})$
and evaluate a quantity, which is given by
\begin{align}
\label{eq:scfbar}
 [F^{\alpha}_{\mathrm{SC}}(L)]^2
 =
 \bar{P}^{\alpha}_{d}(L)
 =
 \frac{1}{N_{\mathrm{far}}} \sum_{|\bm{r}| > \sqrt{2}L/4}
 P^{\alpha}_{d}(\bm{r}).
\end{align}
Here, position $\bm{r}=(x,y)$ runs over all sites within a region $(-L/2,L/2]^2$
and $N_{\mathrm{far}}$ is the number of lattice points satisfying
$|\bm{r}| \in (\sqrt{2}L/4,\sqrt{2}L/2)$.
Then, the SC order parameter in the thermodynamic limit
$F^{\alpha,\infty}_{\mathrm{SC}}$ for layer $\alpha$ is evaluated by
\begin{align}
 \bar{P}^{\alpha,\infty}_{d}
 &=
 \lim_{L\to\infty} \bar{P}^{\alpha}_{d}(L),
\\
 F^{\alpha,\infty}_{\mathrm{SC}}
 &=
 \lim_{L\to\infty} F^{\alpha}_{\mathrm{SC}}(L)
 =
 \sqrt{\bar{P}^{\alpha,\infty}_{d}}.
 \label{eq:FSC}
\end{align}
The linear $1/L$ scaling is assumed during the extrapolation~\cite{ohgoe2020}
because quasiparticle excitations of the $d$-wave SC state at nodal points
exhibit the Dirac-type linear dispersion.

All the physical quantities including the energy are estimated by the standard Monte Carlo sampling. Error bars for a fixed system size are estimated by the statistical errors of each quantity. In the procedure of size extrapolations, we also estimate the error bars coming from the extrapolation.

\section{%
Results at Ambient Pressure
}
\label{sec:ambient}

\begin{figure}[!t]
\centering
\includegraphics[width=0.8\columnwidth]{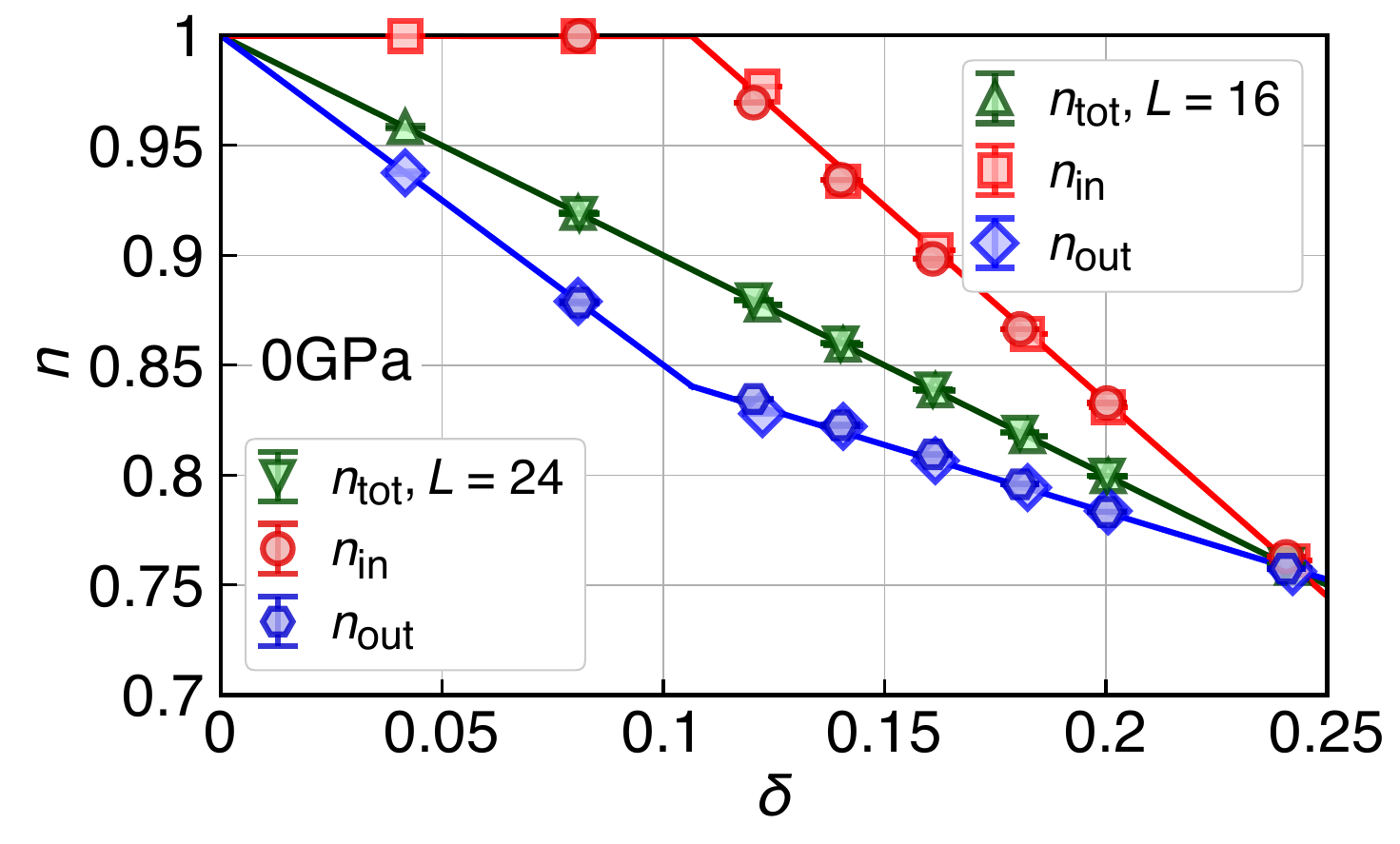}
\caption{%
Inner- and outer-layer hole densities as a function of
total hole density $\delta$ for the \textit{ab initio} Hamiltonian of Hg1223 at ambient pressure.
$n_{\rm tot}$ is the averaged electron density per Cu site, whereas $n_{\rm in}$ and $n_{\rm out}$ are the averaged electron densities per site in the inner and outer plane, respectively.
The IP always has lower hole density.
The plots for $L=16$ and $24$ show negligible size dependence, suggesting that the $L=\infty$ limit is nearly reached.
}
\label{fig:self-doping}
\end{figure}

In this section,
we show the solutions for the ground state of the effective Hamiltonian for Hg1223 at ambient pressure and discuss SC properties competing with AF and stripe states.
A characteristic feature of the multilayer compounds, not present in the single-, double-, and infinite-layer compounds, is self-doping, in which the carrier concentration is self-optimized and varies among layers. Figure~\ref{fig:self-doping} shows how the inner- and outer-layer hole densities, $\delta_{\rm IP}$ and $\delta_{\rm OP}$, respectively, evolve as a function of averaged hole density per layer $\delta=(\delta_{\rm IP}+2\delta_{\rm OP})/3$ for the \textit{ab initio} Hamiltonian of Hg1223 at ambient pressure. From Eq.~\eqref{eq:mu_vs_delta_0GPa}, the IP always has a lower chemical potential of electrons (higher chemical potential of holes) in the realistic carrier density region, so that the IP has a lower hole density.
This is consistent with the experimental indications of the multilayer compounds~\cite{mukuda2012,kurokawa2023}.

\subsection{%
Doping concentration dependence of SC order}
\label{subsec:gs_dope_pressure}

\begin{figure}[!t]
\centering
\includegraphics[width=0.8\columnwidth]{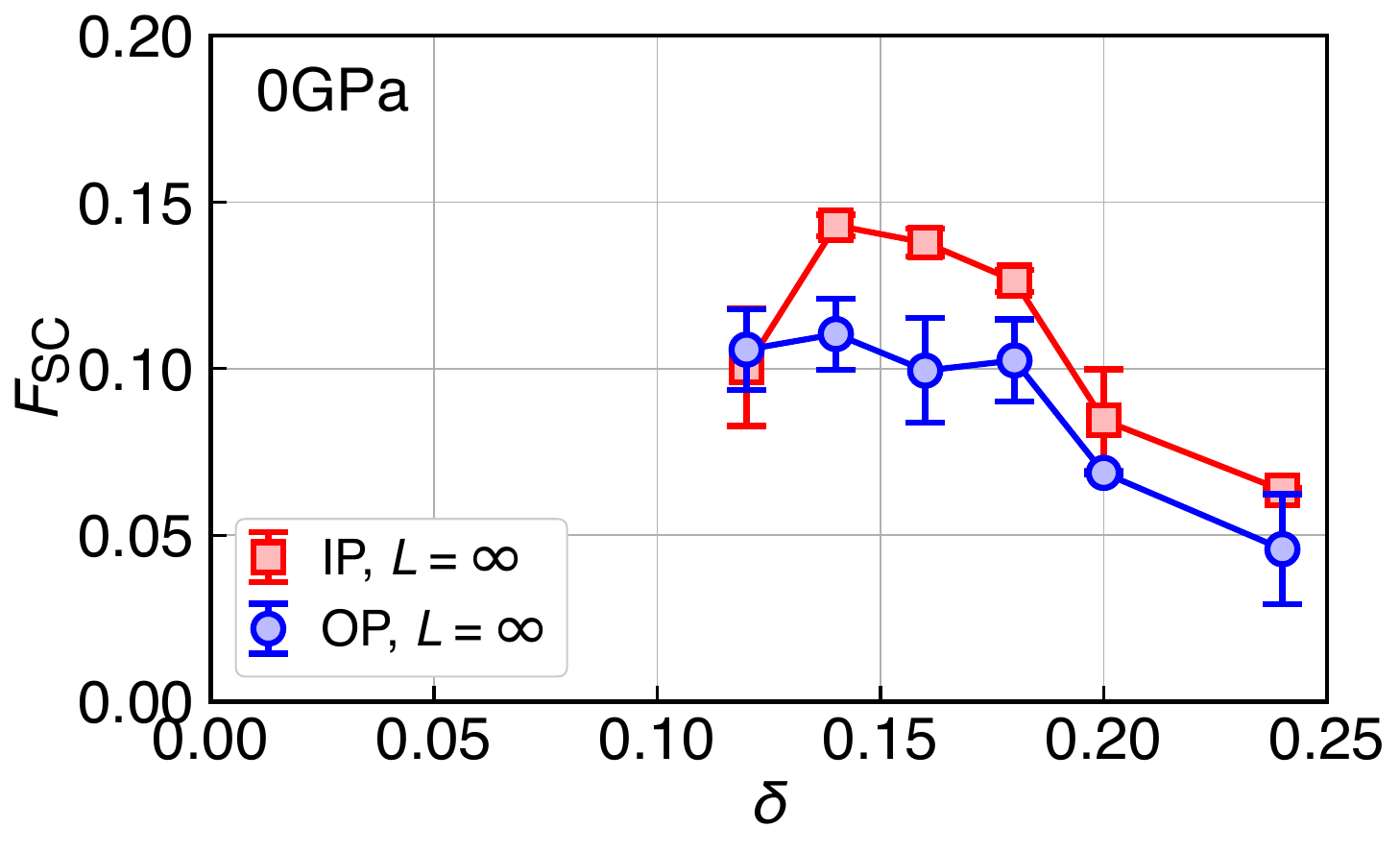}
\caption{%
Doping concentration dependence of the SC order parameter $F_{\rm SC}$ defined in Eq.~(\ref{eq:FSC}) at ambient pressure.
The red and blue lines show $F_{\rm SC}$ for the IP and OP, respectively, after size extrapolation to the thermodynamic limit $L\rightarrow \infty$. The error bars are due to the size extrapolation error.}
\label{fig:FSC_delta_0GPa}
\end{figure}

We show the hole density ($\delta$) dependence of the SC order parameter $F_{\rm SC}$ in the thermodynamic limit at the IP and OP (two OPs show the same behavior)
in Fig.~\ref{fig:FSC_delta_0GPa}.
Hereafter, for $F_{\rm SC}^{\infty}$ obtained after the extrapolation to the thermodynamic limit, we drop $\infty$ as $F_{\rm SC}$.
(The procedure of obtaining $F_{\rm SC}$ from the extrapolation of the SC correlation $P_d$ to the thermodynamic limit is described in Appendices~\ref{App:details_sc_correlation} and \ref{App:size_extrapolate_sc}.)
The dominance of the $d$-wave SC order demonstrates the consistency with the experimental indications. The optimum value of $F_{\rm SC}\sim 0.14$ at $\delta=0.14$--$0.16$ is found in the IP and exceeds the maximum value $\sim 0.13$ of the single-, double-, and infinite-layer compounds reported for Bi2212~\cite{schmid2023}.
The origin of the highest optimum $F_{\rm SC}$ among the cuprates studied here and in Ref.~[\onlinecite{schmid2023}] is mainly attributed to the highest value of $U/|t_1|$ due to the poor screening.
The optimum hole density is roughly consistent with Cu valence measurement~\cite{fukuoka1997,yamamoto2015} and NMR data~\cite{kotegawa2001} by considering its temperature dependence.

\begin{figure}[!t]
\centering
\includegraphics[width=\columnwidth]{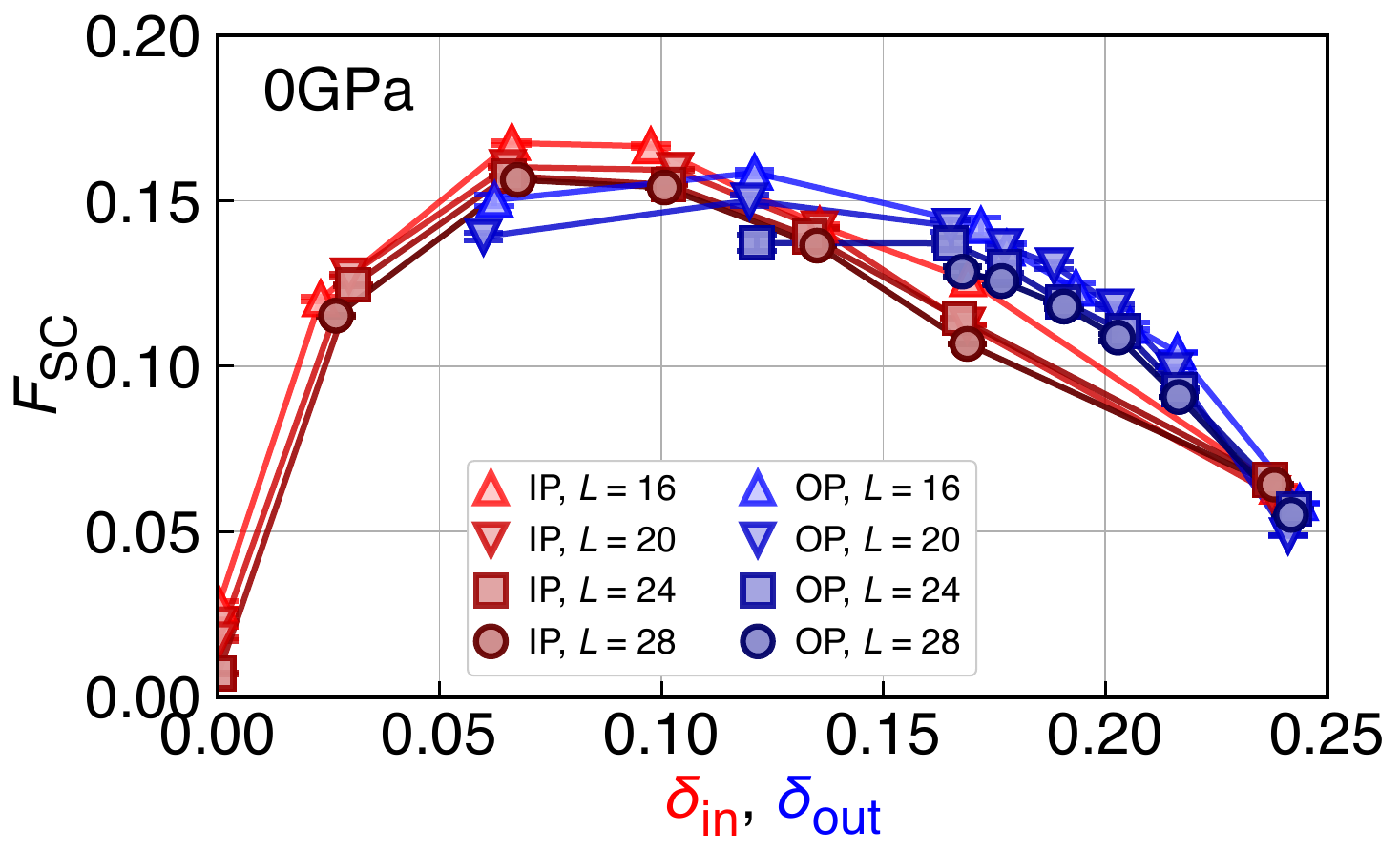}
\caption{%
Size-dependent SC order parameters $F_{\rm SC}$ for the IP and OP at ambient pressure
as a function of hole density $\delta_{\rm IP}$ and $\delta_{\rm OP}$
in each layer for several choices of system sizes. In this figure, $F_{\rm SC}$ is defined as the converged values of $\sqrt{P_d}$ 
at long distances before the size extrapolation.
}
\label{fig:FSC_delta_0GPa_each_layer}
\end{figure}

The IP and OP exhibit a similar doping dependence of the SC dome structure.
In fact, although the Hamiltonian parameters are similar between the IP and OP at ambient pressure except for the chemical potential, $F_{\rm SC}$ shows the maximum at $\delta_{\rm IP}=0.06$ for the IP while at $\delta_{\rm OP}=0.12$ for the OP if we plot as a function of the hole density in each layer as is seen in Fig.~\ref{fig:FSC_delta_0GPa_each_layer}. This shift between the IP and OP cannot be explained if the IP and OP are independent. On the other hand, the shift can be caused by the synchronization between the IP and OP as one sees as a function of the averaged density $\delta$ in Fig.~\ref{fig:FSC_delta_0GPa} suggesting a substantial interlayer proximity effect, which may also enhance the amplitude of $F_{\rm SC}$ at the IP relative to the OP because the IP is sandwiched by the two OPs.
To assess the role of parameter inhomogeneity at $0$\,GPa, we also set the OP parameters equal to the IP parameters and find nearly unchanged doping dependence of the SC order parameters, indicating that the inhomogeneity has little effect on them.
We emphasize that this interlayer proximity effect occurs near ambient pressure, where the Hamiltonian parameters for the IP and OP do not differ significantly.
Under pressure, however, the parameter difference between the two layers becomes more pronounced.
Indeed, at $60$\,GPa, the off-site Coulomb interactions in the IP become substantially stronger than those in the OP (see Table~\ref{tab:selected_parameters_0,30,60GPa}); therefore, the interlayer proximity effect is not strong enough to make the perfect synchronization.

\subsection{%
Coexistence of AF and SC
}
\label{subsec:coexist_af_sc}

The carrier density in the IP is much closer to the half filling
than that in the OP, which tends to stabilize AF order in the IP, while the OP stabilizes SC order. 
Therefore, if the interlayer coupling is switched off by keeping the self-doped hole density, one would expect the segregation and independent survival of the SC order in the OP and the AF order in the IP in an appropriate range of the averaged hole density, because the microscopic coexistence is not seen 
in the numerical result of the single-, double-, and infinite-layer cuprates~\cite{schmid2023}.
However, this segregation is not a microscopic coexistence, but an interlayer phase separation. If a small but nonzero tunneling is introduced between the segregated IP and OP, as in the realistic situation of the multilayer cuprates, the proximity may lead to the coexistence in both planes.
Here, we discuss such a possibility of the microscopic
and homogeneous
coexistence of AF and SC orders at ambient pressure.

To see a typical example, we focus on the case of $8$\%
doping on average.
When we use the \textit{ab initio} Hamiltonian for Hg1223 at ambient pressure, the IP becomes nearly half-filled by the self-doping and exhibits the AF order. 
However, in the underdoped region, the GW calculation tends to be less reliable due to the strong correlation effect of the precursor to the Mott insulator in the real compounds ignored in the GW calculation, which implies that the approach to the Mott insulator in the IP will be slower in reality in accordance with the experimental indications. To mimic this effect, we study the effect of $\mu$ reduced from the GW adjusted value, $\mu=4.41$ to $\mu=3.2$, which makes the difference in $\delta_{\rm IP}$ and $\delta_{\rm OP}$ smaller to $\delta_{\rm IP}=0.07$ and $\delta_{\rm OP}=0.085$ instead of $\delta_{\rm IP}=0$ and $\delta_{\rm OP}=0.12$ for $\mu=4.41$.
(For the detailed analyses regarding the dependence on the chemical potential difference between the IP and OPs, please see Appendix~\ref{App:analysis_coexist_af_sc}.)

\begin{figure}[!t]
\centering
\includegraphics[width=0.8\columnwidth]{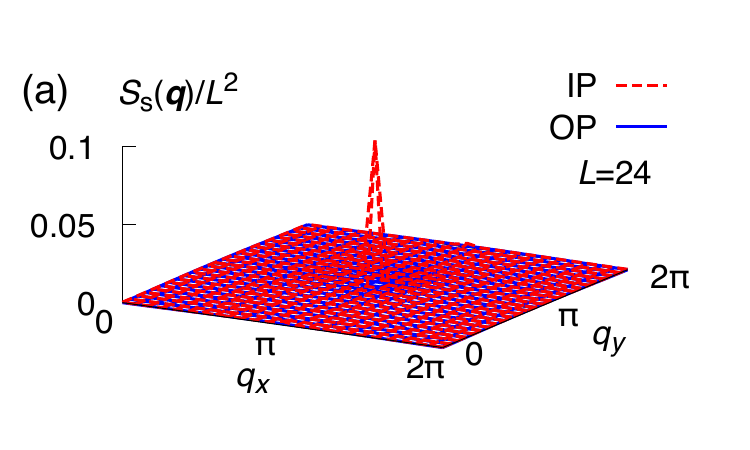}
\\
\includegraphics[width=0.8\columnwidth]{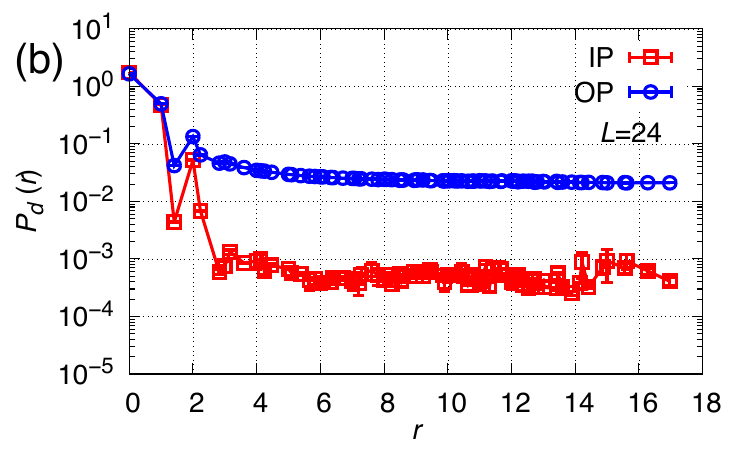}
\caption{%
Spin structure factors defined in Eq.~(\ref{eq:sf}) (a) and SC correlation defined in Eq.~(\ref{eq:scf}) (b) of the IP and OP for $24\times24$ lattice.
The chemical potential difference between the IP and OP is $\mu=3.2$,
which is decreased from the value $\mu=4.41$
for the Hg1223 \textit{ab initio} Hamiltonian at ambient pressure estimated from the GW calculation.
The spin structure shows a prominent peak at $\bm{Q}=(\pi,\pi)$ for the IP, which may grow to the Bragg peak in the thermodynamic limit,
whereas it shows no such peak for the OP.
Both the IP and OP exhibit the SC correlation saturated at long distances.
Because the AF correlation tends to degrade the SC correlation,
the saturated value of the SC correlation
is larger for the OP than for the IP. 
}
\label{fig:coexist_af_sc_correlation}
\end{figure}

\begin{figure}[!t]
\centering
\includegraphics[width=0.9\columnwidth]{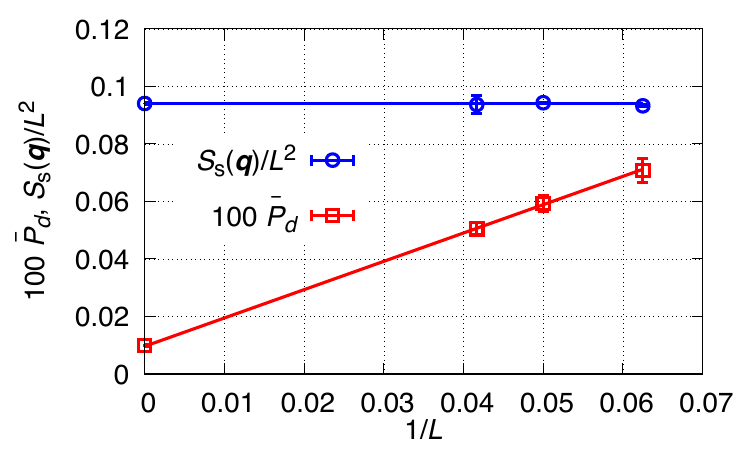}
\caption{%
Size extrapolation of $S_{\mathrm{s}}(\bm{q})/L^2$ and $\bar{P}_d$ defined in Eqs.~\eqref{eq:sf} and \eqref{eq:scfbar}, respectively, for the IP.
At chemical potential $\mu=3.2$ for $8$\% doping concentration,
the AF order defined in Eq.~\eqref{eq:msc} is
$m_{\mathrm{s}}=0.3067(6)$
while the SC order defined in Eq.~\eqref{eq:FSC} is
$F_{\mathrm{SC}}=0.0099(3)$,
supporting the coexistence of AF and SC orders.
}
\label{fig:coexist_af_sc_extrapolation}
\end{figure}

Figure~\ref{fig:coexist_af_sc_correlation} shows the SC and AF correlations in the OP and IP for a $24\times 24$ lattice. The size extrapolation to the thermodynamic limit in Fig.~\ref{fig:coexist_af_sc_extrapolation} indeed supports the coexistence.
(See also Fig.~\ref{fig:ene_cmp_sc_af_after_var_extrapolate}.
The coexisting state has the lowest energy below the uniform SC and uniform AF states. We have not studied the comparison with possible long-period stripe states, which is left for future studies.) The result supports the coexistence at the inner layer in a certain range of parameters by the mechanism of proximity from the robust SC order in the outer layer that lowers the energy from a uniform non-SC AF state.
The coexistence of the AF and SC in multilayer cuprates is overall consistent with experiments~\cite{mukuda2012,kurokawa2023,ideta2025}.

\section{Results under Pressure}
\label{sec:pressure}

In this section, we show the solutions for the ground state under pressure, especially the pressure dependence of the SC order parameter and  $T_{\mathrm{c}}$.

\subsection{%
Pressure dependence of SC order
}
\label{subsec:FSC_pressure}

\begin{figure}[!bt]
\centering
\includegraphics[width=0.8\columnwidth]{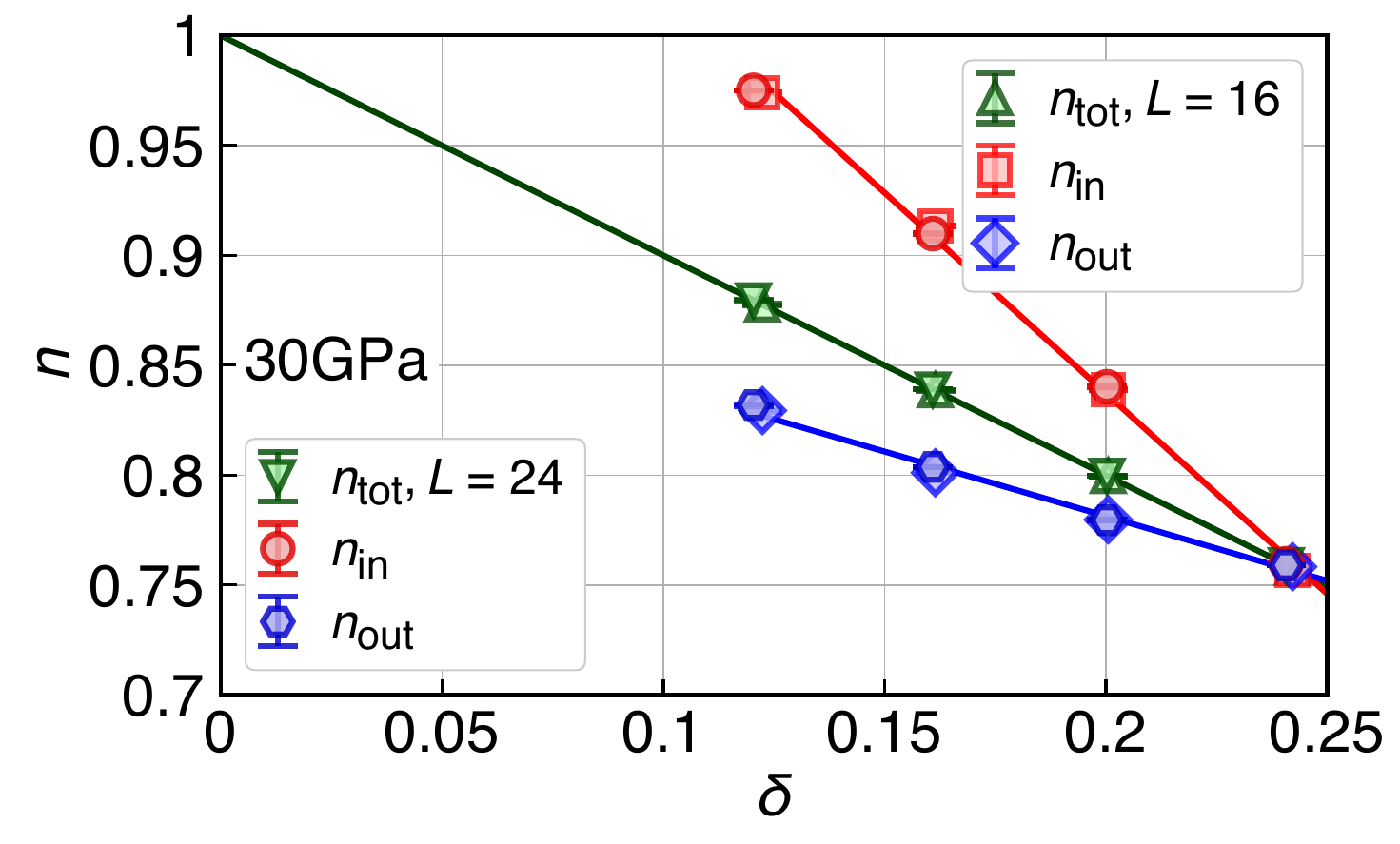}\\
\includegraphics[width=0.8\columnwidth]{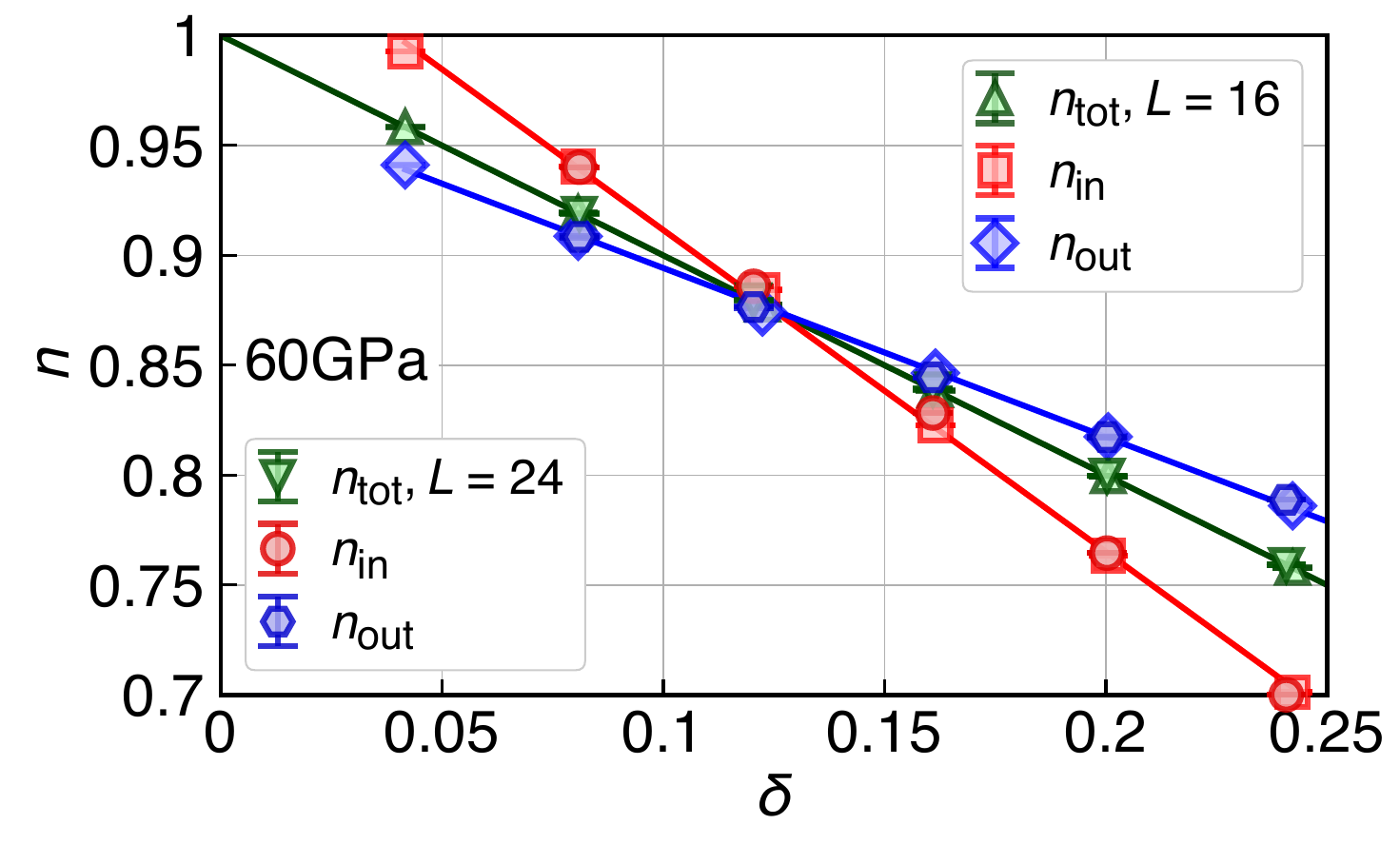}
\caption{%
Inner- and outer-layer hole densities as a function of
total hole density $\delta$ for the \textit{ab initio} Hamiltonian of Hg1223
at $30$\,GPa (a) and $60$\,GPa (b). The comparison between the results at $16 \times 16$ and $24 \times 24$ lattices demonstrates that the size dependence is negligible. 
}
\label{fig:self-doping_3060GPa}
\end{figure}

\begin{figure}[!bt]
\centering
\includegraphics[width=0.8\columnwidth]{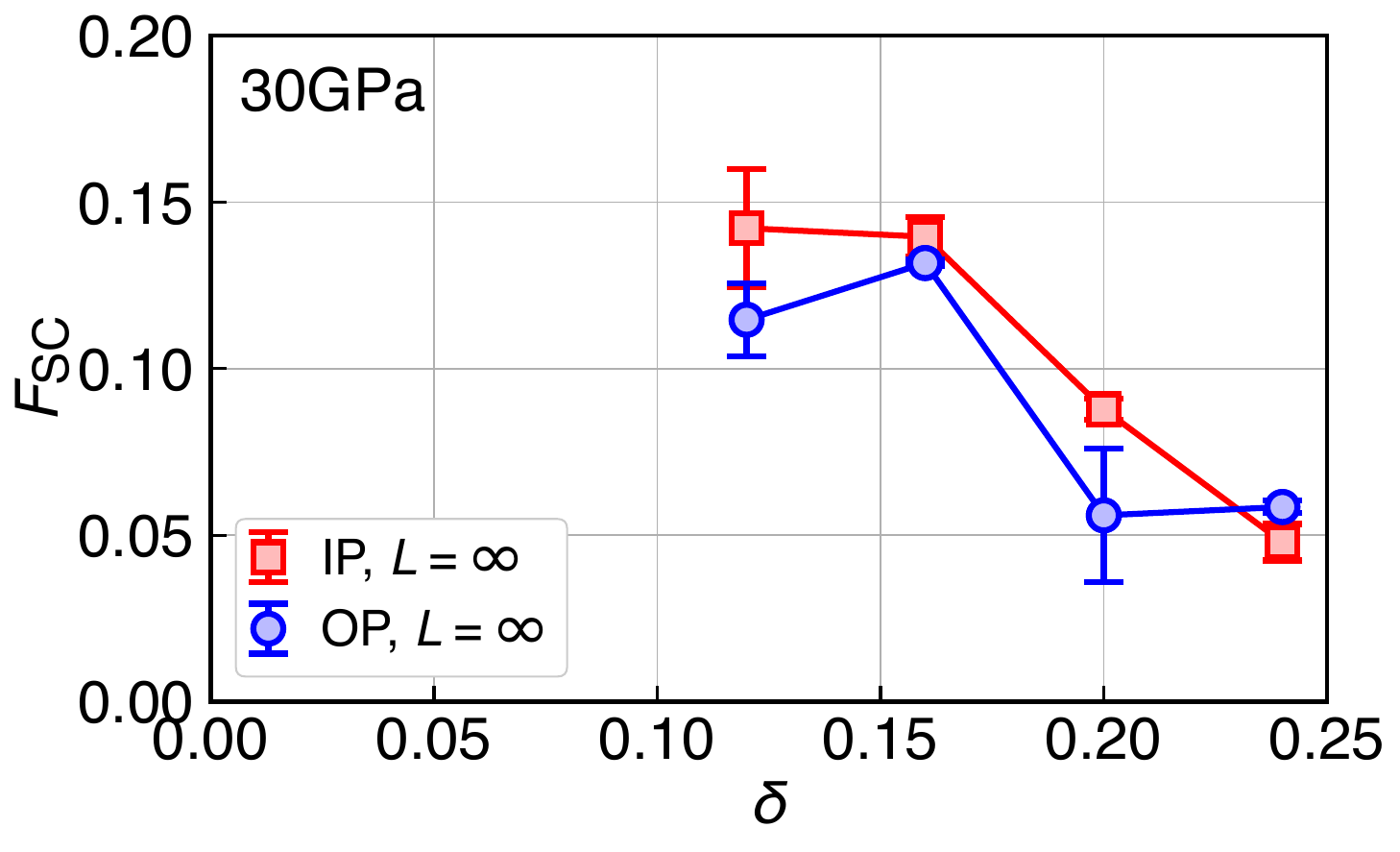}\\
\includegraphics[width=0.8\columnwidth]{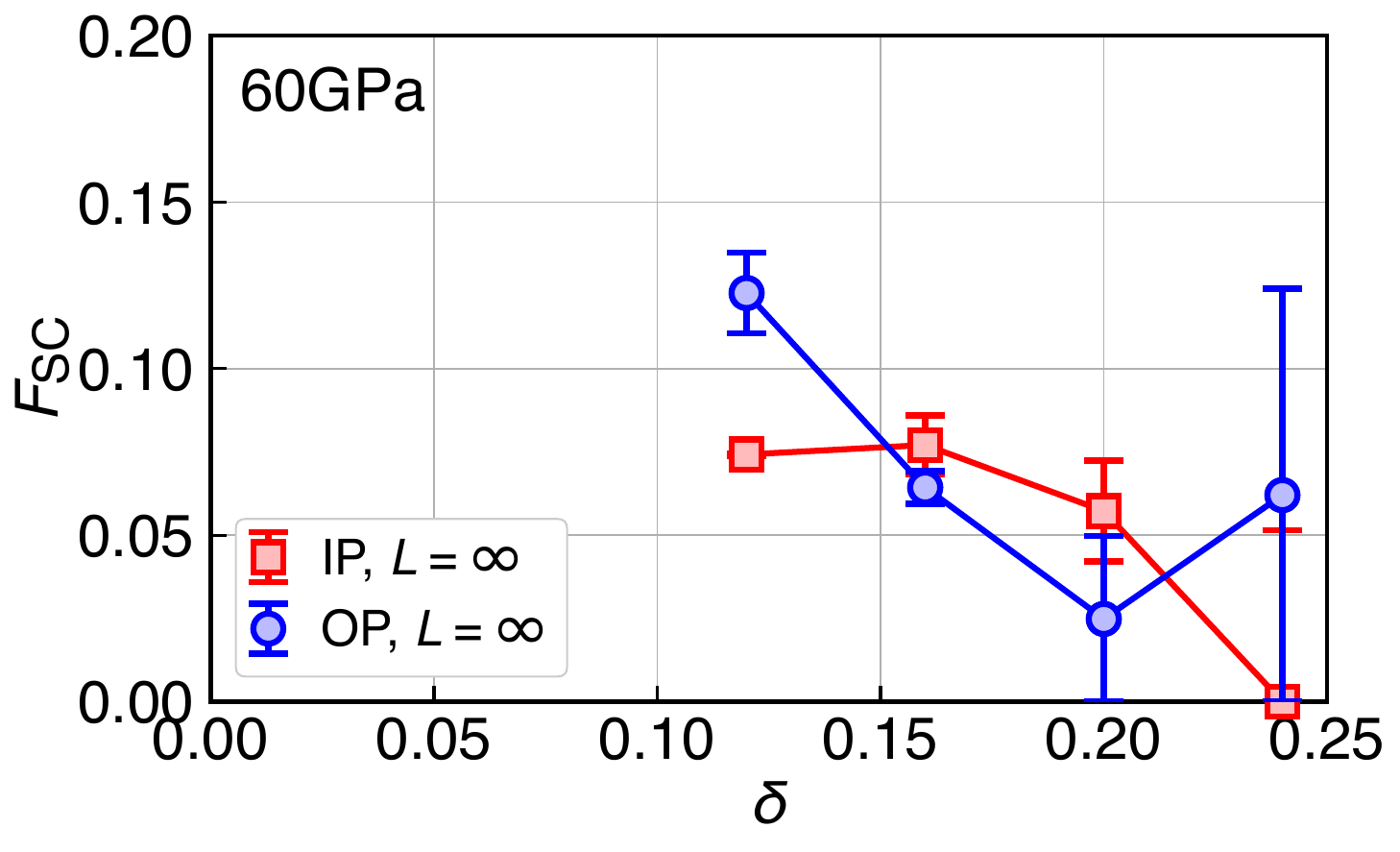}
\caption{%
Doping concentration dependence of the SC order parameter $F_{\rm SC}$ at $30$\,GPa (a) and $60$\,GPa (b) after the size extrapolation to the thermodynamic limit.
The red and blue lines are for the IP and OP, respectively.
}
\label{fig:FSC_delta_3060GPa}
\end{figure}

\begin{figure}[!bt]
\centering
\includegraphics[width=0.8\columnwidth]{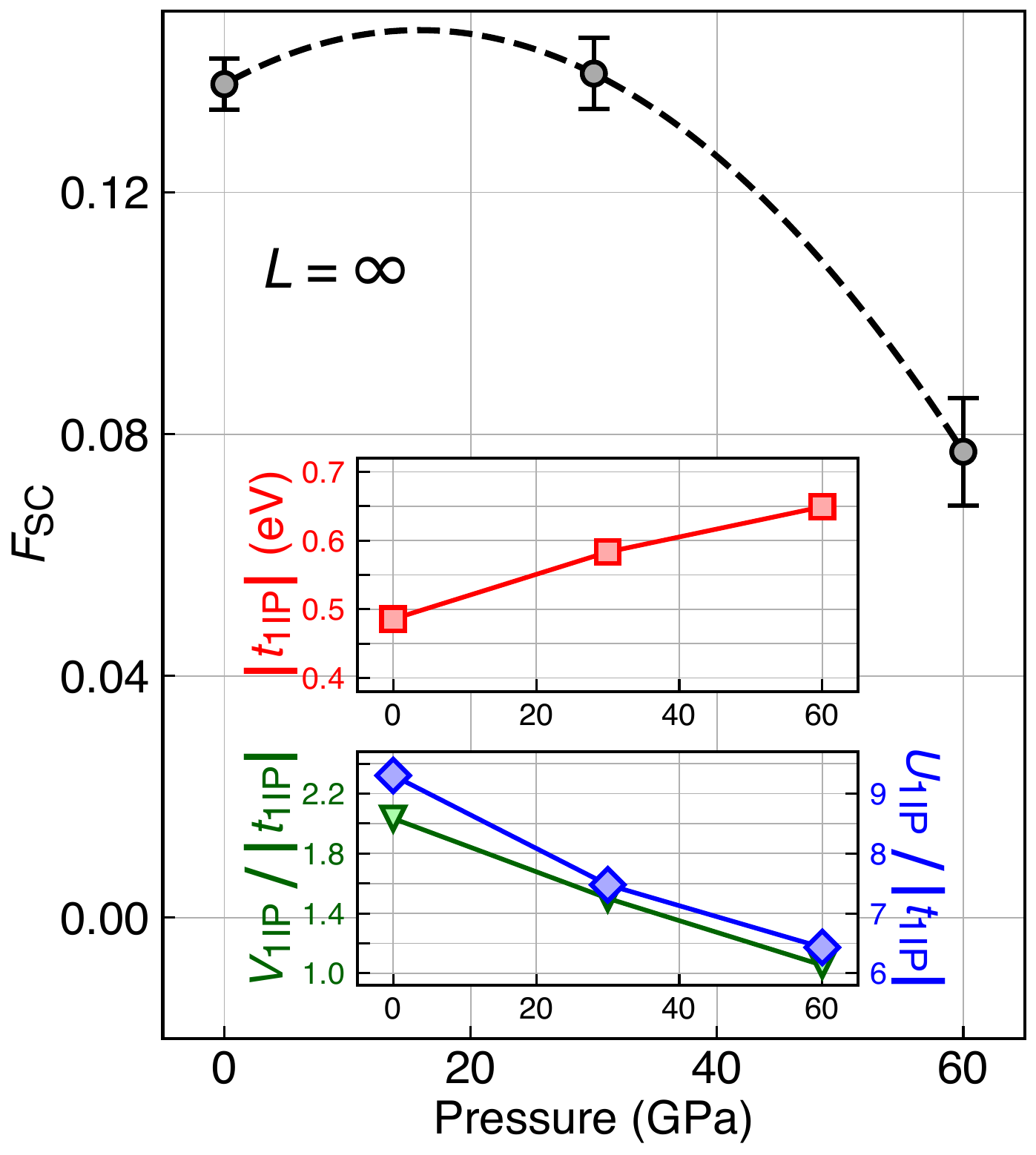}
\caption{%
Pressure dependence of the SC order parameter ($F_\mathrm{SC}$) in the thermodynamic limit after the size extrapolation
for the IP at $\delta=0.16$ of Hg1223. The black dashed curve is a quadratic fitting.
Inset: Pressure dependence of
the nearest-neighbor hopping parameter ($|t_{1\mathrm{IP}}|$)
and the ratio of the onsite Coulomb interaction parameter to
the nearest-neighbor hopping parameter ($U_{\mathrm{IP}}/|t_{1\mathrm{IP}}|$) as well as the ratio of the nearest-neighbor Coulomb interaction parameter to
the nearest-neighbor hopping parameter ($V_{1\mathrm{IP}}/|t_{1\mathrm{IP}}|$),
taken from Ref.~[\onlinecite{moree2024}] and Table~\ref{tab:selected_parameters_0,30,60GPa}.
In this plot,
all the hopping and Coulomb interaction parameters are presented for the IP, where the larger SC order is observed at $\delta=0.16$.
Error bars are estimated from the size extrapolation errors.
}
\label{fig:fsc_t1}
\end{figure}

Figure~\ref{fig:self-doping_3060GPa}
illustrates the $\delta$ dependence of $n_{\rm IP}=1-\delta_{\rm IP}$ and  $n_{\rm OP}=1-\delta_{\rm OP}$ while
Fig.~\ref{fig:FSC_delta_3060GPa}
shows the SC order parameter in the thermodynamic limit,  $F_{\rm SC}$ in the IP and OP at $30$ and $60$\,GPa,
which confirms the dominance of the SC order for $\delta>0.12$
even under pressure. For detailed data for finite sizes and their size extrapolations, see Appendices~\ref{App:details_sc_correlation} and \ref{App:size_extrapolate_sc}. Both at $30$\,GPa and $60$\,GPa, $F_{\rm SC}$ is comparable between the IP and OP while in most cases, the IP shows slightly larger $F_{\rm SC}$. In Fig.~\ref{fig:fsc_t1}, we plot $F_{\rm SC}$ for all the pressures at the hole density $\delta=0.16$, which is the optimal value at ambient pressure. Since $F_{\rm SC}$ is always larger for the IP than for the OP at $\delta=0.16$, we plot only the values for the IP. 
The SC order parameter $F_\mathrm{SC}$ (black circles)
remains nearly constant from $0$\,GPa to $30$\,GPa
and decreases from $30$\,GPa to $60$\,GPa.
We analyze the origin of this trend below.
In the inset of Fig.~\ref{fig:fsc_t1}, the pressure dependence of the Hamiltonian parameters $|t_1|$, $U/|t_1|$ and $V_1/|t_1|$ at the IP are plotted by taking the data from Tables~\ref{tab:parameters_0GPa}, \ref{tab:parameters_30GPa} and \ref{tab:parameters_60GPa}. From $0$ to $60$\,GPa, $|t_1|$ increases, while $U/|t_1|$ and $V_1/|t_1|$ decrease significantly. In particular,  $V_1/|t_1|$ decreases to less than a half.

\begin{figure}[!t]
\centering
\includegraphics[width=0.8\columnwidth]{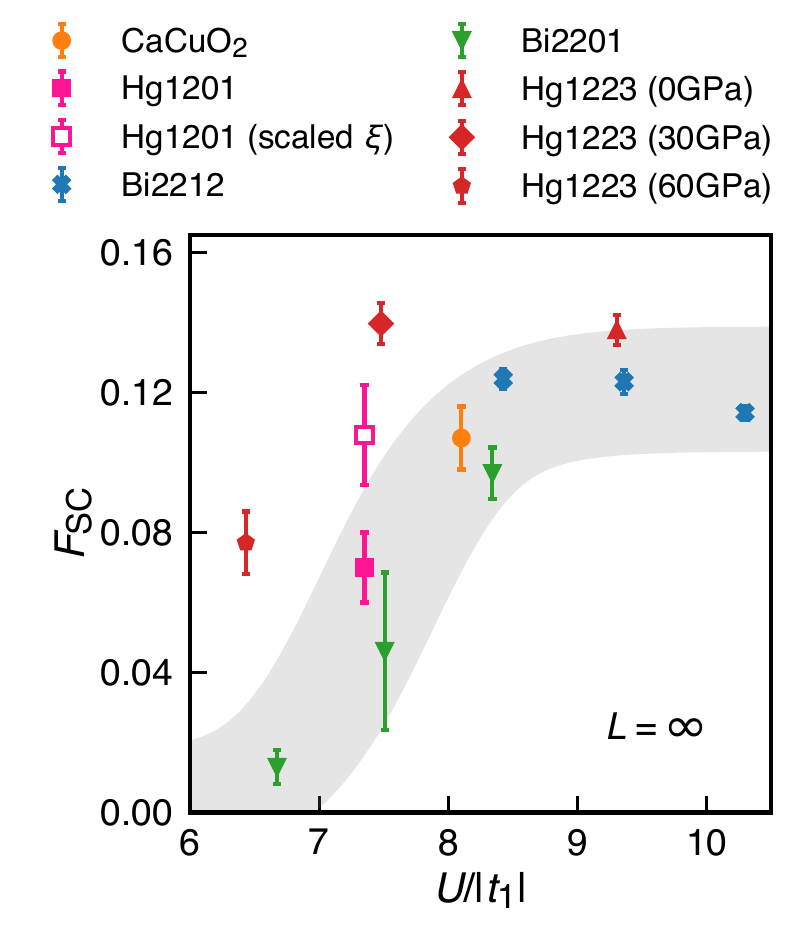}
\caption{%
Comparison of the SC order parameter in $U/|t_1|$ dependence among cuprate SCs in the thermodynamic limit after the size extrapolation.
For Hg1223, the abscissa is the ratio, $U_{\mathrm{IP}}/|t_{1\mathrm{IP}}|$ at the IP.
The data for CaCuO$_2$, Hg1201, Bi2212 and Bi2201 at $\delta\sim 0.15$ are taken from Ref.~[\onlinecite{schmid2023}]. The plot for ``Hg1201 (scaled $\xi$)'' (open magenta square) is obtained by calculating the ground state for the Hg1201 single-layer Hamiltonian (filled magenta square) with reduced off-site interaction $V_i$ by a factor $\xi=0.686$. 
}
\label{fig:compare_cuprate_Fsc}
\end{figure}

We plot $F_{\rm SC}$ at $\delta=0.16$ in Fig.~\ref{fig:compare_cuprate_Fsc} 
for four choices of pressures of Hg1223 at the corresponding values of $U/|t_1|$. We also plot $F_{\rm SC}$ for the infinite-, single-, and double-layer compounds, CaCuO$_2$, Hg1201, Bi2201, and Bi2212 at the optimal doping and ambient pressure taken from Ref.~[\onlinecite{schmid2023}] together.
The data for Hg1223 under pressure clearly deviate from the universal trend for the single-, double-, and infinite-layer compounds at ambient pressure grouped by the shaded belt. 
This implies that $F_{\rm SC}$ is not solely determined by $U/|t_1|$ under pressure. In fact, although the off-site Coulomb interaction is similar among various cuprates at ambient pressure (for instance, the nearest-neighbor Coulomb interaction $V_1\sim 1$\,eV,  see Table~\ref{tab:t1_U_V1_each_compound}), it is decreased substantially under pressure (for instance, $V_1\sim 0.88$\,eV and $0.68$\,eV at $30$\,GPa and $60$\,GPa, respectively) 
(see Appendix~\ref{App:parameters} for details). This reduction is along the same line as the reduction of $U$ under pressure ($U$ for Hg1223 is
$\sim 4.52$, $4.36$, and $4.18$\,eV
%
%
%
at $0$, $30$, and $60$\,GPa, respectively), which is ascribed to the increased screening from the block layers and apical oxygens caused by the shrunk lattice constant under pressure. It shares a common feature observed before in quasi 2D systems~\cite{nakamura2012}, where the reduction of Coulomb interaction caused by the off-plane screening is similar irrespective of the mutual distance (namely, regardless of whether $U, V_1, V_2, ...$)  (and hence the reduction is relatively more significant for larger mutual distance). 

The large enhancement of $F_\mathrm{SC}$ by the reduction of the off-site interaction was actually pointed out in Ref.~[\onlinecite{schmid2023}], and the present result is consistent with it. The gradual pressure dependence of $F_{\rm SC}$ for Hg1223 instead of the steeper reduction in the general trend at ambient pressure is now understood by the interplay of reduced $U$ and $V_1$ (and reduced $V_i$ with $i>2$ as well), where the reduction of $F_{\rm SC}$ by reduced $U/|t_1|$ is partially compensated by a large reduction of the off-site interactions.

To corroborate our understanding of the effect of the reduced off-site interaction under pressure, we calculated for a hypothetical system of modified single-layer Hg1201, where we reduced only the off-site interactions ($V_i, i\geq 1$) from the \textit{ab initio} values of Hg1201 with the factor $\xi=0.686$ as $\xi V_i$ at nearly $15$\% doping concentration ($\delta=0.146$). Here, this factor $\xi$ mimics $V_i/|t_1|$ for pressurized Hg1223 that has $U/|t_1|$ corresponding to Hg1201. The result is plotted as a red open square in Fig.~\ref{fig:compare_cuprate_Fsc}, 
which indeed fits the trend of Hg1223 under pressure. In the multilayer systems, there seems to be an additional cooperative effect from the OP and IP through the proximity to further enhance the SC.

This suggests the importance of controlling the off-site interaction in material design, as is already pointed out in hypothetical parameter search~\cite{schmid2023}. The proposal in Ref.~[\onlinecite{schmid2023}] indeed becomes a reality here as the pressure effect, which works to reduce the effective interaction in the relative ratio to the values for ambient pressure more efficiently for the off-site part than for the onsite one, which helps enhance the superconductivity in Hg1223.
If $U$ could be retained under pressure simultaneously with reduced off-site interaction, $F_{\rm SC}$ could be much more enhanced as we will discuss below.

\subsection{%
Pressure dependence of $T_{\mathrm{c}}$}
\label{sec:Tc_pressure}

It was proposed that $T_{\mathrm{c}}$ at the optimum hole density can be estimated by using the \textit{ab initio} result in the scaling form
\begin{align}
\label{eq:scaling_form}
 T_{\mathrm{c}}\sim 0.16|t_1|F_{\rm SC}    
\end{align}
for the single-, double-, and infinite-layer compounds at ambient pressure~\cite{schmid2023}. Since $F_{\rm SC}$ of Hg1223 under pressure does not follow the common trend of cuprates at ambient pressure and the triple-layer structure contains non-equivalent IP and OP, it is
highly nontrivial whether this scaling  form is also valid in the triple-layer Hg1223
under pressure.

\begin{figure}[!t]
\centering
\includegraphics[width=0.9\columnwidth]{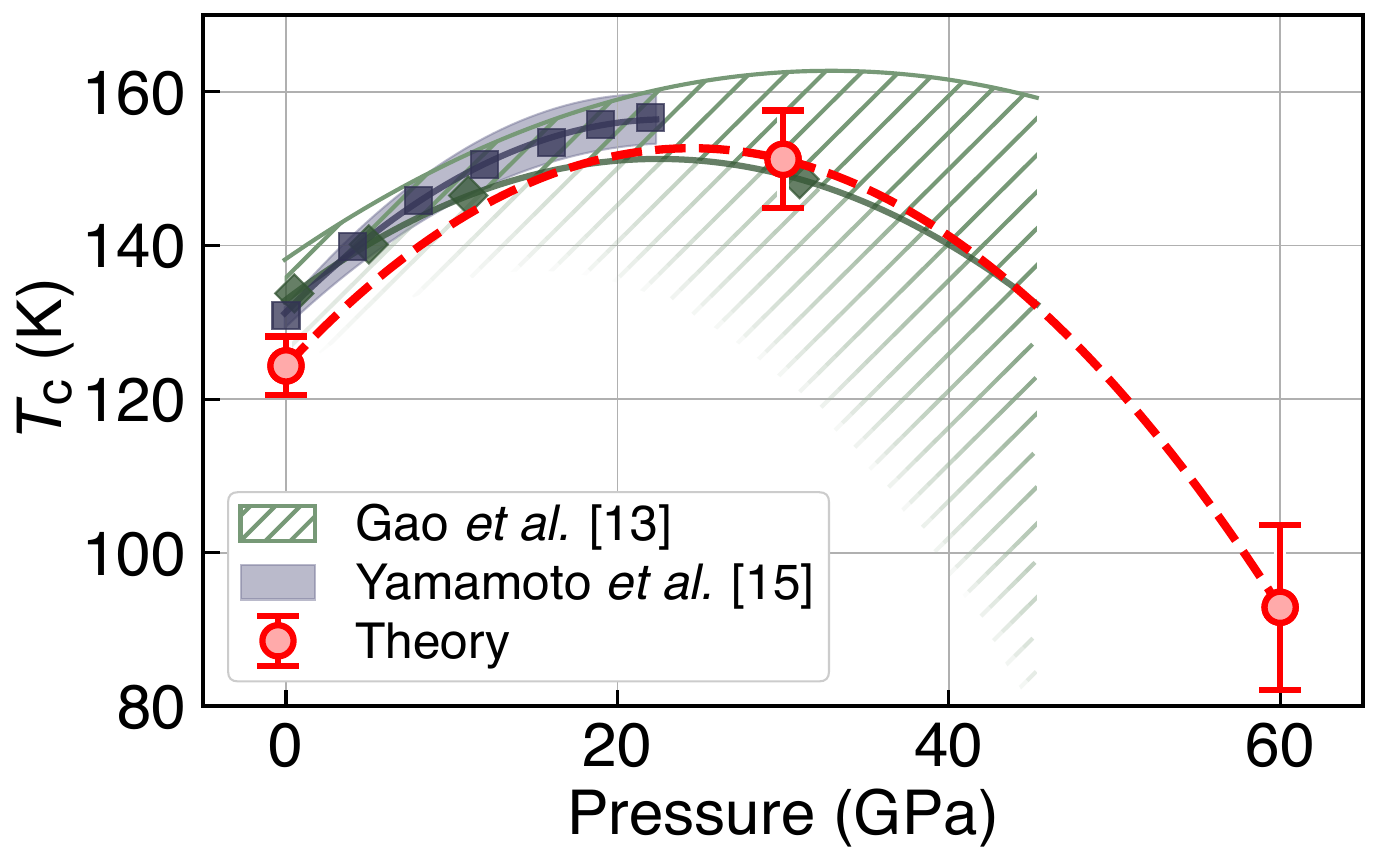}
\caption{%
Comparison of the pressure dependence of $T_{\mathrm{c}}$ between theory and experiment for Hg1223.
We estimate the theoretical $T_{\mathrm{c}}$ (red circles)
at $\delta=0.16$
by using Eq.~\eqref{eq:scaling_form}
for the IP.
The red dashed curve is a guide to the eye obtained by a quadratic fit of the theoretical $T_{\mathrm{c}}$.
Experimental $T_{\mathrm{c}}$ contains uncertainty.
The upper bound of the uncertainty is the onset temperature (onset $T_{\mathrm{c}}$) at which the resistivity $\rho$ starts deviating downwards. The lower bound should, in principle, be the temperature at which the resistivity becomes zero. The trajectory of the inflection point defined by $d^2\rho/dT^2=0$ is widely used as another metric of $T_{\mathrm{c}}$. We show two examples of experimental reports.
The gray shaded belt depicts the experimental $T_{\mathrm{c}}$ at $\sim 16$\% doping concentration reported in Ref.~[\onlinecite{yamamoto2015}] with the upper and lower bounds being the onset and zero-resistivity temperature, respectively.
The black curve with squares shows the inflection-point trajectory.
The greenish hatched belt is from Ref.~[\onlinecite{gao1994}], where the upper bound is the onset and the lower bound drawn as the green thick curve with diamonds represents the inflection-point trajectory. In Ref.~[\onlinecite{gao1994}], the zero resistivity was not reached even at the reported lowest temperature $\sim 80$\,K.
Therefore, the true $T_{\mathrm{c}}$ could be lower than the lower bound of the greenish hatched zone by considering possible non-uniformity and inhomogeneity of pressure in terms of hydrostatic nature above $30$\,GPa (see Sec.~\ref{sec:Tc_pressure}).
The theoretical $T_{\mathrm{c}}$
shows a peak at around $30$\,GPa,
consistently with the experimental $T_{\mathrm{c}}$.
}
\label{fig:tc}
\end{figure}
To test the validity, in Fig.~\ref{fig:tc}, we plot the pressure dependence of $T_{\mathrm{c}}$ estimated by using Eq.~\eqref{eq:scaling_form} together with $t_1$ and the calculated $F_{\rm SC}$ at the IP responsible for the higher $T_{\mathrm{c}}$ and compare it with the experimental
$T_{\mathrm{c}}$
reported in Refs.~[\onlinecite{gao1994}], [\onlinecite{yamamoto2015}], and [\onlinecite{jover1996}].
The experimental and numerical values of $T_{\mathrm{c}}$ agree fairly well at $0$ and $30$\,GPa and they are maximized around $30$\,GPa,
which supports the validity of the scaling form (\ref{eq:scaling_form}) under pressure for the triple-layer compound despite the large deviation of $F_{\rm SC}$ from the trend in the single-, double-, and infinite-layer compounds at ambient pressure.

Note that beyond $30$\,GPa, $T_{\mathrm{c}}$ in the experimental report is largely uncertain, because only one report exists~\cite{gao1994}, where only the onset temperature for the superconductivity is available, and the zero resistivity was not reached at any temperature, implying the inhomogeneity in samples or possible non-uniformity of hydrostatic pressure.

\begin{figure*}[!bt]
\centering
\includegraphics[width=1.3\columnwidth]{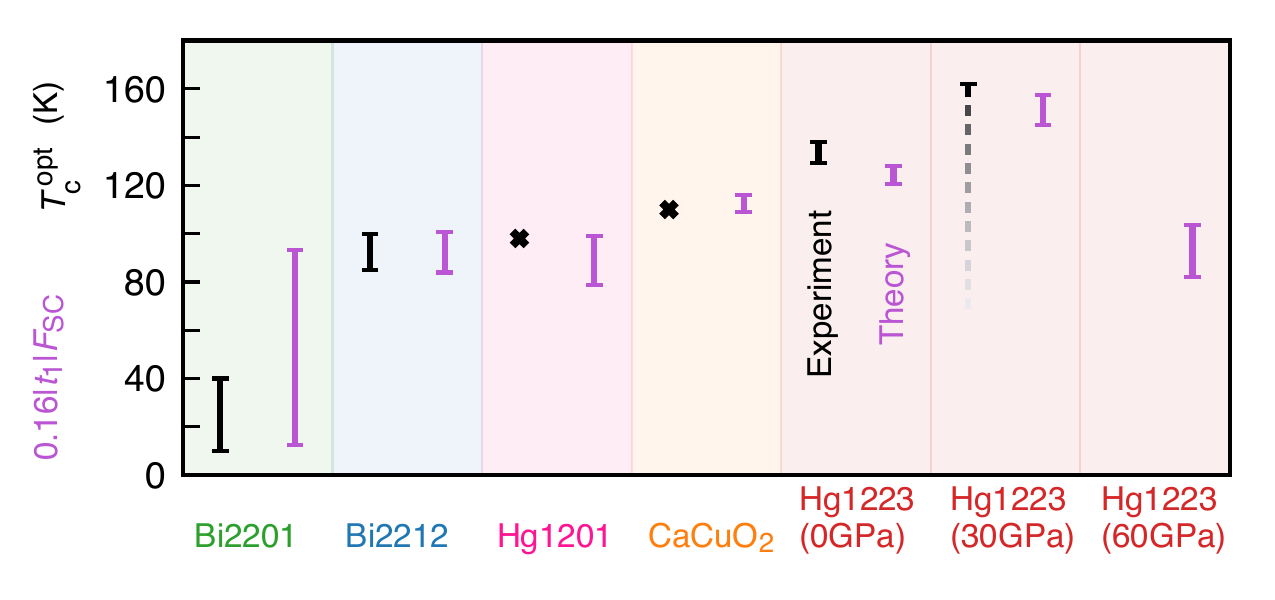}
\caption{%
Materials dependence of SC critical temperature $T_{\mathrm{c}}$: Comparison between the present \textit{ab initio} predictions and experimental observations for triple-layer Hg1223 at ambient and applied pressures is plotted together with single-, double-, and infinite-layer compounds~\cite{schmid2023}. The black bar in the left row for each case indicates experimental optimum $T_{\mathrm{c}}$.
Theoretical estimates are depicted in each right row based on the \textit{ab initio} calculations
with the help of Eq.~\eqref{eq:scaling_form} using calculated $F_{\mathrm{SC}}$ after the size extrapolation to the thermodynamic limit.
The dashed line for Hg1223 ($30$\,GPa) represents an ambiguity in experimental $T_{\mathrm{c}}$ (see Secs.~\ref{sec:Tc_pressure} and \ref{sec:discussion} for more details).
No experimental value of $T_{\mathrm{c}}$ has been reported for Hg1223 at $60$\,GPa.
}
\label{fig:compare_cuprate_Tc}
\end{figure*}
The origin of the $T_{\mathrm{c}}$
peak structure in Hg1223 under
pressure is now understood from the scaling form \eqref{eq:scaling_form}:
With increasing pressure,
the SC order parameter $F_{\rm SC}$ is nearly constant up to $30$\,GPa and then decreases above $30$\,GPa,
whereas the nearest-neighbor hopping parameter $|t_1|$ monotonically
increases. To keep the plateau of  $F_{\rm SC}$, the reduction of off-site Coulomb repulsion plays an important role.
Then the pressure dependence of $T_{\mathrm{c}}$ exhibits a 
maximum
in the intermediate pressure region near $30$\,GPa,
as shown in Fig.~\ref{fig:tc},
because of the interplay between $F_{\rm SC}$ and $|t_1|$.

In Fig.~\ref{fig:compare_cuprate_Tc}, we show the comparison between the experimental $T_{\mathrm{c}}$ and the present estimates for CaCuO$_2$, Hg1201, Bi2201, Bi2212 and Hg1223 at ambient pressure as well as Hg1223 under pressure. The \textit{ab initio} estimates essentially reproduce all the experimental indications. In particular, it captures the experimental trend of the higher $T_{\mathrm{c}}$ of Hg1223 in comparison to others.

\section{%
Effective attraction 
}
\label{sec:effective_attraction}

To understand the mechanism of superconductivity, it is crucially important to uncover the origin of the effective attraction that causes the Cooper pairing. In this section, we report a detailed analysis obtained from the \textit{ab initio} calculations, which should be relevant to both Hg1223 and other cuprate compounds. 

In general, effective static interaction of particles can be estimated from the energy required to add a particle to the system. Namely, the effective interaction energy between the added particle and the particles already existing in the system is measured by the shift of the chemical potential from an $N$-particle system $\mu(N)$ to that of an $N+1$-particle system $\mu(N+1)$, in other words, $\mu(N+1)-\mu(N)\sim\partial\mu/\partial n$. Since $\mu=\frac{\partial E}{\partial n}$ with $n$ and $E$ being the particle density and the energy density, respectively, the effective interaction is given by $\frac{\partial^2 E}{\partial n^2}$. Partial interaction energy can also be defined by decomposing the total energy density $E$ into the sum of partial energies by following the proposal in Ref.~[\onlinecite{schmid2023}]. Here, by decomposing Eq.~(\ref{eq:ham}) into Eqs.~(\ref{eq:ham_ht})-(\ref{eq:ham_hmu}), we focus on the local interaction, $\frac{\partial^2 E_U}{\partial \delta^2}$, where $E_U=\langle {\mathcal H}_U\rangle$ denotes the ground state average of the local interaction.

To estimate the effective interaction at $\delta=\delta_m$ practically,
we fit the onsite Coulomb energy $E_U(\delta)$
by the quadratic function
\begin{align}
 E_U(\delta) = c_0 + c_1 (\delta-\delta_m) + c_2 (\delta-\delta_m)^2,
\end{align}
where $c_0$, $c_1$, and $c_2$ are the fitting parameters.
The coefficient $c_2$ is the effective interaction, and we rewrite it in terms of the effective attraction $g\equiv -c_2$ hereafter.
We typically consider 
$E_U$ at five discrete and adjacent hole densities, say, $\delta_1<\delta_2<\delta_3<\delta_4<\delta_5$ and assign the quadratic fitting parameter $c_2$ around $\delta_m=\delta_3$  as the interaction strength.

\begin{figure}[!t]
\centering
\includegraphics[width=\columnwidth]{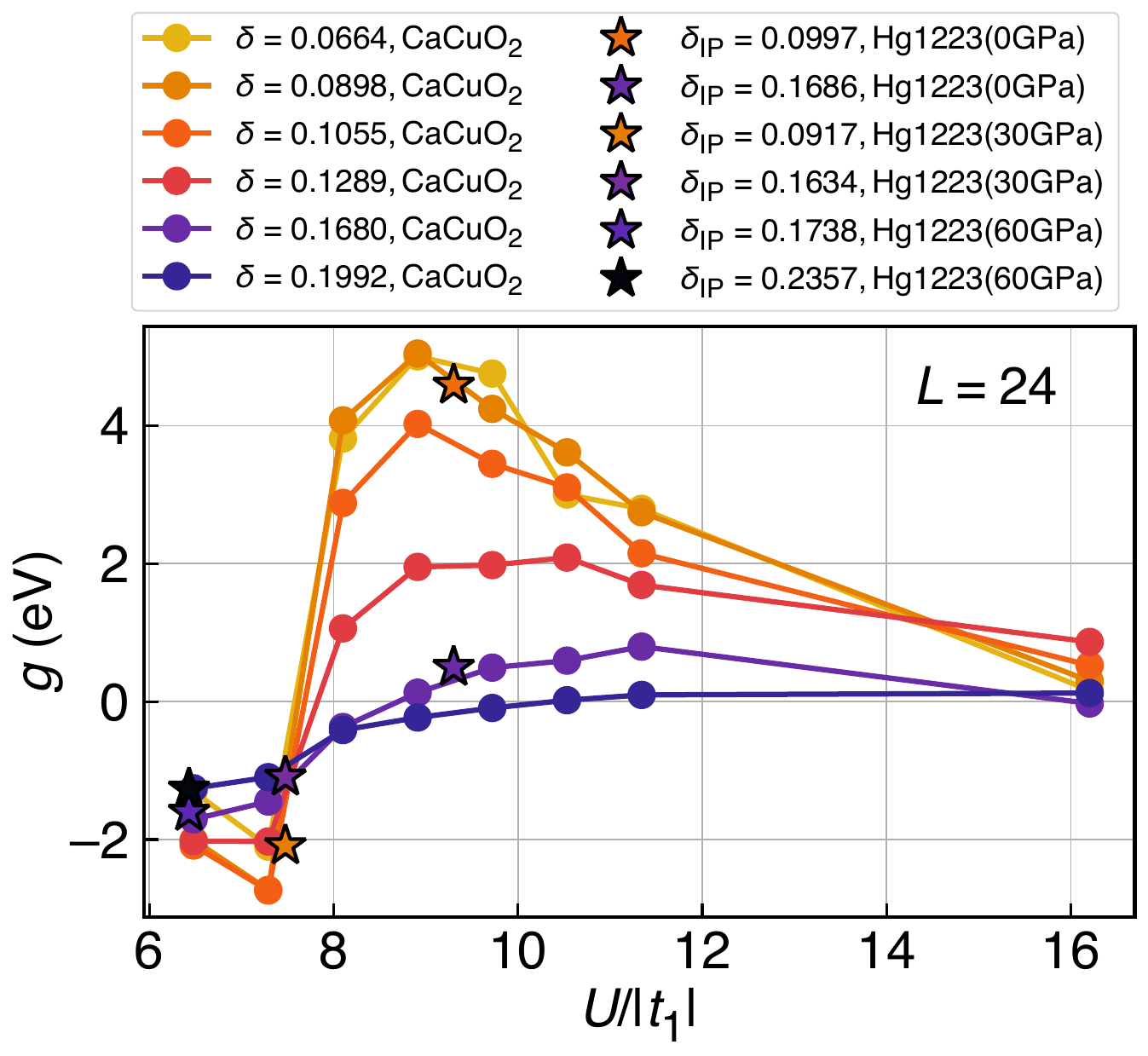}
\caption{%
Coulomb interaction dependence of the estimated attraction $g$ at each doping.
All $g$'s are extracted from data for $L=24$.
The symbol color distinguishes the doping concentration ($\delta$).
For Hg1223, we estimate $g$ from the onsite Coulomb energy $E_U$ of three contiguous doping concentrations in the IP, as shown in Fig.~\ref{fig:EU_vs_delta_cacuo2_hg1223}(b). The corresponding doping concentration $\delta_{\mathrm{IP}}$ is given by the average of the three contiguous doping concentrations.
}
\label{fig:c2_vs_u}
\end{figure}

\begin{figure}[!t]
\centering
\includegraphics[width=\columnwidth]{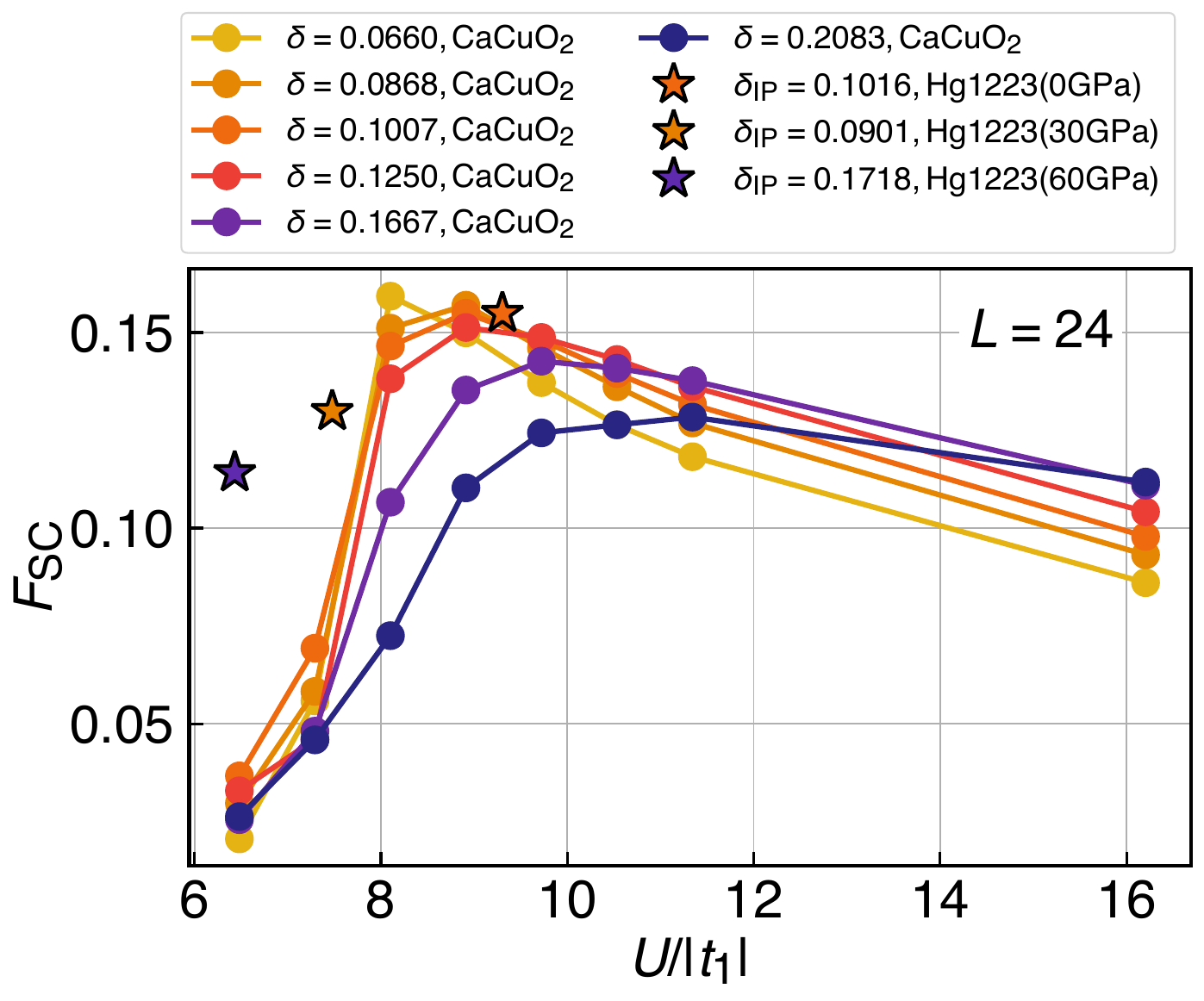}
\caption{%
Coulomb interaction dependence of the SC order parameter $F_{\mathrm{SC}}$ at each doping.
For Hg1223, we show $F_{\mathrm{SC}}$ for a finite-size system ($L=24$) at the optimal doping concentration ($\delta=0.16$).
The corresponding parameters, $U/|t_1|$ and $\delta_{\mathrm{IP}}$ are for the IP.
}
\label{fig:Fsc_vs_u}
\end{figure}

We show $U/|t_1|$ 
dependence of the effective attraction $g$
and the SC order parameter $F_{\rm SC}$ for a number of choices of $\delta$
in Fig.~\ref{fig:c2_vs_u} and~\ref{fig:Fsc_vs_u}, respectively, for Hg1223 at $0$, $30$ and $60$\,GPa. 
In this analysis, $E_U$ is further decomposed into the IP and OP contributions, and we choose the curvature of either the IP or OP that shows the stronger SC order parameter.
As references, we also show  
$g$
and $F_{\rm SC}$ calculated for hypothetical Hamiltonians modified from that of the hole-doped CaCuO$_2$ only by changing the $U$ value by hand from the \textit{ab initio} value $U=4.22$\,eV ($U/|t_1|=8.1$).  

In the case of CaCuO$_2$ and its derivatives,
the effective attraction $g$ shown in Fig.~\ref{fig:c2_vs_u} is enhanced in underdoped regions
and is maximized at the interaction $U/|t_1| \sim 9$ in the realistic range $8\leq U/|t_1|\leq 12$.
The attraction is gradually reduced with increasing doping.
The attraction $g$ for the triple-layer cuprate, Hg1223, in overall trend,
agrees well with that of CaCuO$_2$ at the corresponding values of $U/|t_1|$ and $\delta$.
This means that the effective attraction generated by the onsite
repulsive Coulomb interaction emerges in the same way
in all the single-, double-, triple-, and infinite-layer cuprates.

The SC order parameter $F_{\rm SC}$ is shown in Fig.~\ref{fig:Fsc_vs_u}
for the Hamiltonians for CaCuO$_2$, and its derivatives, which
also shows the maxima at the interaction $U/|t_1| \sim
9$,
just as in the case of $g$. Furthermore, the peak structure is asymmetric around the peak in both $g$ and $F_{\rm SC}$:
The tail is steeper for the weak coupling side $U/|t_1|<9$,
whereas it is broader for the strong-coupling side $U/|t_1|>9$.
The SC order parameter of the triple-layer cuprate, Hg1223,
follows the trend of CaCuO$_2$,
although $F_{\rm SC}$ is higher at higher pressure
thanks to the reduced off-site Coulomb interaction in Hg1223 as is discussed in Sec.~\ref{subsec:FSC_pressure}. 
The overall similar behavior between $g$ and $F_{\rm SC}$ in $U/|t_1|$ and $\delta$ dependencies with a prominent peak around $U/|t_1| \sim 9$ and asymmetric $U/|t_1|$ dependence around the peak strongly supports that the local attraction  $g$ is indeed the origin of the SC order. 

However, there exist two important different trends between $g$ and $F_{\rm SC}$. 1) One is the hole density dependence: $g$ increases strongly with decreasing hole density, while $F_{\rm SC}$ shows much smaller $\delta$ dependence. 
2) The other is that 
$F_{\rm SC}$ shows a clear deviation for Hg1223 under pressure from the trend of modified CaCuO$_2$, namely, substantially enhanced  $F_{\rm SC}$ for Hg1223 at $30$ and $60$\,GPa than the modified CaCuO$_2$ is observed, as is already ascribed to the effect of off-site interaction in Sec.~\ref{subsec:FSC_pressure}. On the other hand, $g$ does not show such an enhancement for Hg1223 under pressure. These discrepancies give us insights into the nature of superconductivity.

On point 1), we note that the attractive interaction induces the SC order, but they are not the same. In fact, it is known experimentally that the SC gap monotonically increases with decreasing hole concentration $\delta$ even in the underdoped region while $T_{\mathrm{c}}$ has a dome structure and decreases to zero toward the Mott insulator~\cite{alldredge2008}. This detachment has the same origin as the difference between $g$ and $F_{\rm SC}$. 
Namely, at least in the underdoped region, $F_{\rm SC}$ must decrease with the carrier density~\cite{sakai2018} and must vanish at the transition to the insulator, while $g$ may remain nonzero. Therefore, $F_{\rm SC}$ is expected to be better represented by $g$ multiplied by $\delta$ at least in the underdoped region.

\begin{figure}[!t]
\centering
\includegraphics[width=\columnwidth]{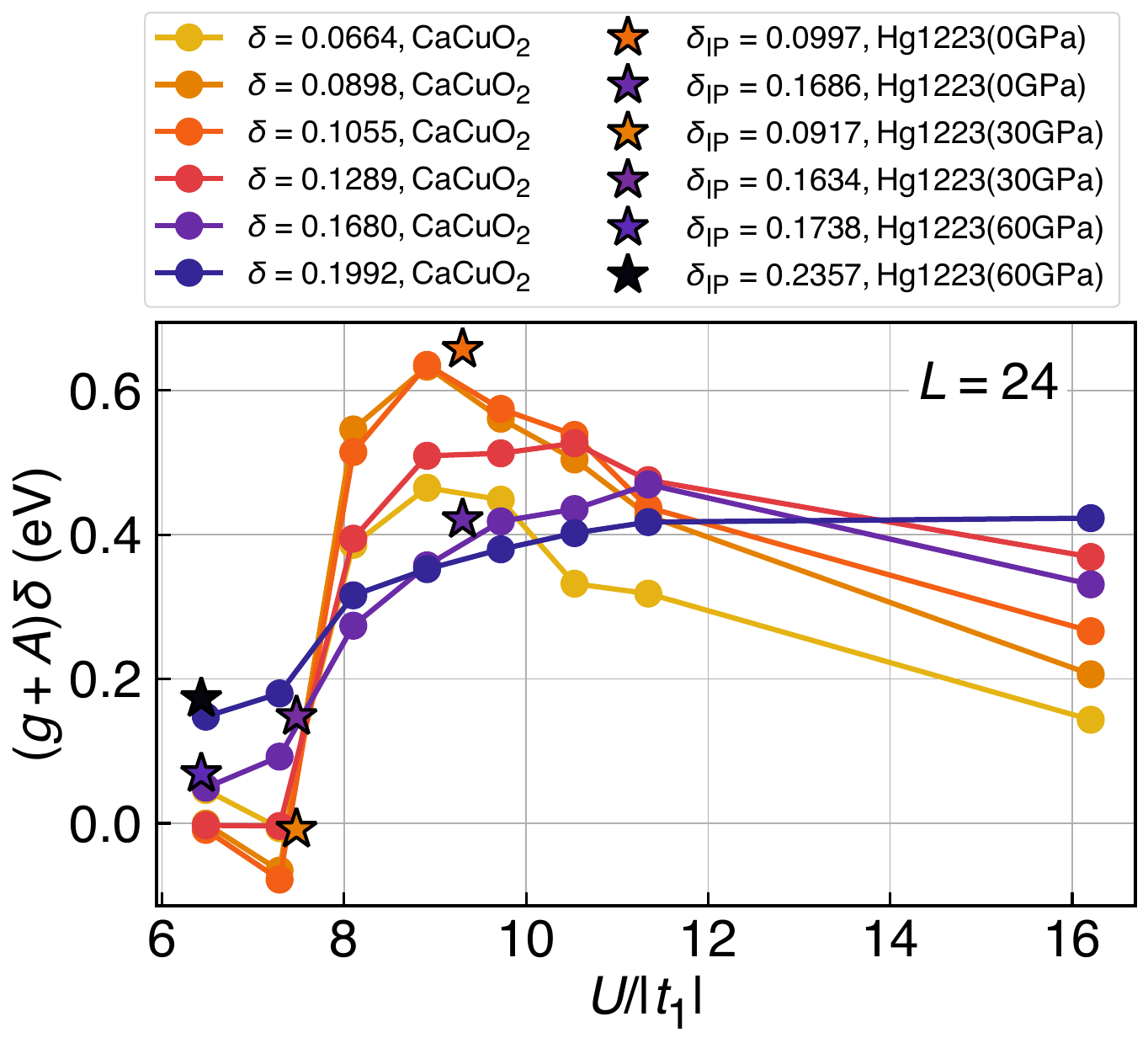}
\caption{%
Product of the attraction $g+A$ and doping $\delta$
as a function of the Coulomb interaction for $L=24$.
The constant $A$ is chosen as $2$.
}
\label{fig:c2_delta_vs_u}
\end{figure}
Based on this observation,
we plot $(g+A)\delta$
in Fig.~\ref{fig:c2_delta_vs_u}, where $A$ is 2. The offset $A$ is considered because even for noninteracting fermions, where the local interaction vanishes as well, $c_2=-g$ must be positive because of the Pauli exclusion principle, and the origin of the local interaction should be located at a negative $g$, which makes the effective attraction $g+A$ with the positive offset $A$. The value of $A$ can be inferred from the value of $c_2$ at small $U$. Figure~\ref{fig:c2_delta_vs_u} shows a better correlation with $F_{\rm SC}$ than Fig.~\ref{fig:c2_vs_u} on the point that the large enhancement in the undoped region is suppressed around the relevant region $U/|t_1|\sim 8$-10.

Let us discuss the point 2).
We have addressed that $F_{\rm SC}$ is dominantly determined by $g$ at ambient pressure. However, it is likely that off-site effective interactions play a certain role, which may show up as the off-site interaction dependence of $A$, namely, $A$ is similar at ambient pressure because the off-site interactions are similar (for instance $V_1\sim 1$ eV throughout the cuprates), while $A$ should increase substantially under pressure due to reduced off-site interaction consistently with the finding in Sec.~\ref{subsec:FSC_pressure}.
This well resolves the point 2).
It implies that 
\begin{equation}
F_{\rm SC}\sim \frac{1}{8}\frac{g^*}{|t_1|}\delta
\label{eq:Fsc_g_delta}
\end{equation}
is roughly satisfied in the realistic region, where the effective total attraction $g^*=g+A$, after including the effect of off-site interaction in $A$, can generally be expressed by the dimensionless scaling function $G$ as $g^*=|t_1|G(\frac{t_2}{|t_1|}, \frac{t_3}{|t_1|},\dots, \frac{U}{|t_1|}, \frac{V_1}{|t_1|},\dots)$. The calculated results indicate that $G$ most dominantly depends on $U/|t_1|$ and secondarily on $V_1/|t_1|$ as $F_{\rm SC}$.

The finding of the close correlation between $F_{\rm SC}$ and the local and instantaneous emergent attraction $g$ looks to support a SC mechanism entirely different from the conventional one mediated by some bosonic glue, such as the dynamical spin or charge fluctuation scenarios. In fact, it introduces an unexplored perspective as we will discuss below. However, we also make a remark on the relation to the spin-fluctuation scenario. The spin fluctuation can be measured by the dynamical spin structure factor
\begin{align}
\label{eq:dsf}
 S_{\mathrm{s}}(\bm{q},\omega)
 =
 \frac{1}{L^2} \int dt \sum_{i,j}
 \langle \bm{S}_{i}(t=0) \cdot \bm{S}_{j}(t) \rangle
 e^{i\bm{q}\cdot(\bm{r}_i-\bm{r}_j)+i\omega t},
\end{align}
while the local Coulomb energy $E_U$ can be rewritten by using the local moment $\langle m^2\rangle=\langle S^2\rangle=\frac{3}{4}\langle (n_{i,\uparrow}-n_{i,\downarrow})^2\rangle=\frac{3}{4}(\langle n_i\rangle-2\langle n_{i,\uparrow}n_{i,\downarrow}\rangle)$ 
as
\begin{align}
\label{eq:EU_dsf}
 E_U=U\left[\frac{N_{\rm e}}{2N_{\rm s}}-\frac{2}{3}\int d\omega d\bm{q} S_{\mathrm{s}}(\bm{q},\omega)\right].
\end{align}
Therefore, the local attraction 
measures the energy-momentum integrated amplitude of the dynamical spin correlation as 
\begin{align}
\label{eq:EU_dsf_2}
 g=\frac{2U}{3}\frac{\partial^2}{\partial\delta^2}\int d\omega d\bm{q} S_{\mathrm{s}}(\bm{q},\omega).
\end{align}

Since $S_{\mathrm{s}}(\bm{q},\omega)$ is extended not only to the so-called spin fluctuation region $\bm{q}\sim(\pi,\pi)$ and paramagnon energy range or superexchange range 
$\omega<J\sim 4t_1^2/U$ 
but also to a much wider range in doped Mott insulators, it makes a very broad structure of $S_{\mathrm{s}}(\bm{q},\omega)$.
In the weak coupling SC materials, the attraction is limited to the characteristic energy range of glue around the Fermi surface, such as the Debye frequency for phonon-mediated SC, while in the present case, a much wider range of energy comparable even to the Mott gap is involved, where the picture of instantaneous attraction
becomes relevant.
In strongly correlated systems, it is often {\it not} appropriate to employ the analysis localized in momentum-energy space; rather, it is better to start from a spatially and temporally local picture.
See Appendix~\ref{App:universality_attraction} for further analyses of attraction.

\section{%
Discussion
}
\label{sec:discussion}

\begin{figure}[!t]
\centering
\includegraphics[width=\columnwidth]{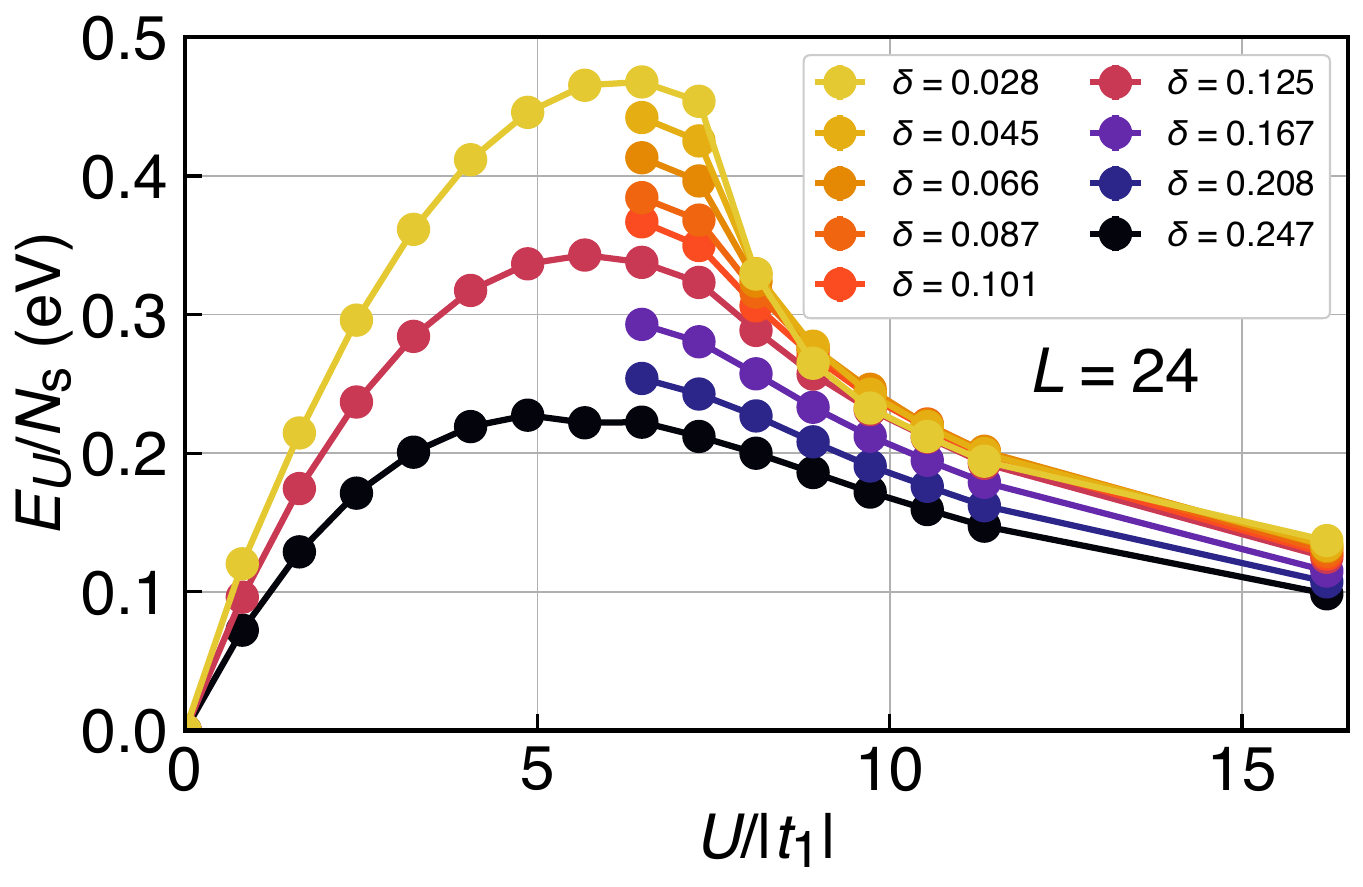}
\caption{%
Onsite Coulomb interaction energy per site ($E_U/N_{\mathrm{s}} = UD$) as a function of $U/|t_1|$
for hole-doped CaCuO$_2$ at $L=24$.
The peak in $E_U/N_{\mathrm{s}}$ appears at $U/|t_1|\sim 6.5$ for any $\delta \in (0,0.25)$.
}
\label{fig:EU_vs_U_cacuo2}
\end{figure}

Here, we discuss why the local emergent attraction arises counterintuitively from the originally strong repulsion term $E_U$. The strong Coulomb repulsion suppresses the double occupation of electrons and induces the AF Mott insulating state at half filling.
The Mott insulator is regarded as the ``vacuum'' of the charge carrier protected by the charge gap.
Nevertheless, the averaged double occupation $D=\langle n_{i\uparrow}n_{i\downarrow}\rangle$ and the corresponding exciton density (the density of the bound doublon and holon) do not vanish unless $U$ becomes infinite.
Keeping in mind this common property of the Mott insulator, and the fact that the emergent attraction generated by the $U$ term shows similar behavior among all the cuprates including Hg1223, we analyze an example of CaCuO$_2$ to elucidate the origin of the attraction.

We show in Fig.~\ref{fig:EU_vs_U_cacuo2}
$U/|t_1|$ dependence of $E_U$ in the case of the modified CaCuO$_2$ Hamiltonians. There exist peaks around $U/|t_1|\sim 6.5$ with inflection points around $U/|t_1|\sim 8$-$9$. Even the simple Hubbard model in the metastable $d$-wave SC state shows a qualitatively common trend, as is shown in Fig.~\ref{fig:EU_vs_U_normal_hubbard} in Appendix~\ref{App:universality_attraction}.  The nonzero $E_U$ in the limit $\delta\rightarrow 0$ may be interpreted as the ``vacuum polarization'' energy, and the Mott insulator at $\delta=0$ is regarded as the ``false vacuum'' particularly in the intermediate $U/|t_1|$ because of large quantum fluctuations. 

This vacuum-polarization structure is in contrast to the band insulator, where the charge density does not fluctuate from zero or two per site, depending on the unoccupied or occupied band, respectively, and the exciton density is strictly zero in the ground state.
In the present case of the Mott insulator, the vacuum polarization (quantum fluctuation) generates a nonzero density of  
doublon-holon pair (exciton) causing high zero-point energy of $E_U$.

The introduction of charge carriers induces the collapse of the Mott insulator (collapse of the gap)
accompanied by the release of the zero-point energy by rapidly reduced doublon-holon pair density and hence $E_U$ as one sees in Fig.~\ref{fig:EU_vs_U_cacuo2} most prominently in the intermediate $U/|t_1|\sim 5$-9. 
This nonlinear reduction of the local energy from the ``false vacuum'' with negative curvature results in the relative local attraction playing the role of 
the origin of the Cooper pairing. This mechanism may be called ``attraction by reduced repulsion''. In the cuprates, $E_U$ peaks around $U/|t_1|\sim 6.5$, and the inflection point around $U/|t_1|\sim 8$-9 in the Mott insulator is indeed the parameter region of the largest attraction $g+A$ and the strongest SC order parameter $F_{\rm SC}$ as is illustrated in Figs.~\ref{fig:c2_delta_vs_u} and \ref{fig:Fsc_vs_u}. The universality of this trend is discussed in Appendix~\ref{App:universality_attraction}.

The $d$-wave SC state changes the sign of the order parameter across the nodal line in the Brillouin zone and prohibits the double occupation of electrons on the same site by symmetry. Therefore, it has a large energy gain if the order develops by reducing $E_U$, which largely contributes to stabilizing the $d$-wave SC state against AF, stripe, and Fermi liquid metal, because these competing states allow the double occupation. By the energy gain, the SC order grows, which is accelerated by the nonlinear reduction of $E_U$ with the negative curvature of $\delta$ dependence generating emergent attraction synergetically in a self-consistent fashion.

A complementary view for the conversion from the repulsion to the attraction in the nodal $d$-wave SC state is known by considering the scattering process arising from the $U$ term containing $Uc^{\dagger}_{k\uparrow}c^{\dagger}_{-k\downarrow}c_{-q\downarrow}c_{q\uparrow}$ with $k\sim(\pi,0)$ and $q\sim(0,\pi)$~\cite{Scalapino_1986}. This generates the momentum transfer $(\pi,\pi)$, which is enhanced if the AF fluctuation is large, while in the mean-field decoupling $\langle c^{\dagger}_{k\uparrow}c^{\dagger}_{-k\downarrow}\rangle \langle c_{-q\downarrow}c_{q\uparrow}\rangle$, the $d$-wave phase change between $(\pi,0)$ and $(0,\pi)$ converts the repulsion $U>0$ into attraction.

\begin{figure}[!t]
\centering
\includegraphics[width=0.8\columnwidth]{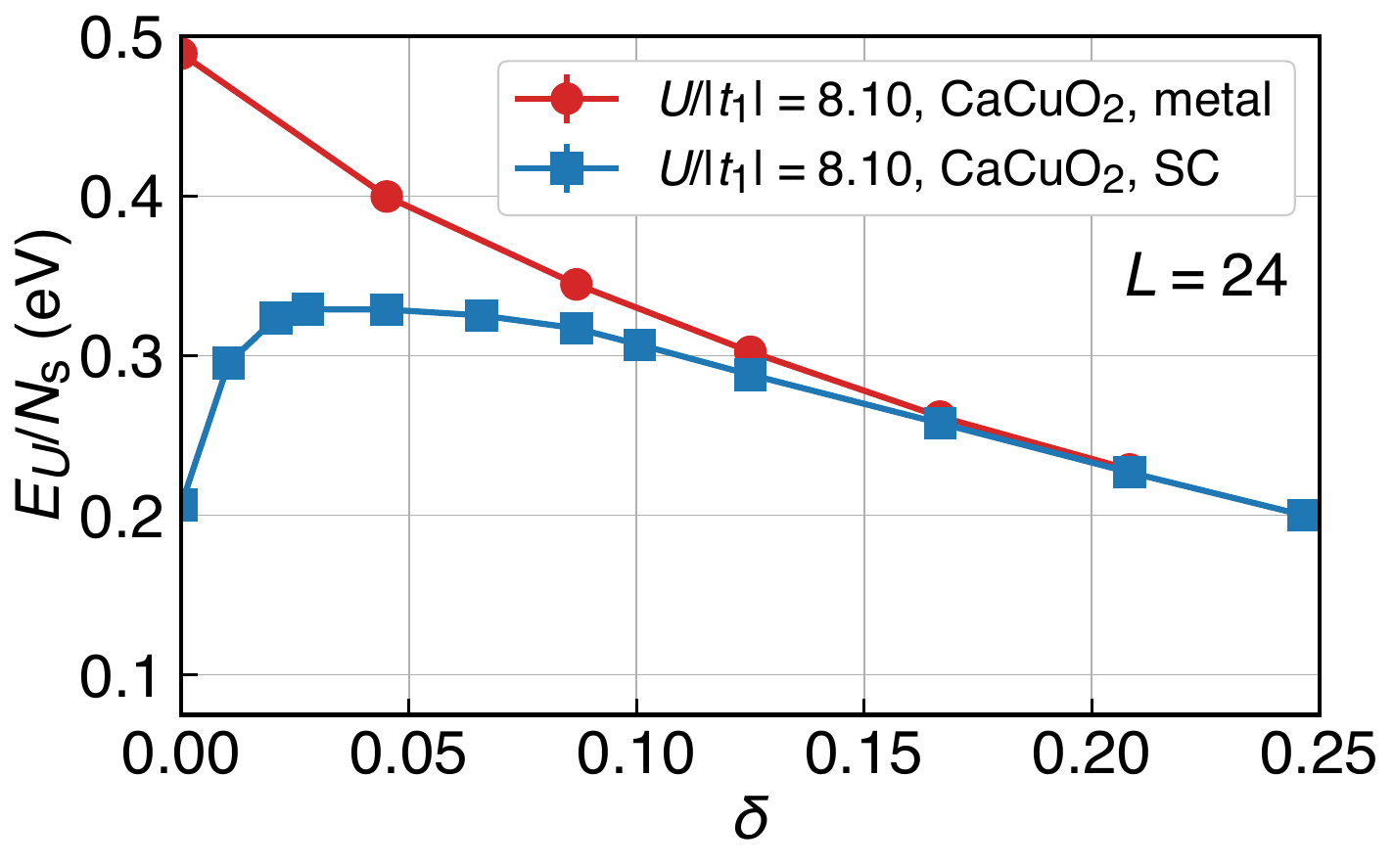}
\caption{%
Doping concentration ($\delta$) dependence of
the onsite Coulomb interaction energy $E_U$ 
for metallic and SC states in CaCuO$_2$ with $L=24$.
}
\label{fig:delta_vs_EU_metal_sc}
\end{figure}

The tight connection between the emergence of attraction and the role of $d$-wave SC is clearly seen in Fig.~\ref{fig:delta_vs_EU_metal_sc}, which shows a sharp contrast of the $\delta$ dependence of $E_U$ between the normal metal and the SC state for CaCuO$_2$. Only the $d$-wave SC state shows the negative curvature and thus the emergent attraction.
Note that the normal metal compared here is different from the metal at temperatures above $T_{\mathrm{c}}$ with the pseudogap.

As shown in Figs.~\ref{fig:tc} and \ref{fig:compare_cuprate_Tc},
the theoretically estimated $T_{\mathrm{c}}$ exhibits a
peak at $\sim 30$\,GPa upon increasing pressure
behaving consistently with the experimental $T_{\mathrm{c}}$.
On the other hand, the estimated $T_{\mathrm{c}}$ is slightly lower for stronger pressure.
One of the possible origins is the inhomogeneity of applied pressure in the experiments, which unsettles the definition of the experimental $T_{\mathrm{c}}$:
For stronger pressure, the experimentally observed onset temperatures ($T_{\mathrm{c,onset}}$) available so far are much higher than the other $T_{\mathrm{c,mid}}$ determined by the inflection point of the temperature derivative of the resistivity.
 
Among a huge number of studies, let us pick up some work, which pays attention to the relation between $T_{\mathrm{c}}$ and materials parameters from general viewpoints on the cuprates, which might have relations to the present work. The importance of the apical oxygen position to control $T_{\mathrm{c}}$ was empirically pointed out in the cuprates in general in \cite{ohta1991}.
Also empirically, parameters of the crystal structure controlling  $T_{\mathrm{c}}$ were discussed~\cite{koblischka2020,al-Ruqaishi2026}.
Several experimental studies on the cuprates reported the relation between $T_{\mathrm{c}}$ and the character of dopant atoms and disorder~\cite{attfield1998,hobou2009}.
The optimum value for the highest $T_{\mathrm{c}}$ was discussed in the $U/|t_1|$ dependence~\cite{nilsson2019,kitatani2023}. Features of the crystal structure to optimize the SC were discussed with the help of machine learning~\cite{moree2024_2}. 
Further studies on the relation between the above work and the present results await future analyses.

On the other hand, many studies pointed out relations between $T_{\mathrm{c}}$ and other physical quantities such as SC energy gaps and AF spin-fluctuation energies. For instance, the Uemura plot focused on the proportionality of $T_{\mathrm{c}}$ and the SC carrier density measured by the $\mu$SR penetration depth in more comprehensive studies on the materials, not only on the cuprates~\cite{uemura1991}.
This essentially shares the common viewpoint with Eq.~(\ref{eq:scaling_form}). It was pointed out that the scaling $T_{\mathrm{c}} \propto F_{\rm SC}$ can be understood from the scaling of the Berezinskii-Kosterlitz-Thouless transition temperature in a weakly interacting 2D Bose gas, $T_{\rm BKT}\propto n_b$~\cite{prokofev2001} by interpreting the superfluid density $F_{\rm SC}$ as the boson density $n_b$ in the BEC picture.
However, it does not tell us how $F_{\rm SC}$ is determined theoretically, which the present work has elucidated for the cuprates, especially with the pressure dependence of Hg1223 by the \textit{ab initio} studies.  In fact, Eq.~(\ref{eq:Fsc_g_delta}) unveils the origin of the SC caused by the instantaneous, local, and emergent attraction.
We note that $F_{\rm SC}$ is quantitatively consistent with the value reported in Ref.~[\onlinecite{uemura1989}] as discussed in Ref.~[\onlinecite{schmid2023}].

There exist several studies on the multilayer cuprates. 
Differentiation in roles of the IP and OP was theoretically emphasized in realizing the high $T_{\mathrm{c}}$ of Hg1223~\cite{luo2023,liu2025}. While the difference in the hole density they presumed is qualitatively similar to the present result, and the interlayer differentiation plays a role through the proximity, the main origin of the higher $T_{\mathrm{c}}$ is found to be ascribed to high $U/|t_1|$ in the present work with quantitative agreement with the experiments. 
 
A recent study~\cite{bacqlabreuil2025} on Hg1223 at ambient pressure, based on the atomic orbital basis combined with the dynamical mean-field framework similarly to the choice of Ref.~[\onlinecite{cui2025}], instead of the bonding-antibonding basis of strongly hybridized Cu$3d_{x^2-y^2}$ and O$2p_{\sigma}$ orbitals employed in the present work, proposed a different view. Ref.~[\onlinecite{bacqlabreuil2025}] obtained similar optimum SC order parameter at ambient pressure between Hg1201 and Hg1223 although the experimental optimum $T_{\mathrm{c}}$ is substantially different ($\sim 98$\,K for Hg1201 and $\sim 134$\,K for Hg1223), while the present work correctly captures this experimental trend of difference. 

The advantage of the present work solving the lattice Hamiltonian is that this approach is able to treat the spatial and temporal quantum fluctuations on equal footing, which also allows treating the severe competitions of different orders and fluctuations as well beyond mean-field approximations.
From the present \textit{ab initio} work, detailed microscopic origins of dependencies on materials, hole density, and pressure have been elucidated by one-to-one correspondence to the \textit{ab initio} parameters of the effective Hamiltonians, which makes the microscopic origins and mechanisms of the materials dependence clear and intuitively transparent.
It also helps in understanding the physical origin of the empirical relations suggested in the literature discussed in the last paragraphs.

We now briefly mention guiding principles for realizing  
higher $T_{\mathrm{c}}$ than the available materials along the line of the mechanism of the cuprate SC.
Since the $T_{\mathrm{c}}$ formula (Eq.~(\ref{eq:scaling_form})) widely holds, the strategy to be employed is to design 1) a wider bandwidth (consequently larger $|t_1|$) and 2) larger $F_{\rm SC}$ simultaneously. 
We learned in the present study that $F_{\rm SC}$ becomes optimized at 2-a) $U/|t_1|\sim 9$ while 2-b) smaller $V_1/|t_1|$ enhances $F_{\rm SC}$.
In real materials, it is not easy to satisfy 1), 2-a) and 2-b) simultaneously, because, for instance, the hydrostatic pressure makes the bandwidth wider while $U/|t_1|$ and $V_1/|t_1|$ (and also $V_2/|t_1|, V_3/|t_1|...$) smaller, which have opposite effects on $F_{\rm SC}$. A possible direction is to search for the materials that have relatively small $V_1/U, V_2/U,...$ within the range of stable SC against the stripe (note that very small $V_i$ generally is subject to making the stripe state more stable) by keeping $U/|t_1|\sim 9$ and large $t_1$. Alternatively, if the condition $U/|t_1|>9$ is satisfied together with not high off-site interactions, pressure is expected to work cooperatively to raise $T_{\mathrm{c}}$. Smaller $V_1/U, V_2/U,...$ and larger $t_1$ could be prepared simultaneously by pressure as the present Hg1223 study demonstrated. Another possibility is to design a surface or interface with an insulating substrate that has a high dielectric constant to screen the off-site interaction. 

\section{%
Summary, Conclusions, and Outlook
}
\label{sec:conclusions}

We have elucidated why the triple-layer Hg1223 shows the highest $T_{\mathrm{c}}$ as compared to the single-, double-, and infinite-layer cuprates at ambient pressure and a further increase in $T_{\mathrm{c}}$ under pressure, together with the mechanism of SC. This has been achieved by examining the ground states of the \textit{ab initio} low-energy
effective Hamiltonians of Hg1223.
To explore the pressure dependence of the superconductivity,
we have investigated the systems under pressure ranging from $0$\,GPa to $60$\,GPa.
By using the VMC method, we have calculated the $d$-wave SC order parameter $F_{\rm SC}$ in the thermodynamic limit and estimated $T_{\mathrm{c}}$ by following the formula (\ref{eq:scaling_form}) and have found the stable and uniform $d$-wave SC phase in the ground state in the hole concentration region $0.1<\delta<0.25$.

Hg1223 at ambient pressure is found to be in a slightly strong coupling region above the optimal value of $U/|t_1|\sim 9$, which is the primary reason for high $T_{\mathrm{c}}$ 
 in comparison to single-, double-, and infinite-layer compounds studied in Ref.~[\onlinecite{schmid2023}], where a high $U/|t_1|$ value (because of large $U$)
originates from poorer screening.
The pressure drives it into the weak coupling region over the optimum $U/|t_1|$, 
which contributes to keeping $F_{\rm SC}$ in the plateau region around the peak at weak pressure $P< 30$\,GPa, while it works to sharply decrease $F_{\rm SC}$ at $P> 30$\,GPa. However, it is partially compensated by the largely decreasing off-site interaction, which works to enhance $F_{\rm SC}$. Then this interplay eventually makes a broad
peak structure in the pressure dependence of $F_{\rm SC}$. 

As for $T_{\mathrm{c}}$,
the pressure makes the increase in $|t_1|$, which contributes to enhancing $T_{\mathrm{c}}$. In addition, the pressure makes a reduction of $U/|t_1|$ as well, contributing to weak enhancement of $T_{\mathrm{c}}$ cooperatively in the region of $U/|t_1|>9$ at low pressures $<30$\,GPa.
On the other hand, higher pressure above $30$\,GPa corresponding to decreasing $U/|t_1|$ below 8 causes a substantial reduction of $T_{\mathrm{c}}$ due to the decay in $F_{\rm SC}$, which becomes the dominant factor that controls $T_{\mathrm{c}}$. This interplay is perfectly consistent with the experimental dome structure.
It demonstrates that one effective route to enhance the superconductivity is to reduce the off-site interaction within the moderate and not too weak range to avoid the stripe phases.
The determination of the boundary between the SC and stripe phases requires careful analysis in the case of weaker off-site interaction and will be left for a subject of future work.

The $T_{\mathrm{c}}$ formula at the optimal carrier density, $T_{\mathrm{c}} \sim 0.16 |t_1|F_{\rm SC}$ established in the single-, double-, and infinite-layer cuprates~\cite{schmid2023} is shown to hold also in the triple-layer cuprates and to reproduce the essential feature of the
pressure dependence, namely, the peak around $30$\,GPa.

We have also found the tendency that, in the underdoped region ($\delta \lesssim 0.1$),
the filling in the IP is much closer to half filling
than that in the OP.
This stabilizes the AF order in the IP simultaneously retaining a weak SC order as the proximity from the OP SC
and enables the coexistence of the SC and AF.

The effective instantaneous attraction for Hg1223 that induces the Cooper pair is found and its origin is identified as the mechanism of ``attraction by reduced repulsion'' emerging in the originally strongly repulsive onsite interaction term, which is common to the single-, double-, and infinite-layer cuprates. This mechanism is very general as is observed also in the SC phase of the Hubbard model. The Cooper pair formed by the instantaneous attraction is in contrast with the weak-coupling BCS SC caused by the retarded interaction with a large coherence length. The SC at the optimum doping of the cuprates does not belong to the BEC regime either, because the attraction is relatively weaker than the kinetic energy scale (namely, effective Fermi energy expected from a large Fermi surface) and the preformed pair is not well developed above $T_{\mathrm{c}}$. In fact, $F_{\rm SC}$ and $T_{\mathrm{c}}$ at the optimum doping rather increase with the increased attraction in the realistic cuprates as Eq.~(\ref{eq:Fsc_g_delta}) suggests, instead of the reduction expected in the typical BEC regime. 

In the underdoped region toward the Mott insulator, Eq.~(\ref{eq:scaling_form}) shares similarity with the BKT transition of a weakly interacting 2D Bose gas in the proportionality of $T_{\mathrm{c}}$ to the superfluid density $F_{\rm SC}$, which is interpreted as the ``boson density''. However, Eq.~(\ref{eq:Fsc_g_delta}) indicates that the ``boson density'' is controlled in a nontrivial fashion by the emergent attractive interaction and the hole density in the present case. Despite the high electron density in the underdoped region, at which one can naively expect a large Fermi surface in metals, the cuprates do not show a diverging electron effective mass by keeping the high density of coherent electrons as a route to the Mott insulator. Instead, the incoherent part becomes dominating toward the Mott insulator, and the effective ``boson'' emerges as Eq.~(\ref{eq:Fsc_g_delta}), if one can ignore the randomness and inhomogeneity in real materials.

A stringent test of the emergence of instantaneous attraction is to experimentally measure Eq.~(\ref{eq:EU_dsf_2}), for instance, by neutron scattering. Energy-momentum-resolved data will unveil the dynamical origin of the attraction.

The emergent attraction may cause the fractionalization of electrons~\cite{schmid2023,imada2019}, which is now supported by various experimental and theoretical studies~\cite{sakai2016,sakai2013,imada2019,imada2021,yamaji2021,singh2022,sakai2025}. To reach a further complete understanding of the superconductivity, it is desired to uncover the nature of the fractionalized hidden fermion splintered from the original electron and examine whether the weakly bound exciton proposed in the literature~\cite{sakai2018,imada2019} is involved in the fractionalization.

There exist other strong-coupling unconventional SCs such as the iron-based SCs~\cite{kamihara2008} and nickel oxide SCs~\cite{li2019,sun2023}. It is an intriguing future subject to clarify whether these families have the common aspect to the cuprates along the line we reported in this paper. The relations to heavy-electron SC in $f$-electron systems and 4$d$-electron SC are also intriguing in a more general perspective of the fractionalization.
Applying the present framework to other trilayer cuprates, such as Bi2223~\cite{chen2010} and compounds containing even more layers, would also be intriguing to examine the broader applicability of the proposed SC mechanism, which is left for future studies.
Although their experimentally observed $T_{\mathrm{c}}$ values are lower and their treatment is beyond the scope of the present work because of the higher computational cost,
multilayer compounds with four or more layers would also be interesting future targets.

In terms of the materials design strategy for higher $T_{\mathrm{c}}$, one can propose to seek materials that allow ``attraction by reduced repulsion'' from the ``false vacuum'', as well as tuning toward optimum onsite repulsion, weaker intersite interaction, and larger bandwidth.
Our findings will stimulate research on designing and optimizing
SCs toward a higher SC critical temperature
by explicitly utilizing the mechanism of the emergence of the attraction from the repulsion.

\begin{acknowledgments}
The authors acknowledge fruitful discussions with
J.-B.\ Mor\'{e}e,
M.\ T.\ Schmid,
and
Y.\ Yamaji.
Electron densities of the IP and OP in the GW approximation were provided by J.-B.\ Mor\'{e}e.
This work was financially supported by
MEXT KAKENHI, Grant-in-Aid for Transformative Research Area
(Grants No.\ JP22H05111 and No.\ JP22H05114).
R.K.\ was supported by
JSPS KAKENHI
(Grants No.\ JP24H00973 and No.\ JP25K07157).
M.I.\ was supported by MEXT as
``Program for Promoting Researches on the Supercomputer Fugaku''
(Simulation for basic science: approaching the new quantum era,
Grant No.\ JPMXP1020230411).
The numerical computations were performed on computers at
the supercomputer Fugaku provided by the RIKEN Center
for Computational Science
and
the Supercomputer Center, Institute for Solid State Physics, University of Tokyo.
\end{acknowledgments}

\section*{Data availability}
The code used in this manuscript is available at Ref.~[\onlinecite{misawa2019}].
The data supporting the findings in this study are available
from the authors upon reasonable request.

\appendix

\section{%
\textit{Ab initio} effective Hamiltonians under pressure
}
\label{App:parameters}

We show the parameters for the \textit{ab initio} low-energy effective Hamiltonian
of Hg1223 at $30$\,GPa and $60$\,GPa in
Tables~\ref{tab:parameters_0GPa}, \ref{tab:parameters_30GPa} and \ref{tab:parameters_60GPa}.
As in the case at ambient pressure,
we use the hopping and interaction parameters that extend over long distances
satisfying $r\le 3\sqrt{2}$.
In Table~\ref{tab:t1_U_V1_each_compound}, we list the core parameters $t_1, U$, and $V_1$ of Hg1223 and compare them with already studied four hole-doped cuprate compounds CaCuO$_2$, Bi2201, Bi2212, and Hg1201 at ambient pressure and at optimum doping.

The onsite effective Coulomb interaction for the present single-band Hamiltonian
has been improved 
by taking into account the energy-level correction between the Cu$d_{x^2-y^2}$ and O$2p_{\sigma}$ atomic orbitals in the downfolding procedure to satisfy the charge consistency,
namely, LRFB as proposed in Refs.~\cite{hirayama2019,moree2022}. 
By this LRFB, the strength of the onsite Coulomb interaction is commonly rescaled by $0.91$-$0.97$~\cite{moree2024}
and the rescaling of other Hamiltonian parameters is small as was elucidated in Ref.~[\onlinecite{moree2024}].
In our VMC calculation, we use the effective Hamiltonian
with this modified onsite Coulomb interaction.
The original onsite Coulomb interactions
$U_{\mathrm{IP}}$ and $U_{\mathrm{OP}}$
are multiplied by $0.95$.
In this manuscript, the symbol $U$ is used for
the onsite Coulomb interaction after the LRFB correction.
Tables~\ref{tab:parameters_0GPa},
\ref{tab:parameters_30GPa}, and \ref{tab:parameters_60GPa}
contain the onsite Coulomb interactions
after the correction by the LRFB.

\begin{table}[!t]
\centering
\caption{%
Parameters for the \textit{ab initio} low-energy effective Hamiltonian
of Hg1223 at $0$\,GPa.
We show the hopping parameter $t_{l}$
(including the chemical potential $\mu$)
and the Coulomb interaction $V_{l}$
(including the onsite one $U$)
within the same IP (IP-IP), those within the same OP (OP-OP),
those between the IP and OP (IP-OP), and those between the different OPs (OP-OP$'$)
up to a distance $r=2\sqrt{2}$.
The chemical potential difference between the IP and the OP is given by $\mu$,
corresponding to the parameterizations,
$\mu_{\mathrm{IP}} = t_{0,\text{IP-IP}} = 0$
and
$\mu_{\mathrm{OP}} = t_{0,\text{OP-OP}} = \mu$
in Eq.~\eqref{eq:ham_hmu}.
The hopping parameters $t_0$ with IP-OP and OP-OP$'$ are
onsite parameters connecting the different planes.
The interaction parameters $V_0$ with IP-IP and OP-OP are
the onsite Coulomb interaction parameters $U_{\mathrm{IP}}$ and $U_{\mathrm{OP}}$, respectively.
These Coulomb interactions in this table are the values after the
LRFB correction~\cite{hirayama2019,moree2022}.
The interaction parameters $V_0$ with IP-OP and OP-OP$'$
are onsite Coulomb interaction parameters connecting the different planes.
All values are given in eV.}
\label{tab:parameters_0GPa}
\begin{tabular}{cccccc}
\hline
\hline
\begin{tabular}{c} index $l$ of \\ hopping $t_l$ \end{tabular} & distance $r$ & IP-IP & OP-OP & IP-OP & OP-OP$'$ \\
\hline
$0$ & $0$ & $\phantom{+}0\phantom{.0000}$ & $\mu$ & $-0.0619$ & $+0.0219$ \\
$1$ & $1$ & $-0.4857$ & $-0.4849$ & $+0.0067$ & $-0.0057$ \\
$2$ & $\sqrt{2}$ & $+0.1326$ & $+0.1300$ & $+0.0107$ & $+0.0019$ \\
$3$ & $2$ & $-0.0503$ & $-0.0444$ & $-0.0042$ & $-0.0012$ \\
$4$ & $\sqrt{5}$ & $+0.0036$ & $+0.0004$ & $+0.0001$ & $-0.0018$ \\
$5$ & $2\sqrt{2}$ & $+0.0024$ & $+0.0050$ & $+0.0014$ & $+0.0027$ \\
$6$ & $3$ & $-0.0108$ & $-0.0113$ & $-0.0005$ & $-0.0015$ \\
$7$ & $\sqrt{10}$ & $-0.0023$ & $-0.0003$ & $+0.0003$ & $-0.0005$ \\
$8$ & $\sqrt{13}$ & $+0.0017$ & $+0.0025$ & $-0.0011$ & $+0.0016$ \\
$9$ & $4$ & $-0.0039$ & $-0.0039$ & $-0.0017$ & $+0.0016$ \\
$10$ & $\sqrt{17}$ & $+0.0003$ & $+0.0007$ & $+0.0003$ & $+0.0007$ \\
$11$ & $3\sqrt{2}$ & $+0.0049$ & $+0.0014$ & $+0.0000$ & $-0.0002$ \\
\hline
\begin{tabular}{c} index $l$ of \\ interaction $V_l$ \end{tabular} & distance $r$ & IP-IP & OP-OP & IP-OP & OP-OP$'$ \\
\hline
$0$ & $0$ & $ 4.5184$ & $ 4.5443$ & $ 0.7987$ & $ 0.3953$ \\
$1$ & $1$ & $ 0.9878$ & $ 0.9785$ & $ 0.5621$ & $ 0.3392$ \\
$2$ & $\sqrt{2}$ & $ 0.5631$ & $ 0.5579$ & $ 0.4375$ & $ 0.2999$ \\
$3$ & $2$ & $ 0.3901$ & $ 0.3854$ & $ 0.3318$ & $ 0.2500$ \\
$4$ & $\sqrt{5}$ & $ 0.3287$ & $ 0.3257$ & $ 0.2930$ & $ 0.2315$ \\
$5$ & $2\sqrt{2}$ & $ 0.2495$ & $ 0.2485$ & $ 0.2317$ & $ 0.1943$ \\
$6$ & $3$ & $ 0.2448$ & $ 0.2437$ & $ 0.2293$ & $ 0.1938$ \\
$7$ & $\sqrt{10}$ & $ 0.2299$ & $ 0.2291$ & $ 0.2163$ & $ 0.1847$ \\
$8$ & $\sqrt{13}$ & $ 0.1994$ & $ 0.1991$ & $ 0.1894$ & $ 0.1661$ \\
$9$ & $4$ & $ 0.1054$ & $ 0.1054$ & $ 0.1006$ & $ 0.0879$ \\
$10$ & $\sqrt{17}$ & $ 0.1012$ & $ 0.1012$ & $ 0.0967$ & $ 0.0848$ \\
$11$ & $3\sqrt{2}$ & $ 0.1740$ & $ 0.1741$ & $ 0.1678$ & $ 0.1494$ \\
\hline
\hline
\end{tabular}
\end{table}

\begin{table}[!t]
\centering
\caption{%
Parameters for the \textit{ab initio} low-energy effective Hamiltonian
of Hg1223 at $30$\,GPa.
The notations of the parameters are the same as in Table~\ref{tab:parameters_0GPa}.
}
\label{tab:parameters_30GPa}
\begin{tabular}{cccccc}
\hline
\hline
\begin{tabular}{c} index $l$ of \\ hopping $t_l$ \end{tabular} & distance $r$ & IP-IP & OP-OP & IP-OP & OP-OP$'$ \\
\hline
$0$ & $0$ & $\phantom{+}0\phantom{.0000}$ & $\mu$ & $-0.0678$ & $+0.0021$ \\
$1$ & $1$ & $-0.5833$ & $-0.5554$ & $+0.0031$ & $-0.0019$ \\
$2$ & $\sqrt{2}$ & $+0.1425$ & $+0.1373$ & $+0.0149$ & $-0.0011$ \\
$3$ & $2$ & $-0.0523$ & $-0.0528$ & $-0.0033$ & $+0.0039$ \\
$4$ & $\sqrt{5}$ & $+0.0038$ & $+0.0066$ & $-0.0019$ & $-0.0016$ \\
$5$ & $2\sqrt{2}$ & $+0.0023$ & $+0.0009$ & $+0.0033$ & $+0.0012$ \\
$6$ & $3$ & $-0.0051$ & $-0.0080$ & $-0.0023$ & $+0.0003$ \\
$7$ & $\sqrt{10}$ & $-0.0003$ & $-0.0028$ & $+0.0003$ & $-0.0014$ \\
$8$ & $\sqrt{13}$ & $-0.0016$ & $+0.0013$ & $+0.0002$ & $+0.0013$ \\
$9$ & $4$ & $-0.0010$ & $-0.0006$ & $-0.0018$ & $-0.0010$ \\
$10$ & $\sqrt{17}$ & $+0.0009$ & $-0.0024$ & $+0.0008$ & $+0.0018$ \\
$11$ & $3\sqrt{2}$ & $+0.0015$ & $+0.0014$ & $+0.0010$ & $-0.0001$ \\
\hline
\begin{tabular}{c} index $l$ of \\ interaction $V_l$ \end{tabular} & distance $r$ & IP-IP & OP-OP & IP-OP & OP-OP$'$ \\
\hline
$0$ & $0$ & $ 4.3615$ & $ 4.4986$ & $ 0.7573$ & $ 0.3696$ \\
$1$ & $1$ & $ 0.8752$ & $ 0.8056$ & $ 0.4939$ & $ 0.2957$ \\
$2$ & $\sqrt{2}$ & $ 0.4683$ & $ 0.4430$ & $ 0.3651$ & $ 0.2502$ \\
$3$ & $2$ & $ 0.3115$ & $ 0.3014$ & $ 0.2663$ & $ 0.2045$ \\
$4$ & $\sqrt{5}$ & $ 0.2573$ & $ 0.2574$ & $ 0.2327$ & $ 0.1850$ \\
$5$ & $2\sqrt{2}$ & $ 0.1917$ & $ 0.1959$ & $ 0.1806$ & $ 0.1534$ \\
$6$ & $3$ & $ 0.1860$ & $ 0.1912$ & $ 0.1778$ & $ 0.1519$ \\
$7$ & $\sqrt{10}$ & $ 0.1750$ & $ 0.1801$ & $ 0.1674$ & $ 0.1449$ \\
$8$ & $\sqrt{13}$ & $ 0.1498$ & $ 0.1544$ & $ 0.1452$ & $ 0.1290$ \\
$9$ & $4$ & $ 0.0790$ & $ 0.0818$ & $ 0.0772$ & $ 0.0683$ \\
$10$ & $\sqrt{17}$ & $ 0.0763$ & $ 0.0789$ & $ 0.0741$ & $ 0.0659$ \\
$11$ & $3\sqrt{2}$ & $ 0.1298$ & $ 0.1342$ & $ 0.1270$ & $ 0.1153$ \\
\hline
\hline
\end{tabular}
\end{table}

\begin{table}[]
\centering
\caption{%
Parameters for the \textit{ab initio} low-energy effective Hamiltonian
of Hg1223 at $60$\,GPa.
The notations of the parameters are the same as in Table~\ref{tab:parameters_0GPa}.
}
\label{tab:parameters_60GPa}
\begin{tabular}{cccccc}
\hline
\hline
\begin{tabular}{c} index $l$ of \\ hopping $t_l$ \end{tabular} & distance $r$ & IP-IP & OP-OP & IP-OP & OP-OP$'$ \\
\hline
$0$ & $0$ & $\phantom{+}0\phantom{.0000}$ & $\mu$ & $-0.0739$ & $+0.0194$ \\
$1$ & $1$ & $-0.6492$ & $-0.5812$ & $+0.0024$ & $-0.0049$ \\
$2$ & $\sqrt{2}$ & $+0.1509$ & $+0.1274$ & $+0.0141$ & $+0.0004$ \\
$3$ & $2$ & $-0.0578$ & $-0.0406$ & $-0.0060$ & $+0.0019$ \\
$4$ & $\sqrt{5}$ & $+0.0013$ & $+0.0018$ & $-0.0028$ & $-0.0016$ \\
$5$ & $2\sqrt{2}$ & $+0.0028$ & $-0.0004$ & $+0.0037$ & $+0.0020$ \\
$6$ & $3$ & $-0.0069$ & $-0.0096$ & $+0.0003$ & $+0.0028$ \\
$7$ & $\sqrt{10}$ & $+0.0009$ & $-0.0014$ & $-0.0007$ & $+0.0011$ \\
$8$ & $\sqrt{13}$ & $-0.0007$ & $+0.0019$ & $+0.0007$ & $-0.0008$ \\
$9$ & $4$ & $-0.0034$ & $+0.0002$ & $-0.0019$ & $-0.0024$ \\
$10$ & $\sqrt{17}$ & $+0.0010$ & $-0.0001$ & $-0.0000$ & $-0.0004$ \\
$11$ & $3\sqrt{2}$ & $+0.0005$ & $+0.0015$ & $+0.0002$ & $-0.0024$ \\
\hline
\begin{tabular}{c} index $l$ of \\ interaction $V_l$ \end{tabular} & distance $r$ & IP-IP & OP-OP & IP-OP & OP-OP$'$ \\
\hline
$0$ & $0$ & $ 4.1751$ & $ 4.2381$ & $ 0.6922$ & $ 0.3058$ \\
$1$ & $1$ & $ 0.6842$ & $ 0.5832$ & $ 0.3593$ & $ 0.2068$ \\
$2$ & $\sqrt{2}$ & $ 0.3156$ & $ 0.2878$ & $ 0.2393$ & $ 0.1566$ \\
$3$ & $2$ & $ 0.1859$ & $ 0.1743$ & $ 0.1540$ & $ 0.1110$ \\
$4$ & $\sqrt{5}$ & $ 0.1419$ & $ 0.1429$ & $ 0.1267$ & $ 0.0967$ \\
$5$ & $2\sqrt{2}$ & $ 0.0908$ & $ 0.0970$ & $ 0.0850$ & $ 0.0682$ \\
$6$ & $3$ & $ 0.0870$ & $ 0.0921$ & $ 0.0826$ & $ 0.0675$ \\
$7$ & $\sqrt{10}$ & $ 0.0783$ & $ 0.0840$ & $ 0.0747$ & $ 0.0613$ \\
$8$ & $\sqrt{13}$ & $ 0.0598$ & $ 0.0655$ & $ 0.0581$ & $ 0.0488$ \\
$9$ & $4$ & $ 0.0331$ & $ 0.0359$ & $ 0.0321$ & $ 0.0275$ \\
$10$ & $\sqrt{17}$ & $ 0.0309$ & $ 0.0337$ & $ 0.0298$ & $ 0.0256$ \\
$11$ & $3\sqrt{2}$ & $ 0.0453$ & $ 0.0505$ & $ 0.0444$ & $ 0.0376$ \\
\hline
\hline
\end{tabular}
\end{table}

\begin{table}[!t]
\centering
\caption{%
Selected important parameters for the \textit{ab initio} low-energy effective Hamiltonians
of the IP of Hg1223 at $0$\,GPa, $30$\,GPa, and $60$\,GPa
and those of
CaCuO$_2$, HgBa$_2$CuO$_4$ (abbreviated as Hg1201), Bi$_2$Sr$_2$CaCu$_2$O$_8$ (Bi2212), and Bi$_2$Sr$_2$CuO$_6$ (Bi2201) at ambient pressure.
We show the nearest-neighbor hopping parameter $t_1$,
the onsite Coulomb interaction $U$,
and the nearest-neighbor off-site Coulomb interaction $V_1$.
All values are given in eV.
}
\label{tab:t1_U_V1_each_compound}
\begin{tabular}{cccc}
\hline
\hline
  & $t_1$ & $U$ & $V_1$ \\
\hline
 Hg1223, IP,  $0$\,GPa & $-0.4857$ & $4.5184$ & $0.9878$ \\
 Hg1223, IP, $30$\,GPa & $-0.5833$ & $4.3615$ & $0.8752$ \\
 Hg1223, IP, $60$\,GPa & $-0.6492$ & $4.1751$ & $0.6842$ \\
 CaCuO$_2$ & $-0.521$ & $4.221$ & $0.969$ \\
 Hg1201    & $-0.544$ & $3.999$ & $1.002$ \\
 Bi2212    & $-0.452$ & $4.226$ & $0.915$ \\
 Bi2201    & $-0.527$ & $4.393$ & $1.030$ \\
\hline
\hline
\end{tabular}
\end{table}

\section{%
Estimation of chemical potential difference
between the IP and OPs
}
\label{App:estimate_mu}

The carrier densities in the IP and OPs during the VMC calculations
should be adjusted to reproduce those obtained by GW calculations along the same idea employed for multi-orbital systems as was established in the procedure of LRFB~\cite{hirayama2019}.
This can be achieved by controlling the chemical potential difference
between the IP and OP
under the condition of a fixed total number of electrons
for each system size in the canonical ensemble. Namely, this adjustment is accomplished by performing the VMC calculation of the \textit{ab initio} effective Hamiltonian (\ref{eq:ham}) by shifting the chemical potential difference $\mu$ until the inner and outer carrier densities meet the GW result, for instance, at $\delta=0$ and 0.2~\cite{moree_unpublished}. When the GW result is not available, we estimated it from the interpolation of the nearby available results of generalized gradient approximation (GGA) combined with its correction estimated from the difference between the GGA and GW charge densities. The carrier concentrations of the IP and OP thus obtained
are shown in Figs.~\ref{fig:self-doping} and \ref{fig:self-doping_3060GPa} 
and the corresponding chemical potential difference $\mu$ is given by Eqs.~(\ref{eq:mu_vs_delta_0GPa})-(\ref{eq:mu_vs_delta_60GPa}). The physical origin of positive $\mu$ seems to come from the larger positive charge surrounding the OP than the inner plane caused by the ions including the block layer, which is not taken into account by the VMC calculations. It requires the positive $\mu$ correction. Under pressure, a closer distance between the IP and OPs makes their chemical potential difference weaker, as is expected.

\section{%
Procedure of energy-variance extrapolation of the energy
}
\label{App:variance_extrapolate_sc}

In strongly correlated electron systems, the ground-state candidate states described by the variational wave functions often energetically compete with one another.
To fairly determine the true ground state, it is desirable to estimate the eigenenergy of each state by extrapolating the energy to the zero-variance limit.
The variance
$(\delta E)^2 = (\langle H^2 \rangle - \langle H \rangle^2)/\langle H \rangle^2$
can be systematically reduced by improving the quality of the wave functions.
For this purpose, one can increase the number of variational parameters, such as those in the RBM wave functions,
and apply the power-Lanczos method to the optimized wave functions.
The variance usually gets smaller as we improve the wave functions through this prescribed order. When the variational state is sufficiently close to the ground state,
the energy difference 
between the ground state and the variational wave function 
becomes proportional to the energy variance~\cite{imada2000}.
The ground state is identified as the candidate state that achieves the lowest zero-variance energy after the linear extrapolation.
This energy-variance extrapolation
was shown to be useful
and possess a wide versatility as discussed in prior research~\cite{wu2024}.
Furthermore, the proportionality constant was shown to be nearly universal when the energy origin is set at the infinite temperature
irrespective of the choices of the model Hamiltonian and the lattice structure~\cite{wu2024}.
Given its high degree of general applicability, this approach has been applied to various complex systems.
For instance, the method has been applied to determine the ground states among
competing stripe states of different periods,
as well as SC and AF states, in various Hubbard-like models and \textit{ab initio} Hamiltonians and demonstrated its usefulness~\cite{ido2018,ohgoe2020,schmid2023}.
For an example of the variance extrapolation, see Appendix \ref{App:analysis_coexist_af_sc}.

\section{%
Details of the SC correlation
}
\label{App:details_sc_correlation}

\begin{figure}[h!]
\centering
\includegraphics[width=\columnwidth]{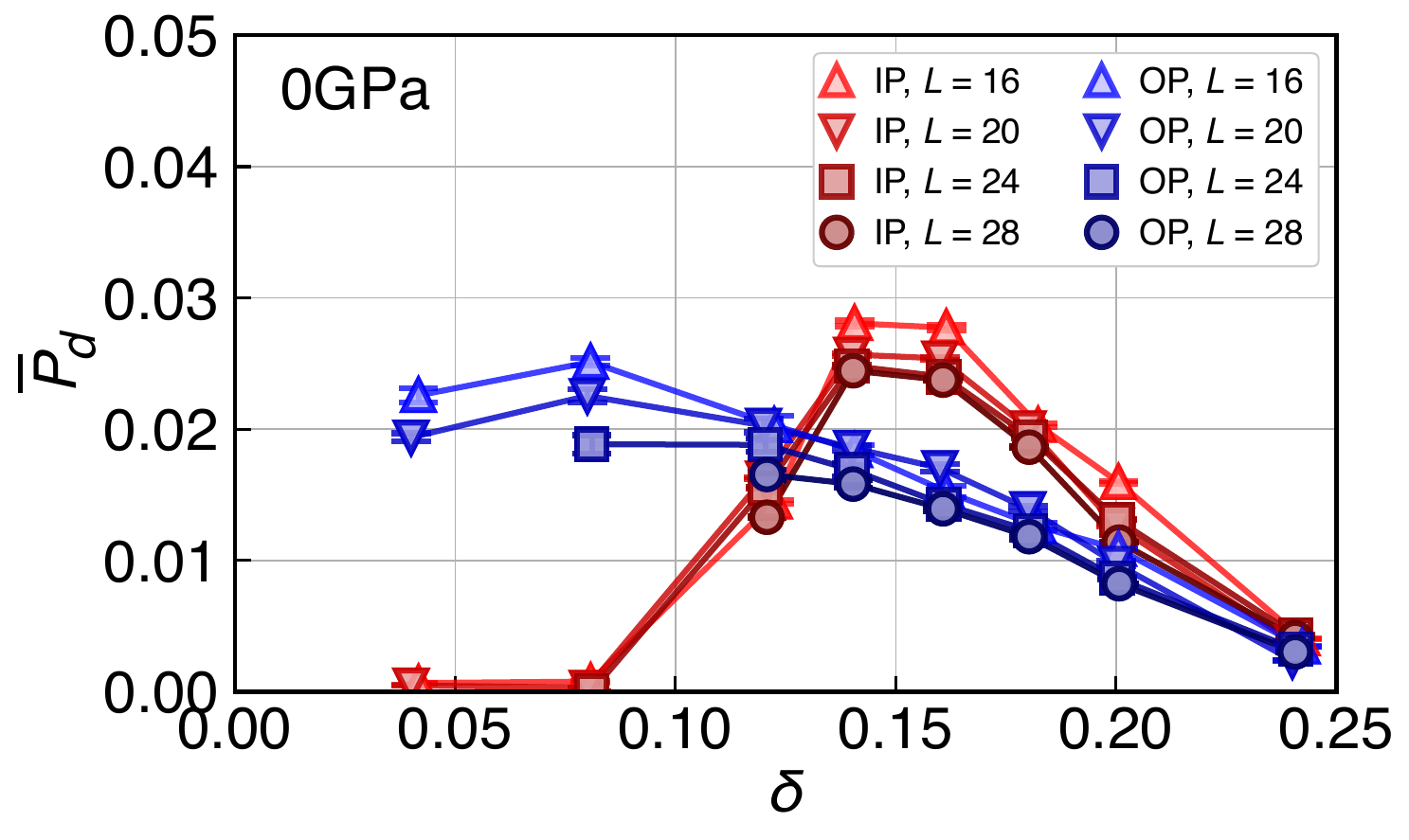}
\caption{%
Doping concentration dependence of the SC correlation averaged over long distances at $0$\,GPa.
We observe dome structures both in the IP and OPs.
}
\label{fig:dome_0GPa}
\end{figure}

\begin{figure}[h!]
\centering
\includegraphics[width=\columnwidth]{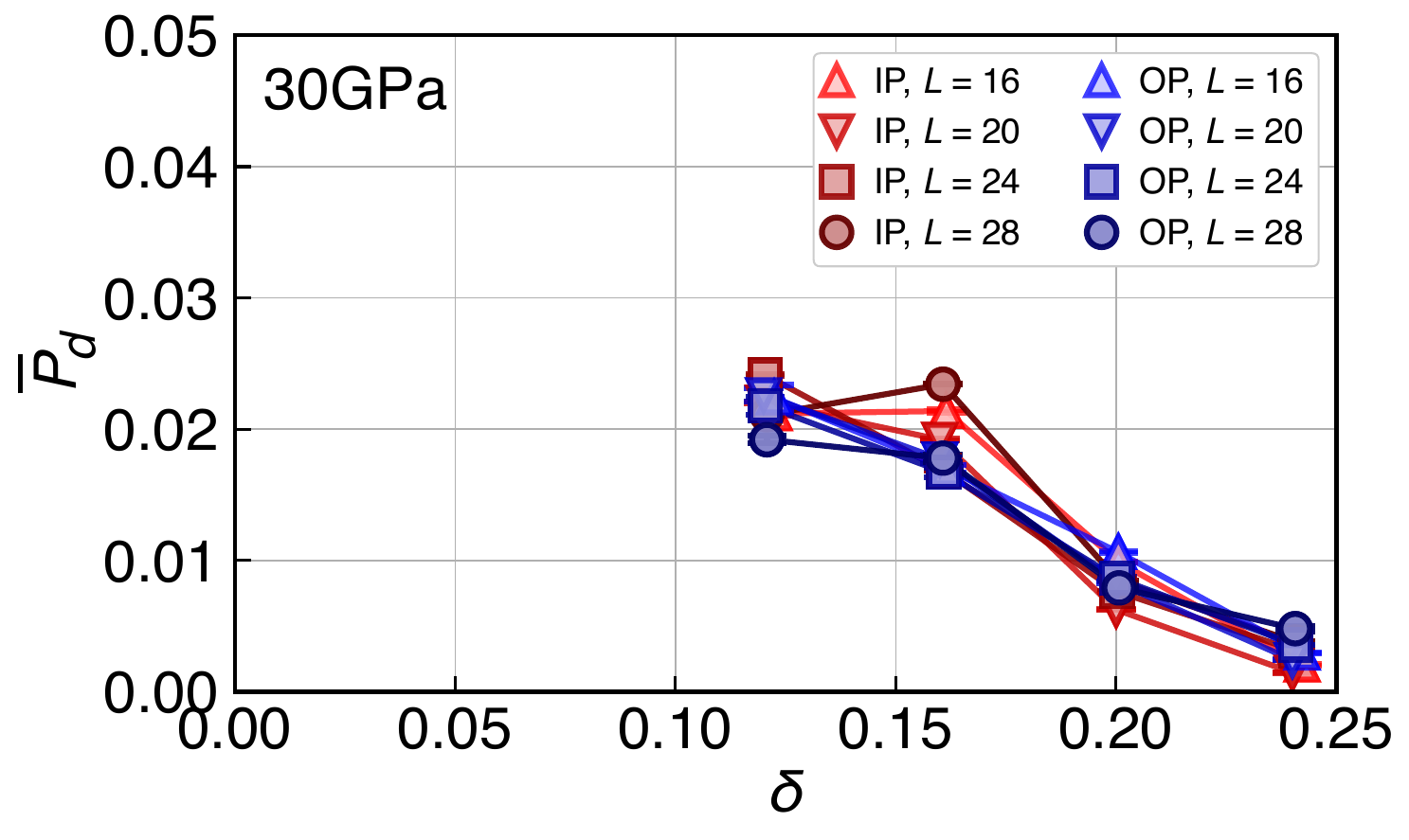}
\caption{%
Doping concentration dependence of the SC correlation averaged over long distances at $30$\,GPa.
}
\label{fig:dome_30GPa}
\end{figure}

\begin{figure}[h!]
\centering
\includegraphics[width=\columnwidth]{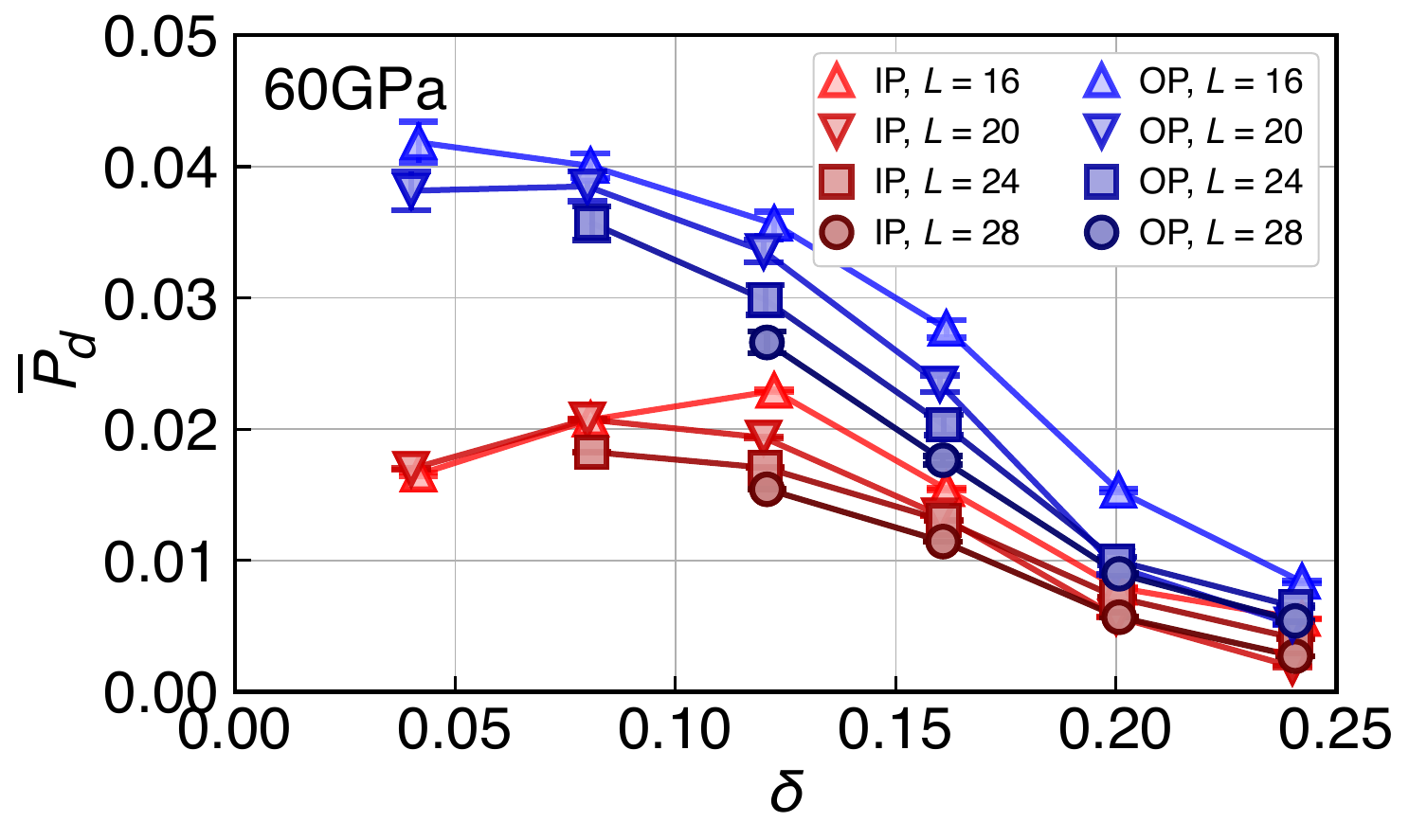}
\caption{%
Doping concentration dependence of the SC correlation averaged over long distances at $60$\,GPa.
}
\label{fig:dome_60GPa}
\end{figure}

We show the doping concentration dependencies of the SC correlation $\bar{P}_d$ averaged over long distances in
Figs.~\ref{fig:dome_0GPa}, \ref{fig:dome_30GPa}, and \ref{fig:dome_60GPa} with size dependence for calculated finite-size lattices.
The characteristics of the dome structures between the IP and OP vary as a function of pressure.

\section{%
Procedure of size extrapolation for the SC correlation function
}
\label{App:size_extrapolate_sc}
\begin{figure}[h!]
\centering
\includegraphics[width=\columnwidth]{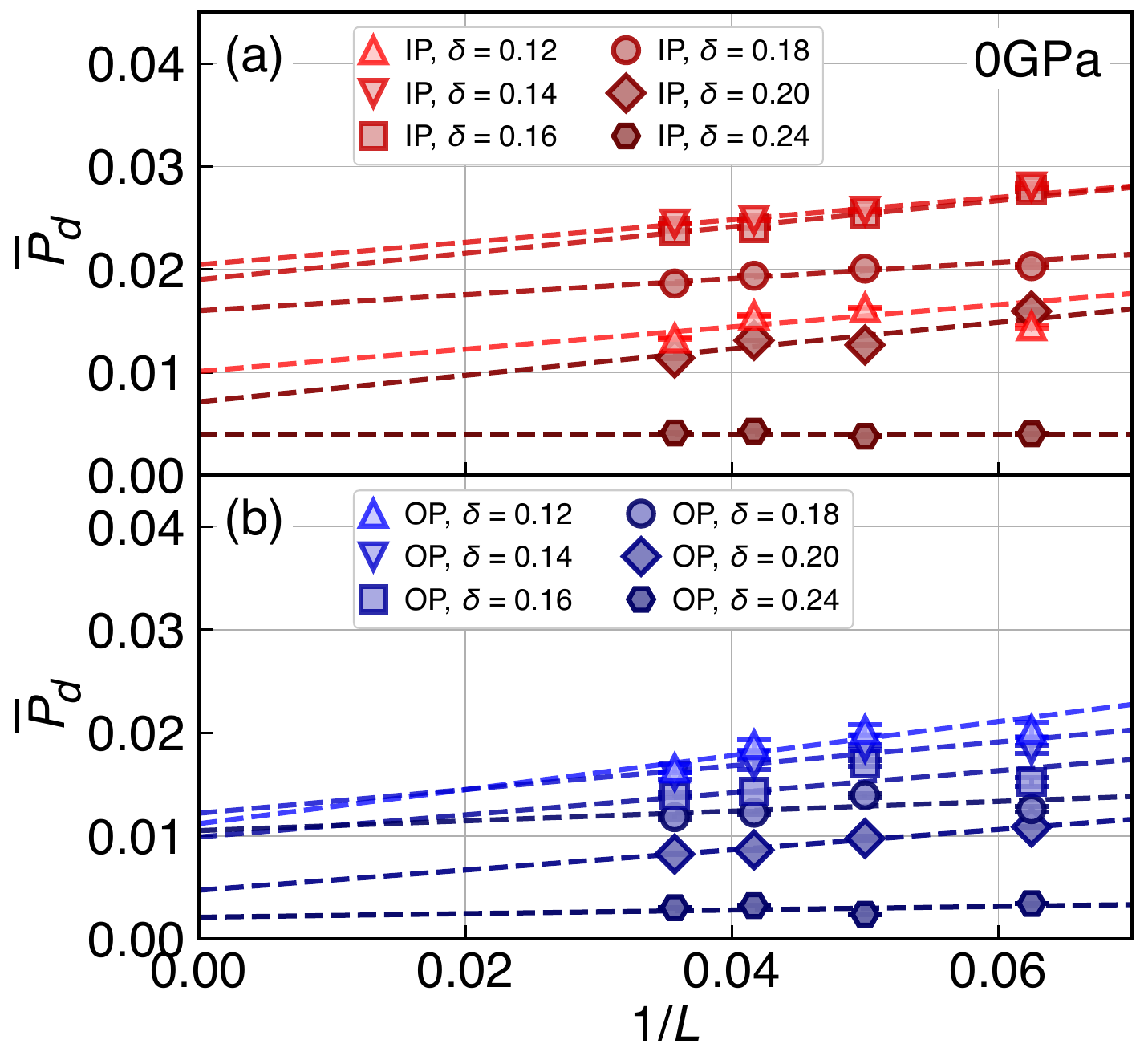}
\caption{%
Size extrapolation of the SC correlation averaged over long distances at $0$\,GPa.
We show the extrapolation for the (a) IP and (b) OPs.
}
\label{fig:pdr_vs_L_0GPa}
\end{figure}

\begin{figure}[h!]
\centering
\includegraphics[width=\columnwidth]{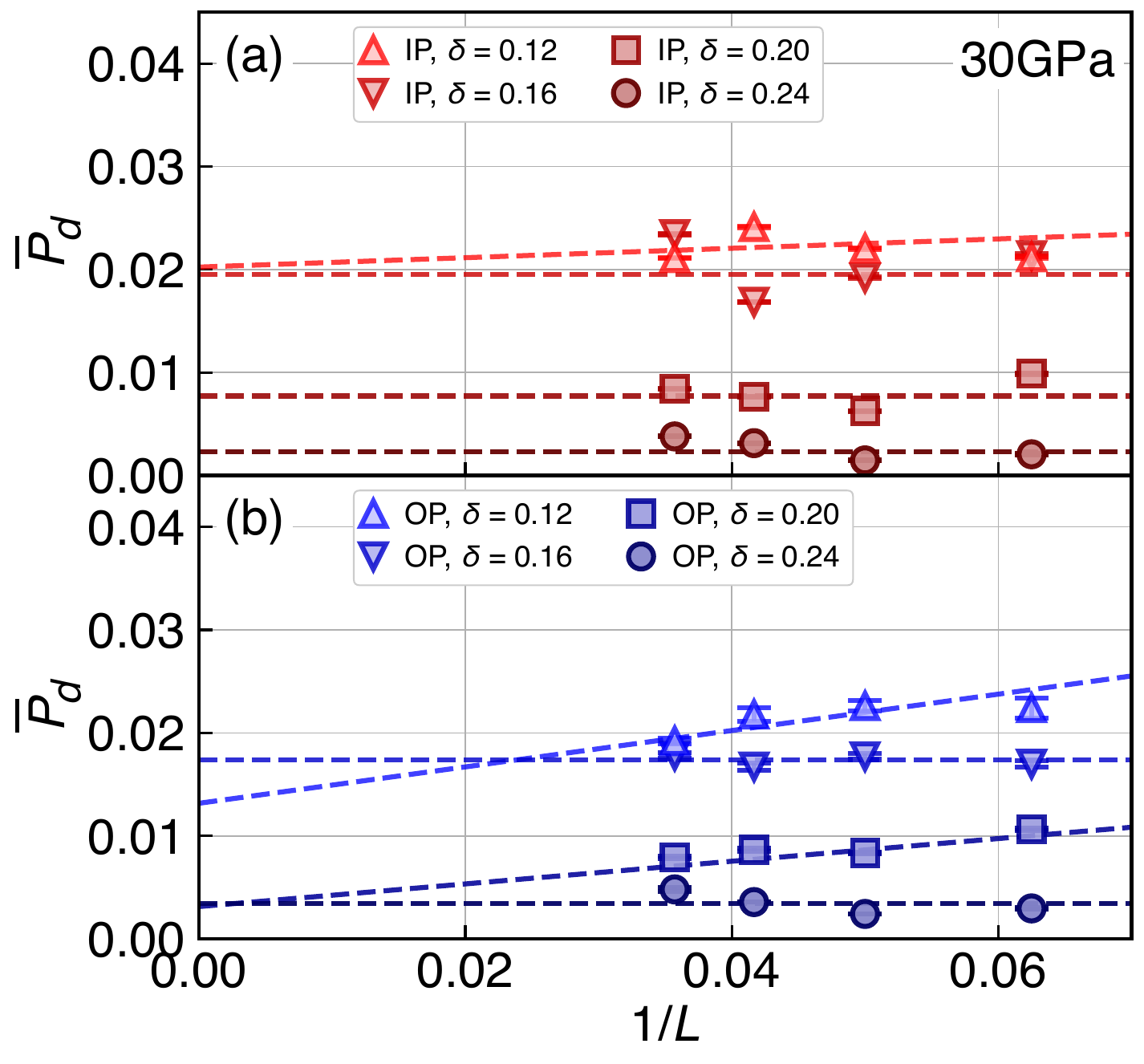}
\caption{%
Size extrapolation of the SC correlation averaged over long distances at $30$\,GPa.
We show the extrapolation for the (a) IP and (b) OPs.
}
\label{fig:pdr_vs_L_30GPa}
\end{figure}

\begin{figure}[h!]
\centering
\includegraphics[width=\columnwidth]{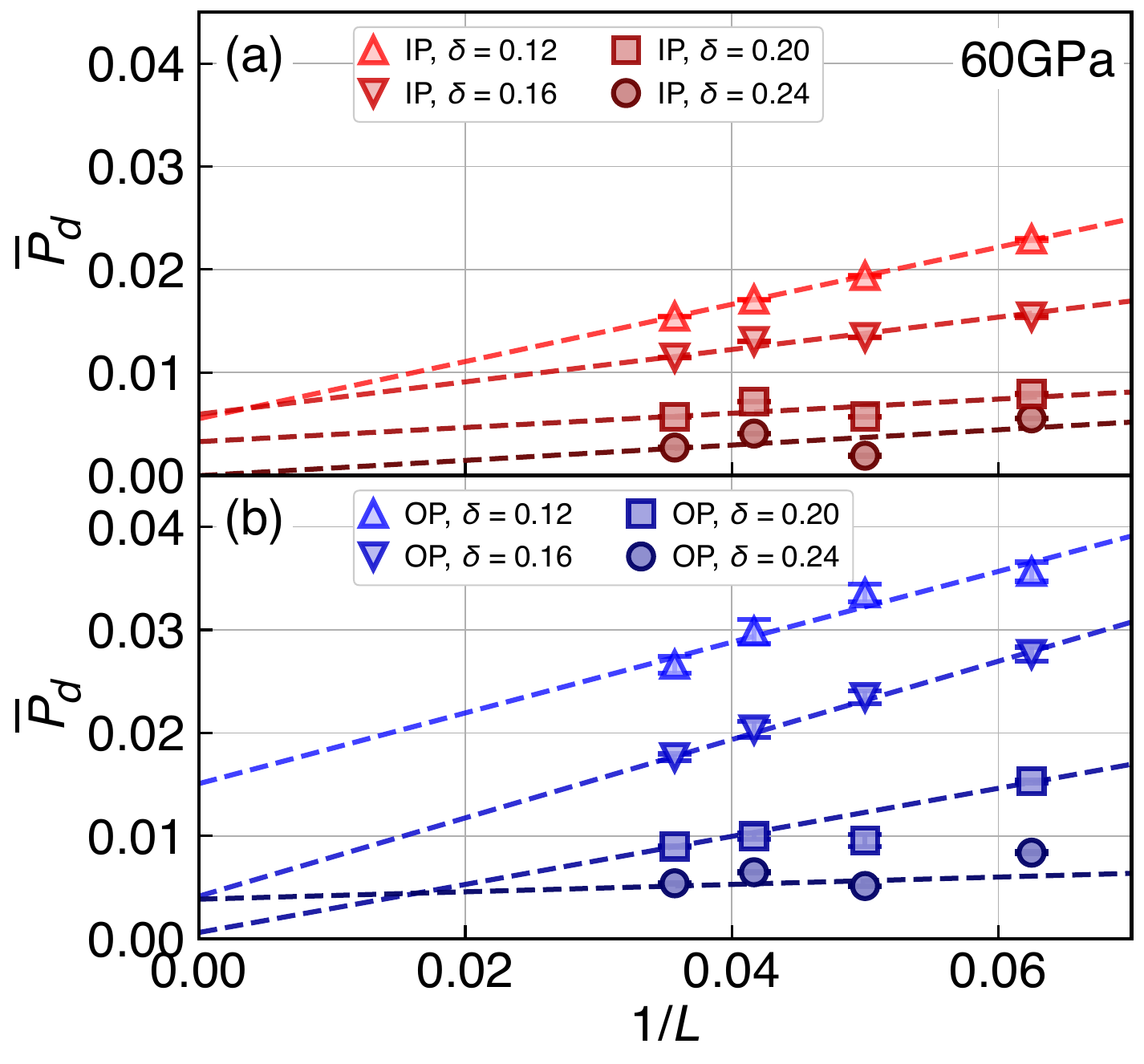}
\caption{%
Size extrapolation of the SC correlation averaged over long distances at $60$\,GPa.
We show the extrapolation for the (a) IP and (b) OPs.
}
\label{fig:pdr_vs_L_60GPa}
\end{figure}

By using the SC correlations $\bar{P}_d$ averaged over long distances obtained in Appendix~\ref{App:details_sc_correlation}, we extrapolate them to the thermodynamic limit.
We summarize this size extrapolation
in Figs.~\ref{fig:pdr_vs_L_0GPa}, \ref{fig:pdr_vs_L_30GPa}, and \ref{fig:pdr_vs_L_60GPa}.
Because the $d$-wave SC order possesses a gapless linear excitation, the size dependence of the SC correlation should be proportional to $1/L$ with $L$ being the length of the lattice.
All the data points are well fitted by this scaling formula.
The SC order parameter $F_{\rm SC}$ in the thermodynamic limit is obtained from the square root as $F_{\rm SC}=\sqrt{\bar{P}_d}$.

\section{%
Analysis of coexisting AF and SC orders
}
\label{App:analysis_coexist_af_sc}

\begin{figure}[!t]
\centering
\includegraphics[width=\columnwidth]{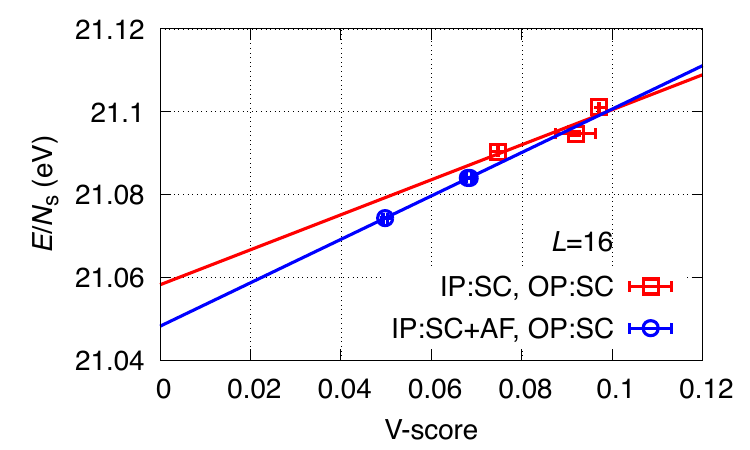}
\caption{%
Variance extrapolation of energies of ground-state candidate states.
The abscissa is chosen as the V-score~\cite{wu2024}, which is defined by $N_{\mathrm{s}}(\langle H^2 \rangle - \langle H \rangle^2)/(\langle H \rangle - E_{\infty})^2$ with $E_{\infty}$ being the infinite-temperature energy.
The states are obtained for $L=16$ at $8$\% doping
with the chemical potential difference $\mu=3.2$.
We show the energy of the pair-product wave function,
that of the RBM wave function with $\alpha=8$,
and that after the one-step Lanczos method, which are aligned in the plots from the right to left in this order.
At $\mu=3.2$, the energy of the state exhibiting coexistence of
SC and AF orders in the IP is lower than
that of the state with inner-layer SC order.
Note that the slope of the variance extrapolation is universal regardless of the details of the Hamiltonian, and the slope on a logarithmic scale in the present system
[e.g., for the SC+AF state of the pair-product wave function, $\ln\frac{\langle H \rangle - E_0}{E_{\infty} - E_0} - \ln\mathrm{(V\text{-}score)} \sim \ln(0.0269)-\ln(0.0685)\sim -0.935$, with $\langle H \rangle = 16192.51$, $E_{\infty}=17186.87$, and the extrapolated energy being $E_0=16165.02$] falls within the range of the slope fluctuations reported in previous studies~\cite{wu2024}.
}
\label{fig:ene_cmp_sc_af_after_var_extrapolate}
\end{figure}

Here we show a realistic example of the coexistence.
Since a state with the coexisting AF and SC orders is severely competing with the purely SC state, we show the variance extrapolation of energies discussed in Appendix \ref{App:variance_extrapolate_sc} for the ground-state candidate states in Fig.~\ref{fig:ene_cmp_sc_af_after_var_extrapolate}.
The data are for the system size $L=16$ at $8$\% doping with the chemical potential difference between the IP and OP being $\mu=3.2$,
which is
smaller than the value corresponding to the \textit{ab initio} 
Hamiltonian with the GW-based estimate of the LRFB.
Here, the chemical potential shift $\mu$ is adjusted by hand so that the hole doping concentration in the IP becomes nonzero ($\delta_{\mathrm{IP}}=0.07$ in the present case) and should be regarded as a parameter search to explore the possibility of the coexistence in a physically relevant doped system.
The IP is closer to half filling than the OP in this parameter region, and thus, the AF correlation is 
more developed in the IP. 
Ground-state candidate states include the state exhibiting SC order in both IP and OP, the state exhibiting AF order in both IP and OP,
and the state exhibiting SC order in the OP, but coexisting SC and AF orders in the IP.
Although the variance extrapolation in Fig.~\ref{fig:ene_cmp_sc_af_after_var_extrapolate} shows some uncertainty, the similar slopes of the two competing states support the stability of the extrapolation, and 
the ground state exhibits coexisting SC and AF orders over a certain parameter range.

\section{%
Universality of attraction
}
\label{App:universality_attraction}

\begin{figure}[!t]
\centering
\includegraphics[width=\columnwidth]{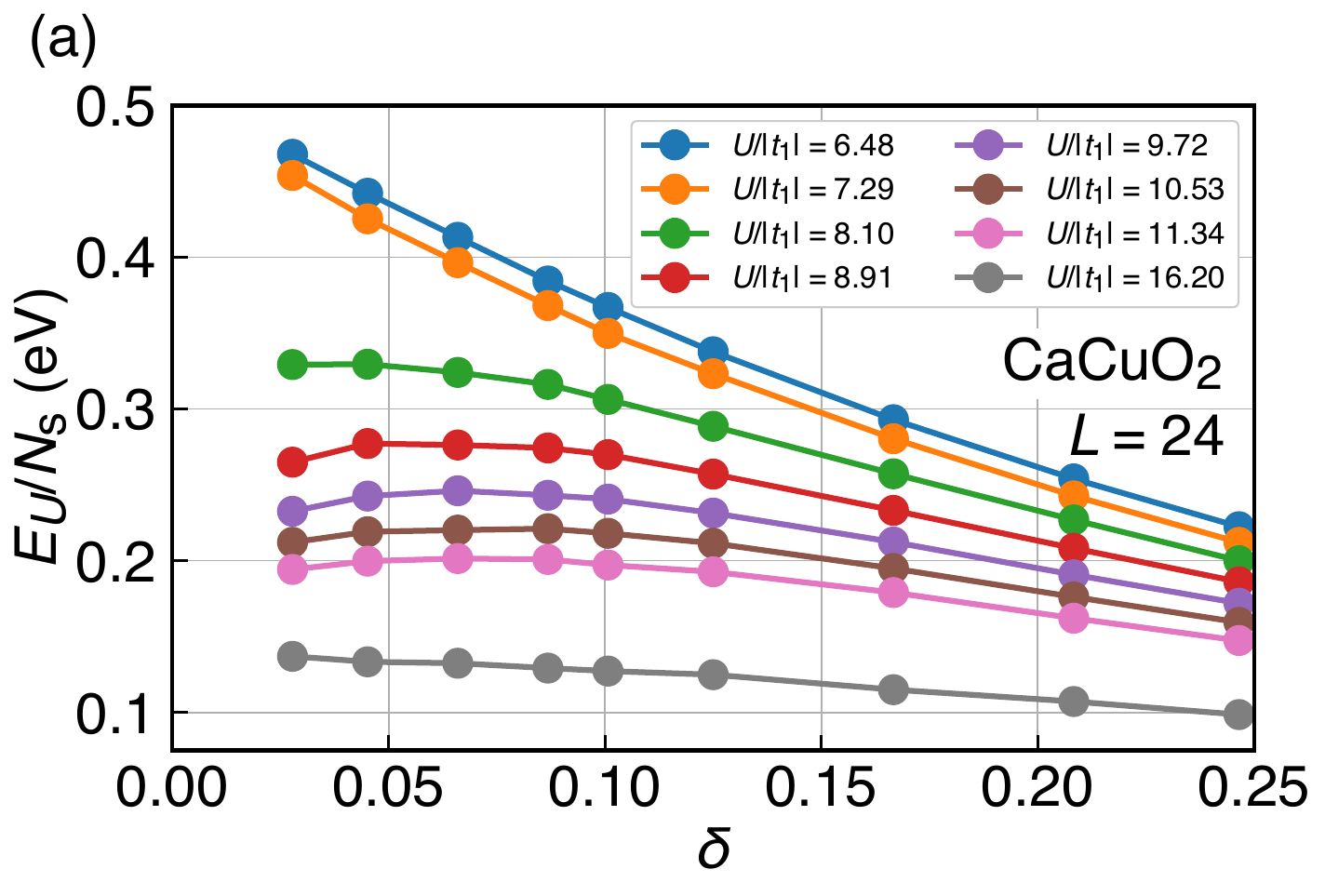}
\\
\includegraphics[width=\columnwidth]{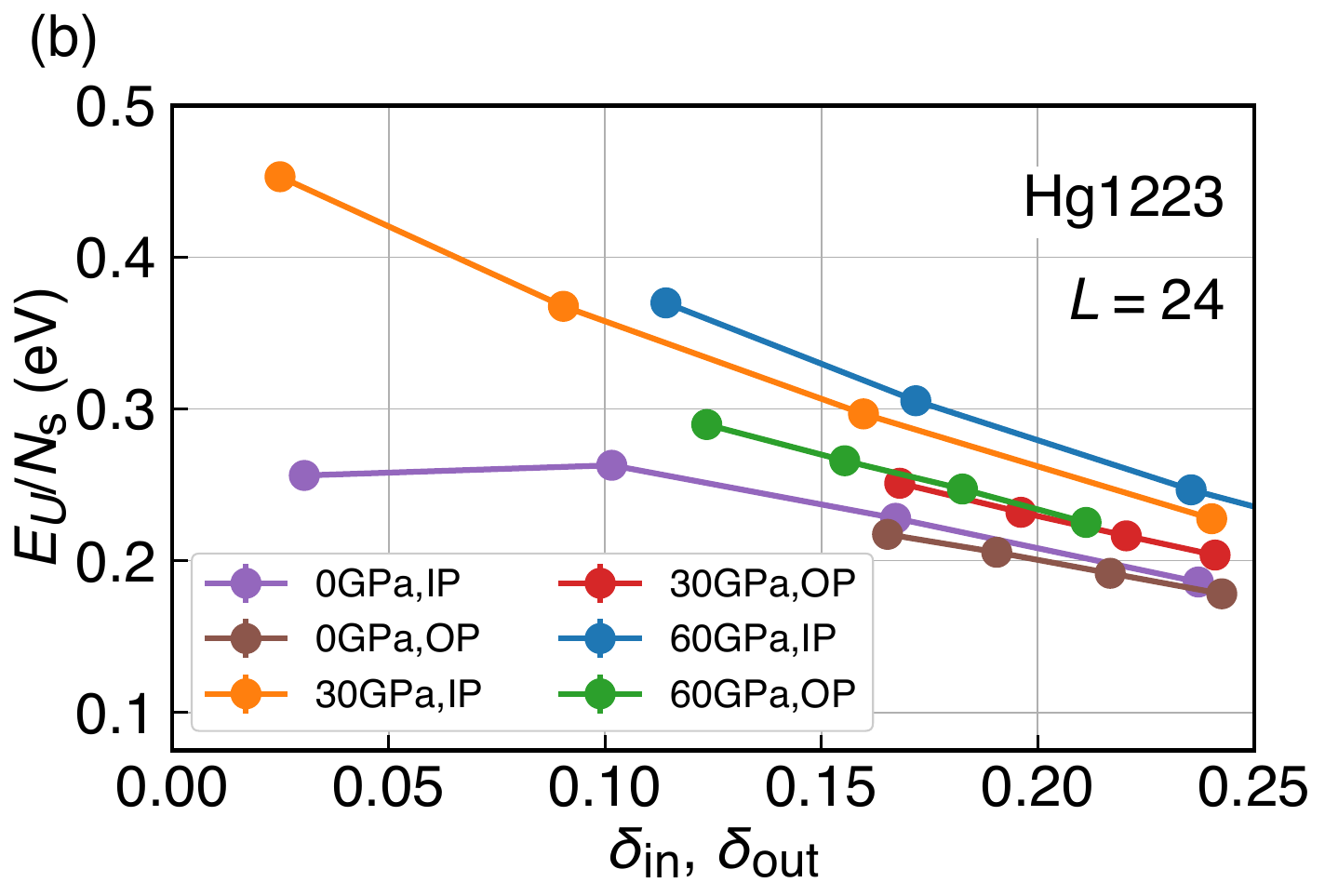}
\\
\includegraphics[width=\columnwidth]{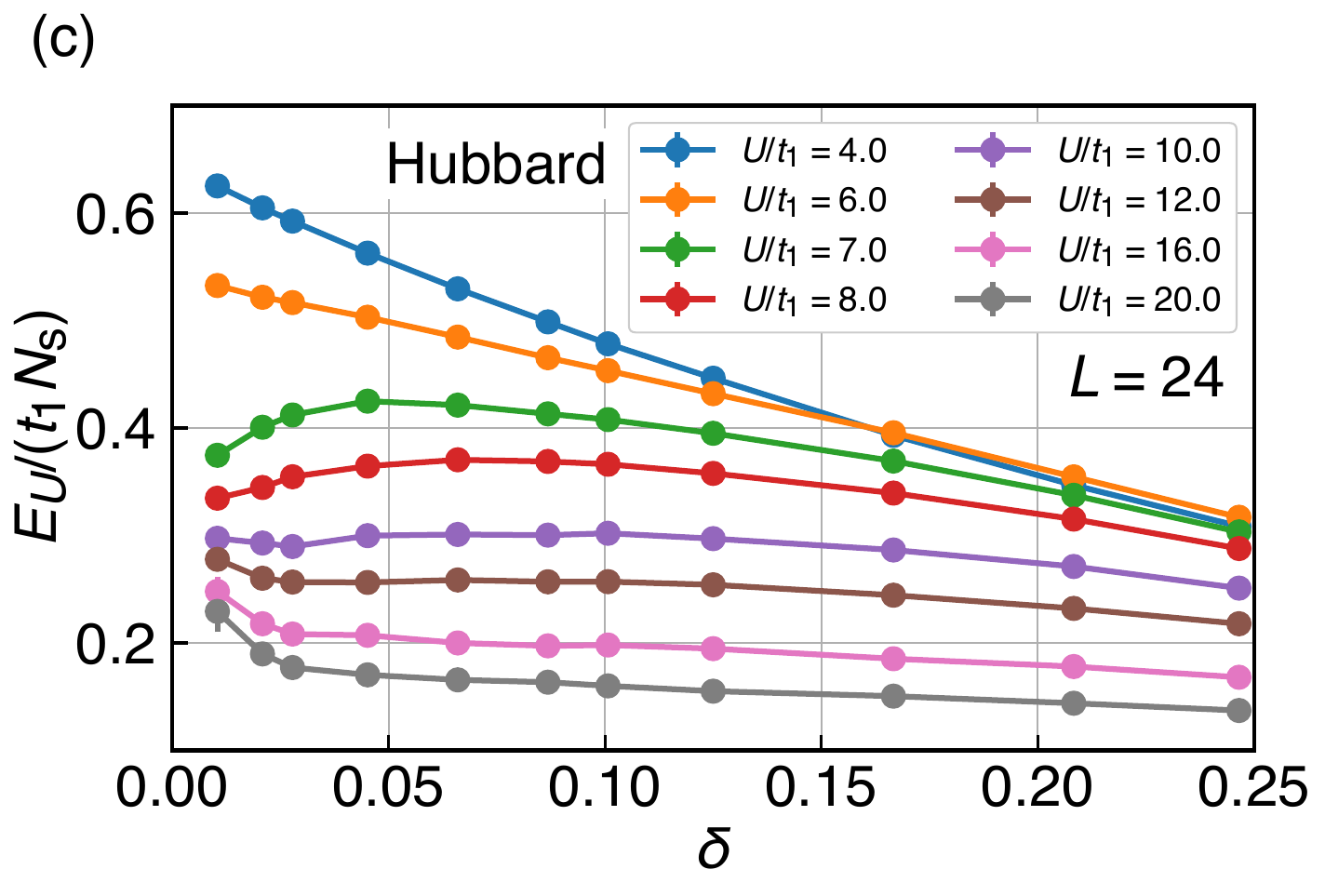}
\caption{%
Doping concentration ($\delta$) dependence of
the onsite Coulomb interaction energy $E_U$ as a function of $U/|t_1|$
for CaCuO$_2$ (a),
Hg1223 (b),
and the simple Hubbard model (c) with the size $L=24$.
The energy $E_U$ commonly shows the concave upward $\delta$ dependence for weaker $U/|t_1|$, whereas it exhibits the concave downward $\delta$ dependence indicating attractive interaction for stronger $U/|t_1|$.
In the case of the simple Hubbard model, we also observe the concave upward $\delta$ dependence (namely, repulsive interaction) for very large $U$ in the underdoped region.
}
\label{fig:EU_vs_delta_cacuo2_hg1223}
\end{figure}

\begin{figure}[!t]
\centering
\includegraphics[width=\columnwidth]{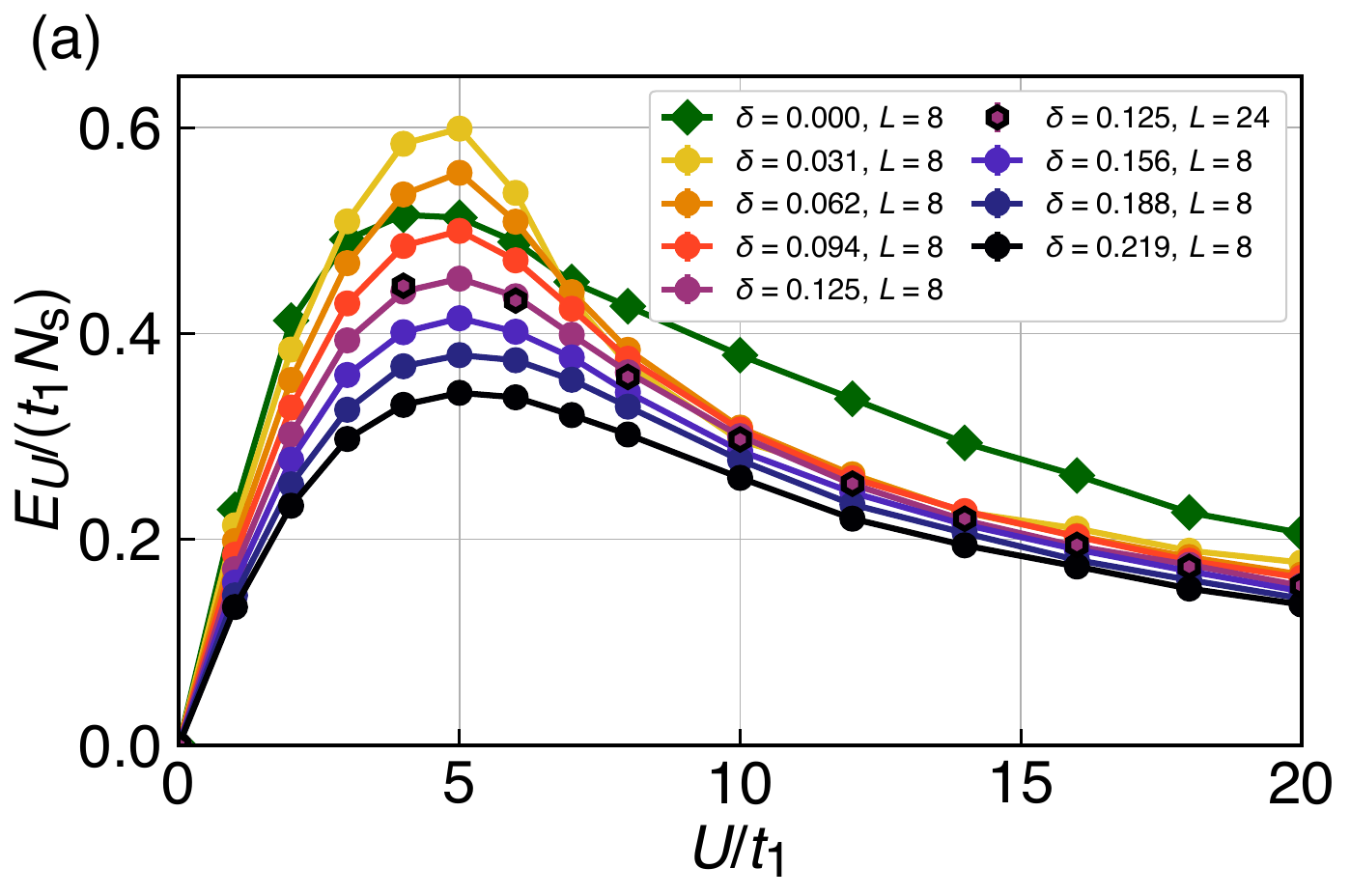}
\\
\includegraphics[width=\columnwidth]{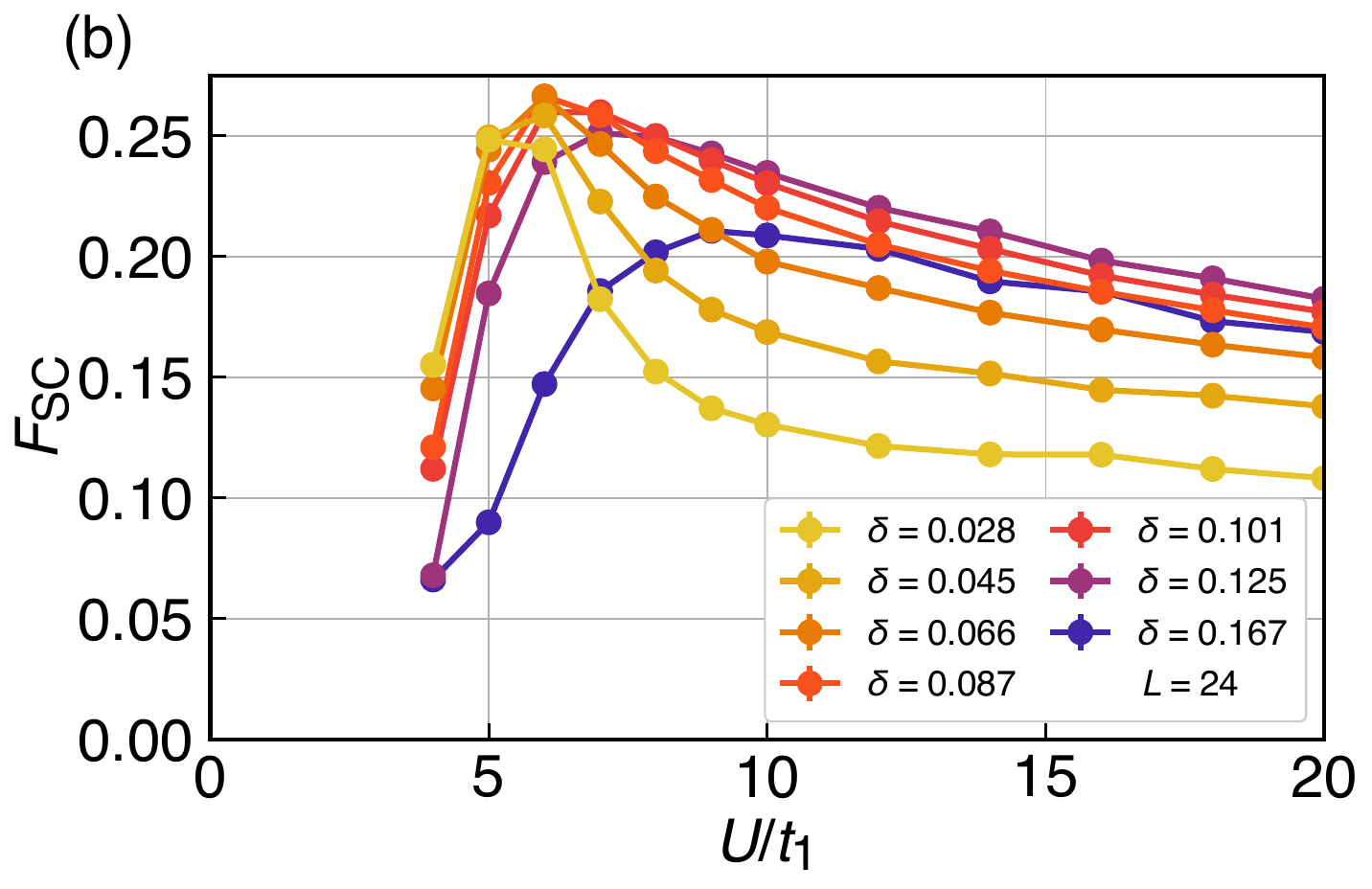}
\\
\includegraphics[width=\columnwidth]{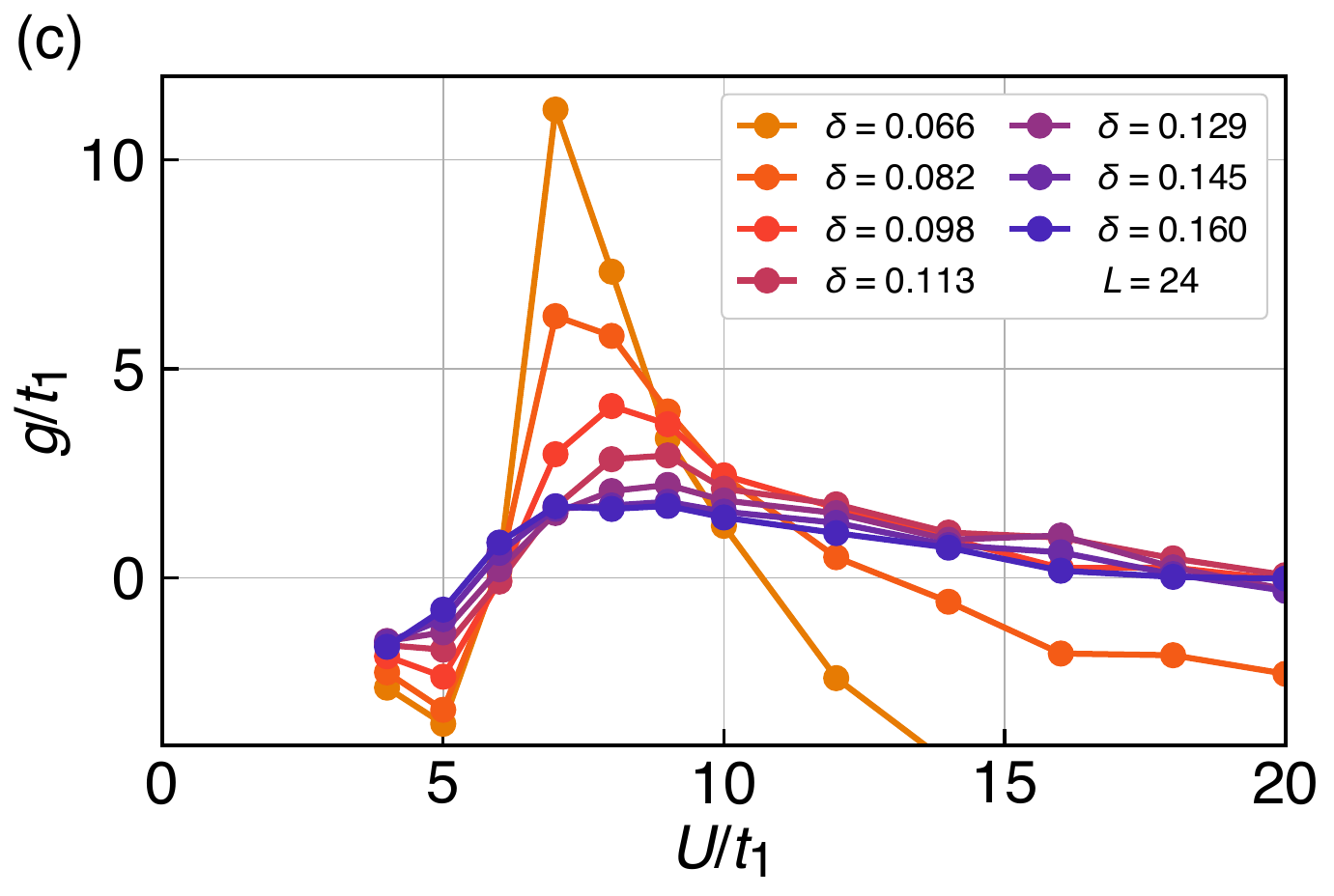}
\caption{%
Onsite interaction ($U$) dependence of
the onsite Coulomb interaction energy $E_U$ (a),
the SC order parameter $F_{\mathrm{SC}}$ (b),
and the estimated attraction $g$ (c)
at various choices of $\delta$
for the simple Hubbard model on a square lattice.
The data of the AF states are shown at half filling ($\delta=0$),
whereas those of the SC states are presented for nonzero doping concentrations ($\delta>0$).
The onsite Coulomb interaction energies, $E_U$'s per site, are nearly the same between system sizes $L=8$ and $L=24$ and they are close to the values in the thermodynamic limit.
For $F_{\mathrm{SC}}$ and $g$, we plot selected data for $L=24$.
All the physical quantities exhibit a peak at $U\sim 5\text{--}8$, which gradually increases upon doping for  $F_{\mathrm{SC}}$.
The peak intensities of both $F_{\mathrm{SC}}$ and $g$ diminish with increasing the doping concentration.
}
\label{fig:EU_vs_U_normal_hubbard}
\end{figure}

\begin{figure}[!t]
\centering
\includegraphics[width=\columnwidth]{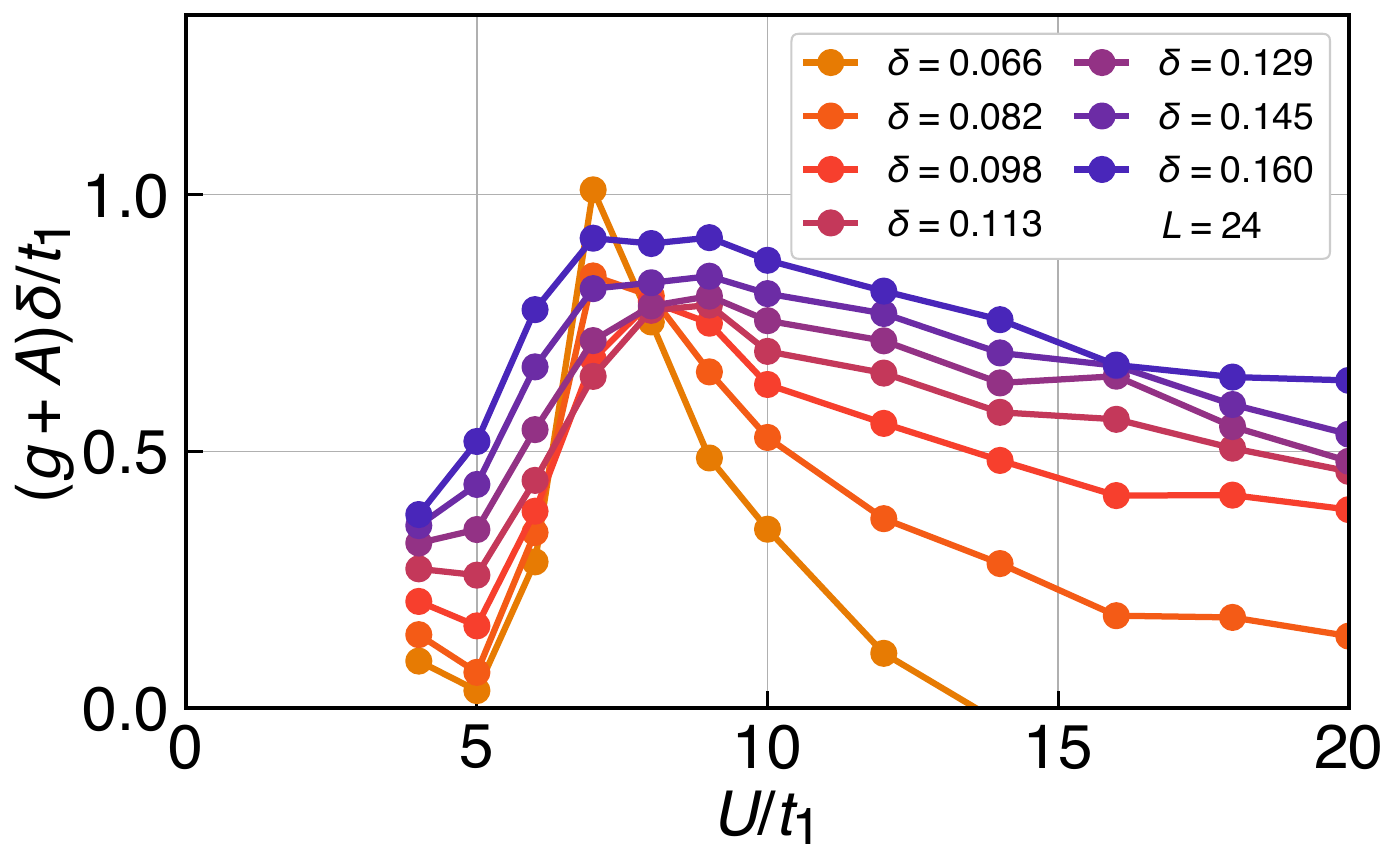}
\caption{%
Onsite interaction ($U$) dependence of the product of the effective attraction $g$ and the doping concentration $\delta$ for the simple Hubbard model on a square lattice with the size $L=24$.
Here we choose the constant shift of $g$ as $A=4$,
which corresponds to the minimum value of $g$ for a weak interaction region,
as shown in Fig.~\ref{fig:EU_vs_U_normal_hubbard}(c).
The quantity $(g+A)\delta$ mostly reproduces the interaction dependence of $F_{\mathrm{SC}}$ in Fig.~\ref{fig:EU_vs_U_normal_hubbard}(b).
For an underdoped region, the effective attraction from the onsite Coulomb interaction is relatively weaker under strong interaction,
and instead, the effective attraction driven by the kinetic energy plays an important role, as shown in Fig.~\ref{fig:g_from_Et1}.
}
\label{fig:gdelta_vs_U_normal_hubbard}
\end{figure}

Figures~\ref{fig:EU_vs_delta_cacuo2_hg1223}
show the doping concentration ($\delta$) dependence of the onsite Coulomb interaction energy $E_U$ per site as a function of $U/|t_1|$ for CaCuO$_2$ (a) and Hg1223 (b).
For Hg1223, the doping concentration dependencies are 
presented separately for the IP and OP.
Below $U/|t_1|=8$, the onsite Coulomb interaction energy for CaCuO$_2$ exhibits the concave upward $\delta$ dependence with the positive curvature, corresponding to nearly vanishing
effective attraction and weak SC order.
Above $U/|t_1|=8$, the $\delta$ dependence of $E_U$ turns concave downward, inducing the effective attraction that may stabilize superconductivity.
When $U/|t_1|$ becomes very large ($U/|t_1|\gtrsim 16$), the onsite Coulomb interaction energy decays almost linearly in $\delta$, suppressing the effective attraction.

The $\delta$ dependence of $E_U$ in Hg1223 follows the same trend as that of CaCuO$_2$.
Near the optimal doping ($16$\%), $T_{\mathrm{c}}$ is dominated by the SC order in IP.
As the pressure increases, the parameter $U/|t_1|$ in IP decreases from $9.30$ to $6.43$ (see Tables~\ref{tab:parameters_0GPa}--\ref{tab:parameters_60GPa}).
Reflecting this reduction in $U/|t_1|$, the $\delta_{\mathrm{IP}}$ dependence of $E_{U}$ of IP, as shown in Fig.~\ref{fig:EU_vs_delta_cacuo2_hg1223}(b), changes from concave downward to concave upward as the pressure increases.
This observation suggests a reduction in the effective attraction due to pressure, consistent with the decrease in the SC order at high pressure (see Fig.~\ref{fig:fsc_t1}).

To examine the universality of the pairing mechanism ascribed to the emergent local attraction generated by $E_U$, we have also calculated the local attraction $g$ and $F_{\rm SC}$ for the simple Hubbard model, which has only the nearest-neighbor hopping $t_1$ and the local repulsion $U$ in the metastable uniform $d$-wave SC state. (Note that the ground state has the charge inhomogeneous stripe order, and here, we focus on a metastable excited state that shows the SC order.)
For the case of the simple Hubbard model, we take the energy unit by $t_1$.

The similar trend in the $\delta$ dependence of $E_U$ already appears in the simple Hubbard model without any further neighbor Coulomb interaction and hopping parameters [see Fig.~\ref{fig:EU_vs_delta_cacuo2_hg1223}(c)]. Because the \textit{ab initio} Hamiltonians include substantial off-site Coulomb interaction parameters, the screening effects push the onsite Coulomb interaction that maximizes SC order to a stronger value than the Hubbard model. Indeed, as shown in Fig.~\ref{fig:EU_vs_U_normal_hubbard}, the Hubbard model exhibits the SC-order peaks at $U/|t_1|\sim 6$-8, whereas the cuprate Hamiltonians display the SC-order peaks at $U/|t_1|\sim 8$-10. Consistent with the interaction dependence of the SC order, the estimated effective attraction in the simple Hubbard model also peaks at $U/|t_1|\sim$ 7-8, which is slightly weaker than the peak interaction for the \textit{ab initio} Hamiltonians. The peak of the SC order around $U/|t_1|\sim 6$-8 is in overall agreement with earlier studies of the Hubbard model~\cite{yokoyama2013,kitatani2023}. The asymmetric peak structures both for $F_{\rm SC}$ and $g$ in Fig.~\ref{fig:EU_vs_U_normal_hubbard} with a long tail at large $U$ are similar to the case of the cuprates in Figs.~\ref{fig:Fsc_vs_u} and \ref{fig:c2_vs_u}. The interaction that maximizes $F_{\rm SC}$ gradually increases in strength upon doping in addition to the reduction of the peak value. 
This trend is also similar to the cuprates demonstrated in Fig.~\ref{fig:Fsc_vs_u} and can commonly be understood from the screening by the doped carrier, which diminishes effective $U$. The core mechanism of this attraction in the cuprates is, therefore, likely to be already contained in the simple Hubbard model.

\begin{figure}[!t]
\centering
\includegraphics[width=\columnwidth]{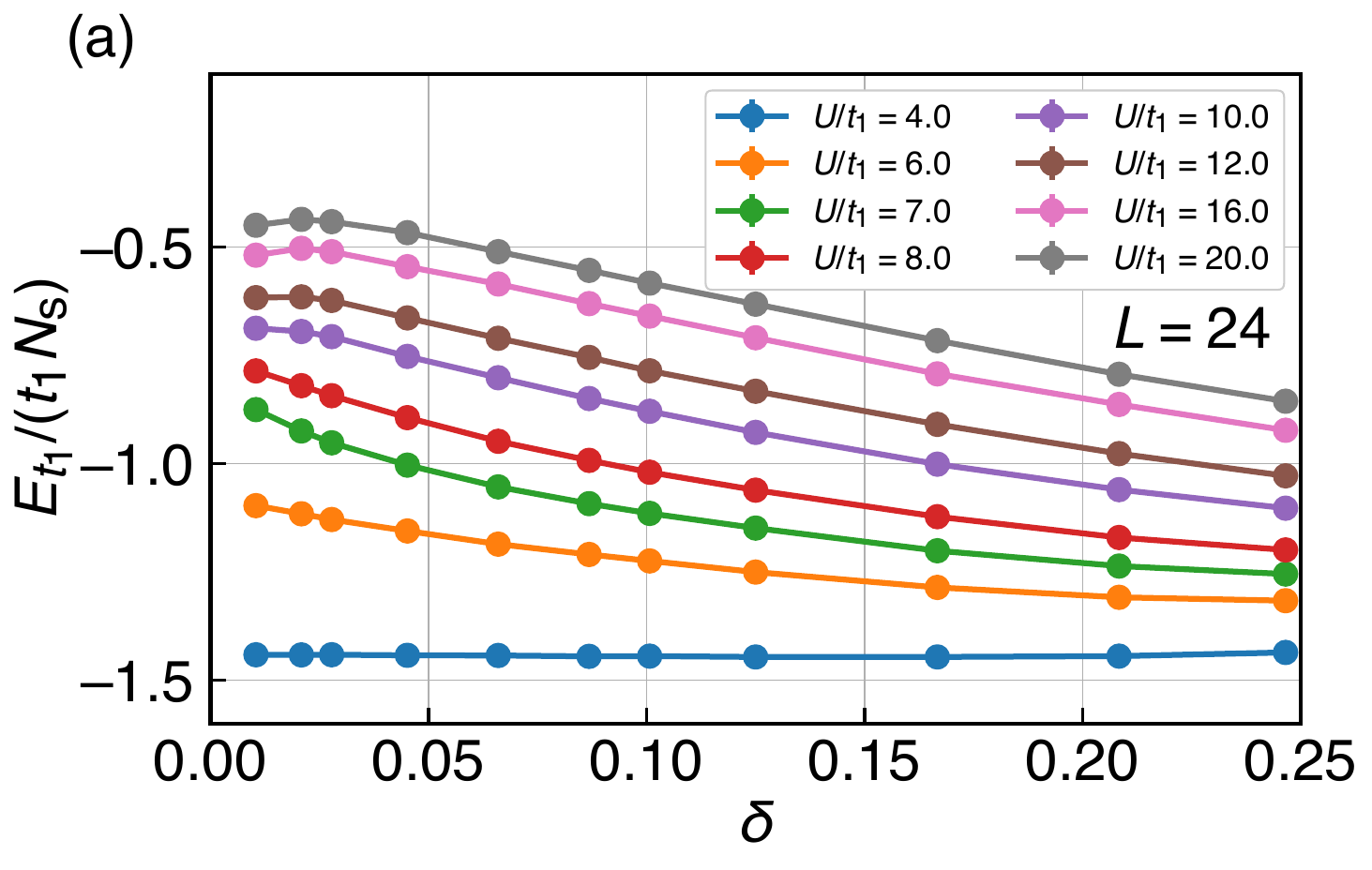}
\\
\includegraphics[width=\columnwidth]{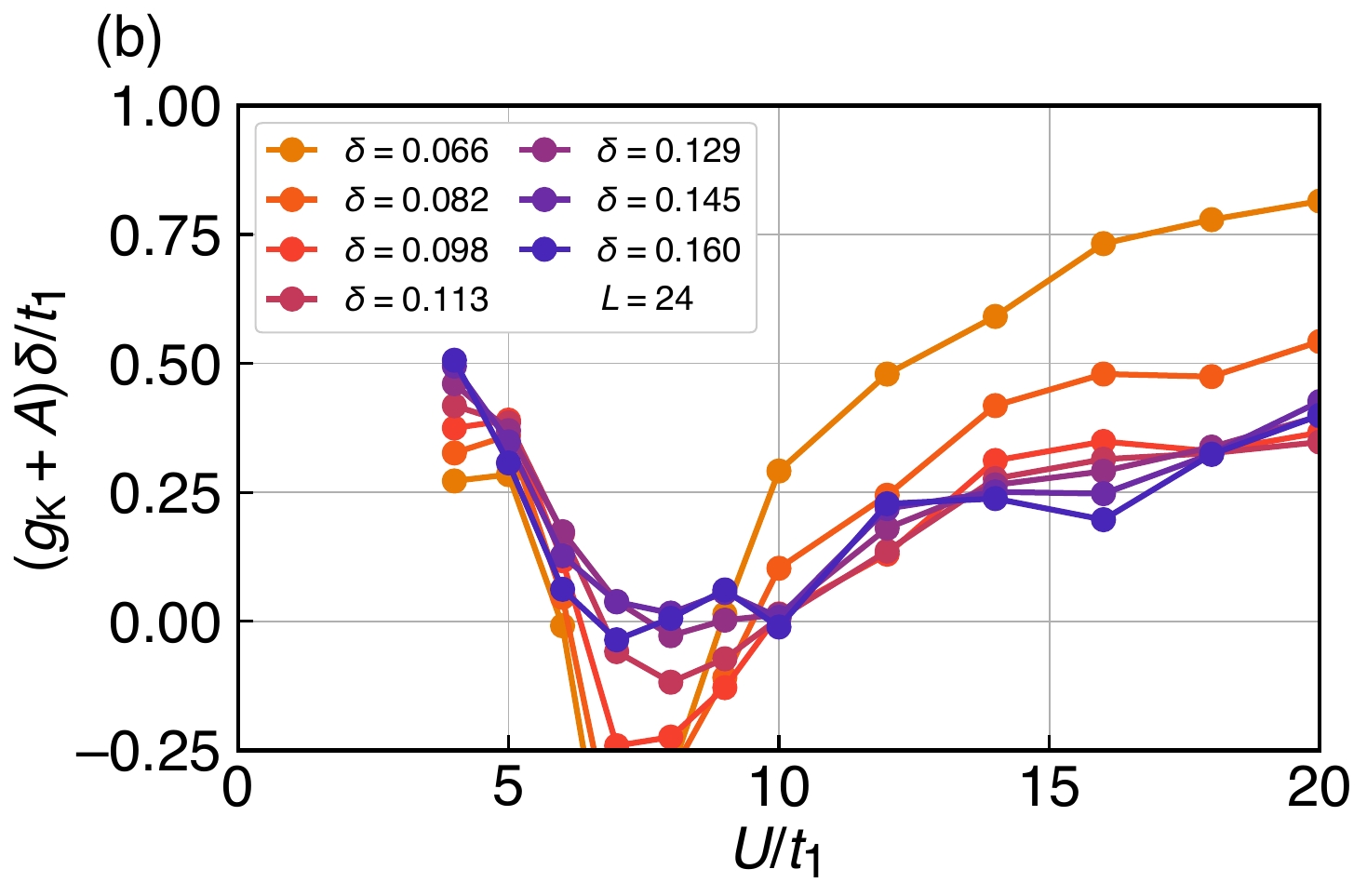}
\caption{%
Doping concentration ($\delta$) dependence of
the kinetic energy $E_{t_1}$ as a function of $U$
for the simple Hubbard model (a) with the size $L=24$.
Onsite interaction, $U$ dependence of the product of the doping concentration $\delta$ and the effective attraction $g_{\mathrm{K}}$ extracted from $E_{t_1}$ (b).
Here we choose the constant shift of $g_{\mathrm{K}}$ as $A=4$
as in Fig.~\ref{fig:gdelta_vs_U_normal_hubbard}.
The effective attraction driven by the kinetic energy emerges in an underdoped region when the Coulomb interaction is very strong.
}
\label{fig:g_from_Et1}
\end{figure}

Note that the effective
attraction also emerges in the kinetic energy in the strong-interaction limit, as was suggested in a previous study~\cite{yokoyama2013}.
To examine this, we also investigate the energy of the SC state in the simple Hubbard model on a square lattice without any off-site Coulomb interactions and further-neighbor hoppings.
When the onsite Coulomb interaction is very large ($U/|t_1|\sim 20$) and the system is underdoped (typically, $\delta<0.03$), we find the concave downward doping dependence of the nearest-neighbor hopping energy [see Fig.~\ref{fig:g_from_Et1}(a)], consistently with the emergent attraction caused by the kinetic energy [see Fig.~\ref{fig:g_from_Et1}(b)].
However, we do not observe such behavior in the cases of \textit{ab initio} low-energy effective Hamiltonians.
It supports
that the SC order in the real cuprate materials is mainly governed by the effective instantaneous attraction associated with the onsite Coulomb interaction energy rather than the kinetic energy.

\bibliographystyle{apsrev4-2_modified}
\bibliography{references.bib}

@article{momma2011,
  author = "Momma, Koichi and Izumi, Fujio",
  title = "{{\it VESTA3} for three-dimensional visualization of crystal, volumetric and morphology data}",
  journal = "J. Appl. Crystallogr.",
  year = "2011",
  volume = "44",
  number = "6",
  pages = "1272--1276",
  month = "Dec",
  doi = {10.1107/S0021889811038970},
  url = {https://doi.org/10.1107/S0021889811038970},
}

@article{finger1994,
  title = {Crystal chemistry of HgBa$_2$CaCu$_2$O$_{8+\delta}$ and HgBa$_2$Ca$_2$Cu$_3$O$_{8+\delta}$ single-crystal x-ray diffraction results},
  journal = {Physica C},
  volume = {226},
  number = {3},
  pages = {216-221},
  year = {1994},
  issn = {0921-4534},
  doi = {https://doi.org/10.1016/0921-4534(94)90197-X},
  url = {https://www.sciencedirect.com/science/article/pii/092145349490197X},
  author = {L.W Finger and R.M Hazen and R.T Downs and R.L Meng and C.W Chu},
}

@article{downs2003,
  author = {Downs, R. T. and Hall-Wallace, M.},
  title = {The American Mineralogist Crystal Structure Database},
  journal = {Am. Mineral.},
  year = {2003},
  volume = {88},
  pages = {247-250},
  url = {https://pubs.geoscienceworld.org/msa/ammin/article-abstract/88/1/247/43886/The-American-Mineralogist-crystal-structure},
}

@article{grazulis2009,
  author = "Gra{\v{z}}ulis, Saulius and Chateigner, Daniel and Downs, Robert T. and Yokochi, A. F. T. and Quir{\'{o}}s, Miguel and Lutterotti, Luca and Manakova, Elena and Butkus, Justas and Moeck, Peter and Le Bail, Armel",
  title = "{Crystallography Open Database {--} an open-access collection of crystal structures}",
  journal = "J. Appl. Crystallogr.",
  year = "2009",
  volume = "42",
  number = "4",
  pages = "726--729",
  month = "Aug",
  doi = {10.1107/S0021889809016690},
  url = {https://doi.org/10.1107/S0021889809016690},
}

@article{vaitkus2023,
  author    = {Vaitkus, Antanas and Merkys, Andrius and Sander, Thomas and Quir\'{o}s, Miguel and Thiessen, Paul A. and Bolton, Evan E. and Gra\v{z}ulis, Saulius},
  title     = {A workflow for deriving chemical entities from crystallographic data and its application to the {C}rystallography {O}pen {D}atabase},
  journal   = {J. Cheminform.},
  year      = {2023},
  volume    = {15},
  pages     = {123},
  number    = {1},
  month     = {Dec},
  doi       = {10.1186/s13321-023-00780-2},
  url       = {https://doi.org/10.1186/s13321-023-00780-2},
  publisher = {Springer Science and Business Media LLC},
}

@article{schmid2023,
  title = {Superconductivity Studied by Solving Ab Initio Low-Energy Effective Hamiltonians for Carrier Doped ${\mathrm{CaCuO}}_{2}$, ${\mathrm{Bi}}_{2}{\mathrm{Sr}}_{2}{\mathrm{CuO}}_{6}$, ${\mathrm{Bi}}_{2}{\mathrm{Sr}}_{2}{\mathrm{CaCu}}_{2}{\mathrm{O}}_{8}$, and ${\mathrm{HgBa}}_{2}{\mathrm{CuO}}_{4}$},
  author = {Schmid, Michael Thobias and Mor\'ee, Jean-Baptiste and Kaneko, Ryui and Yamaji, Youhei and Imada, Masatoshi},
  journal = {Phys. Rev. X},
  volume = {13},
  issue = {4},
  pages = {041036},
  numpages = {23},
  year = {2023},
  month = {Nov},
  publisher = {American Physical Society},
  doi = {10.1103/PhysRevX.13.041036},
  url = {https://link.aps.org/doi/10.1103/PhysRevX.13.041036}
}

@article{moree2024,
  title = {Dome structure in pressure dependence of superconducting transition temperature for ${\mathrm{HgBa}}_{2}{\mathrm{Ca}}_{2}{\mathrm{Cu}}_{3}{\mathrm{O}}_{8}$: Studies by ab initio low-energy effective Hamiltonian},
  author = {Mor\'ee, Jean-Baptiste and Yamaji, Youhei and Imada, Masatoshi},
  journal = {Phys. Rev. Res.},
  volume = {6},
  issue = {2},
  pages = {023163},
  numpages = {22},
  year = {2024},
  month = {May},
  publisher = {American Physical Society},
  doi = {10.1103/PhysRevResearch.6.023163},
  url = {https://link.aps.org/doi/10.1103/PhysRevResearch.6.023163}
}

@article{gao1994,
  title = {Superconductivity up to 164 K in ${\mathrm{HgBa}}_{2}$${\mathrm{Ca}}_{\mathit{m}\mathrm{\ensuremath{-}}1}$${\mathrm{Cu}}_{\mathit{m}}$${\mathrm{O}}_{2\mathit{m}+2+\mathrm{\ensuremath{\delta}}}$ (m=1, 2, and 3) under quasihydrostatic pressures},
  author = {Gao, L. and Xue, Y. Y. and Chen, F. and Xiong, Q. and Meng, R. L. and Ramirez, D. and Chu, C. W. and Eggert, J. H. and Mao, H. K.},
  journal = {Phys. Rev. B},
  volume = {50},
  issue = {6},
  pages = {4260--4263},
  numpages = {0},
  year = {1994},
  month = {Aug},
  publisher = {American Physical Society},
  doi = {10.1103/PhysRevB.50.4260},
  url = {https://link.aps.org/doi/10.1103/PhysRevB.50.4260}
}

@article{moree2022,
  title = {Ab initio low-energy effective Hamiltonians for the high-temperature superconducting cuprates ${\mathrm{Bi}}_{2}{\mathrm{Sr}}_{2}{\mathrm{CuO}}_{6},$ ${\mathrm{Bi}}_{2}{\mathrm{Sr}}_{2}{\mathrm{CaCu}}_{2}{\mathrm{O}}_{8},$ ${\mathrm{HgBa}}_{2}{\mathrm{CuO}}_{4},$ and ${\mathrm{CaCuO}}_{2}$},
  author = {Mor\'ee, Jean-Baptiste and Hirayama, Motoaki and Schmid, Michael Thobias and Yamaji, Youhei and Imada, Masatoshi},
  journal = {Phys. Rev. B},
  volume = {106},
  issue = {23},
  pages = {235150},
  numpages = {22},
  year = {2022},
  month = {Dec},
  publisher = {American Physical Society},
  doi = {10.1103/PhysRevB.106.235150},
  url = {https://link.aps.org/doi/10.1103/PhysRevB.106.235150}
}

@article{hirayama2019,
  title = {Effective Hamiltonian for cuprate superconductors derived from multiscale ab initio scheme with level renormalization},
  author = {Hirayama, Motoaki and Misawa, Takahiro and Ohgoe, Takahiro and Yamaji, Youhei and Imada, Masatoshi},
  journal = {Phys. Rev. B},
  volume = {99},
  issue = {24},
  pages = {245155},
  numpages = {21},
  year = {2019},
  month = {Jun},
  publisher = {American Physical Society},
  doi = {10.1103/PhysRevB.99.245155},
  url = {https://link.aps.org/doi/10.1103/PhysRevB.99.245155}
}

@article{tahara2008,
  author = {Tahara, Daisuke and Imada, Masatoshi},
  title = {Variational Monte Carlo Method Combined with Quantum-Number Projection and Multi-Variable Optimization},
  journal={J. Phys. Soc. Jpn.},
  volume = {77},
  number = {11},
  pages = {114701},
  year = {2008},
  doi = {10.1143/JPSJ.77.114701},
}

@article{misawa2019,
  title = {mVMC--Open-source software for many-variable variational Monte Carlo method},
  journal = {Comput. Phys. Commun.},
  volume = {235},
  pages = {447-462},
  year = {2019},
  issn = {0010-4655},
  doi = {https://doi.org/10.1016/j.cpc.2018.08.014},
  url = {https://www.sciencedirect.com/science/article/pii/S0010465518303102},
  author = {Takahiro Misawa and Satoshi Morita and Kazuyoshi Yoshimi and Mitsuaki Kawamura and Yuichi Motoyama and Kota Ido and Takahiro Ohgoe and Masatoshi Imada and Takeo Kato},
}

@article{nomura2017,
  title = {Restricted Boltzmann machine learning for solving strongly correlated quantum systems},
  author = {Nomura, Yusuke and Darmawan, Andrew S. and Yamaji, Youhei and Imada, Masatoshi},
  journal = {Phys. Rev. B},
  volume = {96},
  issue = {20},
  pages = {205152},
  numpages = {8},
  year = {2017},
  month = {Nov},
  publisher = {American Physical Society},
  doi = {10.1103/PhysRevB.96.205152},
  url = {https://link.aps.org/doi/10.1103/PhysRevB.96.205152}
}

@article{nomura2021,
  title = {Dirac-Type Nodal Spin Liquid Revealed by Refined Quantum Many-Body Solver Using Neural-Network Wave Function, Correlation Ratio, and Level Spectroscopy},
  author = {Nomura, Yusuke and Imada, Masatoshi},
  journal = {Phys. Rev. X},
  volume = {11},
  issue = {3},
  pages = {031034},
  numpages = {19},
  year = {2021},
  month = {Aug},
  publisher = {American Physical Society},
  doi = {10.1103/PhysRevX.11.031034},
  url = {https://link.aps.org/doi/10.1103/PhysRevX.11.031034}
}

@article{gutzwiller1963,
  title = {Effect of Correlation on the Ferromagnetism of Transition Metals},
  author = {Gutzwiller, Martin C.},
  journal = {Phys. Rev. Lett.},
  volume = {10},
  issue = {5},
  pages = {159--162},
  numpages = {0},
  year = {1963},
  month = {Mar},
  publisher = {American Physical Society},
  doi = {10.1103/PhysRevLett.10.159},
  url = {https://link.aps.org/doi/10.1103/PhysRevLett.10.159}
}

@article{jastrow1955,
  title = {Many-Body Problem with Strong Forces},
  author = {Jastrow, Robert},
  journal = {Phys. Rev.},
  volume = {98},
  issue = {5},
  pages = {1479--1484},
  numpages = {0},
  year = {1955},
  month = {Jun},
  publisher = {American Physical Society},
  doi = {10.1103/PhysRev.98.1479},
  url = {https://link.aps.org/doi/10.1103/PhysRev.98.1479}
}

@article{capello2005,
  title = {Variational Description of Mott Insulators},
  author = {Capello, Manuela and Becca, Federico and Fabrizio, Michele and Sorella, Sandro and Tosatti, Erio},
  journal = {Phys. Rev. Lett.},
  volume = {94},
  issue = {2},
  pages = {026406},
  numpages = {4},
  year = {2005},
  month = {Jan},
  publisher = {American Physical Society},
  doi = {10.1103/PhysRevLett.94.026406},
  url = {https://link.aps.org/doi/10.1103/PhysRevLett.94.026406}
}

@article{yokoyama1990,
  title={Variational Monte-Carlo Studies of Hubbard Model. III. Intersite Correlation Effects},
  author={Hisatoshi Yokoyama and Hiroyuki Shiba},
  journal={J. Phys. Soc. Jpn.},
  volume={59},
  number={10},
  pages={3669-3686},
  year={1990},
  doi={10.1143/JPSJ.59.3669},
}

@article{sorella2001,
  title = {Generalized Lanczos algorithm for variational quantum Monte Carlo},
  author = {Sorella, Sandro},
  journal = {Phys. Rev. B},
  volume = {64},
  issue = {2},
  pages = {024512},
  numpages = {16},
  year = {2001},
  month = {Jun},
  publisher = {American Physical Society},
  doi = {10.1103/PhysRevB.64.024512},
  url = {https://link.aps.org/doi/10.1103/PhysRevB.64.024512}
}

@article{carleo2017,
  author = {Giuseppe Carleo and Matthias Troyer},
  title = {Solving the Quantum Many-Body Problem with Artificial Neural Networks},
  journal = {Science},
  volume = {355},
  number = {6325},
  pages = {602-606},
  year = {2017},
  doi = {10.1126/science.aag2302},
  URL = {https://www.science.org/doi/abs/10.1126/science.aag2302}
}

@article{ohgoe2020,
  title = {Ab initio study of superconductivity and inhomogeneity in a Hg-based cuprate superconductor},
  author = {Ohgoe, Takahiro and Hirayama, Motoaki and Misawa, Takahiro and Ido, Kota and Yamaji, Youhei and Imada, Masatoshi},
  journal = {Phys. Rev. B},
  volume = {101},
  issue = {4},
  pages = {045124},
  numpages = {11},
  year = {2020},
  month = {Jan},
  publisher = {American Physical Society},
  doi = {10.1103/PhysRevB.101.045124},
  url = {https://link.aps.org/doi/10.1103/PhysRevB.101.045124}
}

@article{ido2018,
  title = {Competition among various charge-inhomogeneous states and $d$-wave superconducting state in Hubbard models on square lattices},
  author = {Ido, Kota and Ohgoe, Takahiro and Imada, Masatoshi},
  journal = {Phys. Rev. B},
  volume = {97},
  issue = {4},
  pages = {045138},
  numpages = {9},
  year = {2018},
  month = {Jan},
  publisher = {American Physical Society},
  doi = {10.1103/PhysRevB.97.045138},
  url = {https://link.aps.org/doi/10.1103/PhysRevB.97.045138}
}

@article{ido2022,
  title={Unconventional dual 1D--2D quantum spin liquid revealed by ab initio studies on organic solids family},
  author={Ido, Kota and Yoshimi, Kazuyoshi and Misawa, Takahiro and Imada, Masatoshi},
  journal={npj Quantum Mater.},
  volume={7},
  number={1},
  pages={48},
  year={2022},
  publisher={Nature Publishing Group UK London},
  doi={10.1038/s41535-022-00452-8}
}

@article{yamamoto2015,
  title = {High pressure effects revisited for the cuprate superconductor family with highest critical temperature},
  volume = {6},
  pages = {8990},
  ISSN = {2041-1723},
  url = {http://dx.doi.org/10.1038/ncomms9990},
  DOI = {10.1038/ncomms9990},
  number = {1},
  journal = {Nat. Commun.},
  publisher = {Springer Science and Business Media LLC},
  author = {Yamamoto,  Ayako and Takeshita,  Nao and Terakura,  Chieko and Tokura,  Yoshinori},
  year = {2015},
  month = dec 
}

@article{jover1996,
  title = {Pressure dependence of the superconducting critical temperature of ${\mathrm{HgBa}}_{2}$${\mathrm{Ca}}_{2}$${\mathrm{Cu}}_{3}$${\mathrm{O}}_{8+\mathit{y}}$ and ${\mathrm{HgBa}}_{2}$${\mathrm{Ca}}_{3}$${\mathrm{Cu}}_{4}$${\mathrm{O}}_{10+\mathit{y}}$ up to 30 GPa},
  author = {Jover, D. Tristan and Wijngaarden, R. J. and Wilhelm, H. and Griessen, R. and Loureiro, S. M. and Capponi, J.-J. and Schilling, A. and Ott, H. R.},
  journal = {Phys. Rev. B},
  volume = {54},
  issue = {6},
  pages = {4265--4275},
  numpages = {0},
  year = {1996},
  month = {Aug},
  publisher = {American Physical Society},
  doi = {10.1103/PhysRevB.54.4265},
  url = {https://link.aps.org/doi/10.1103/PhysRevB.54.4265}
}

@article{imada2000,
  author =  {Imada, Masatoshi and Kashima, Tsuyoshi},
  journal = {J. Phys. Soc. Jpn.},
  pages = {2723},
  title = {Path-Integral Renormalization Group Method for Numerical Study of Strongly Correlated Electron Systems},
  volume = {69},
  year = {2000},
  doi = {10.1143/JPSJ.69.2723},
  URL = { https://doi.org/10.1143/JPSJ.69.2723},
}

@article{kashima2001,
  author = {Kashima, Tsuyoshi and Imada, Masatoshi},
  journal = {J. Phys. Soc. Jpn.},
  pages = {2287},
  title = {{\it Path-Integral Renormalization Group Method for Numerical Study on Ground States of Strongly Correlated Electron Systems}},
  volume = {70},
  year = {2001},
  doi = {10.1143/JPSJ.70.2287},
  URL = {   https://doi.org/10.1143/JPSJ.70.2287}
}

@article{uemura1989,
  title = {Universal Correlations between ${T}_{c}$ and $\frac{{n}_{s}}{{m}^{*}}$ (Carrier Density over Effective Mass) in High-${T}_{c}$ Cuprate Superconductors},
  author = {Uemura, Y. J. and Luke, G. M. and Sternlieb, B. J. and Brewer, J. H. and Carolan, J. F. and Hardy, W. N. and Kadono, R. and Kempton, J. R. and Kiefl, R. F. and Kreitzman, S. R. and Mulhern, P. and Riseman, T. M. and Williams, D. Ll. and Yang, B. X. and Uchida, S. and Takagi, H. and Gopalakrishnan, J. and Sleight, A. W. and Subramanian, M. A. and Chien, C. L. and Cieplak, M. Z. and Xiao, Gang and Lee, V. Y. and Statt, B. W. and Stronach, C. E. and Kossler, W. J. and Yu, X. H.},
  journal = {Phys. Rev. Lett.},
  volume = {62},
  issue = {19},
  pages = {2317--2320},
  numpages = {0},
  year = {1989},
  month = {May},
  publisher = {American Physical Society},
  doi = {10.1103/PhysRevLett.62.2317},
  url = {https://link.aps.org/doi/10.1103/PhysRevLett.62.2317}
}

@article{sakai2018,
  title = {Direct connection between Mott insulators and $d$-wave high-temperature superconductors revealed by continuous evolution of self-energy poles},
  author = {Sakai, Shiro and Civelli, Marcello and Imada, Masatoshi},
  journal = {Phys. Rev. B},
  volume = {98},
  issue = {19},
  pages = {195109},
  numpages = {14},
  year = {2018},
  month = {Nov},
  publisher = {American Physical Society},
  doi = {10.1103/PhysRevB.98.195109},
  url = {https://link.aps.org/doi/10.1103/PhysRevB.98.195109}
}

@article{yamaji2021,
  title = {Hidden self-energies as origin of cuprate superconductivity revealed by machine learning},
  author = {Yamaji, Youhei and Yoshida, Teppei and Fujimori, Atsushi and Imada, Masatoshi},
  journal = {Phys. Rev. Res.},
  volume = {3},
  issue = {4},
  pages = {043099},
  numpages = {37},
  year = {2021},
  month = {Nov},
  publisher = {American Physical Society},
  doi = {10.1103/PhysRevResearch.3.043099},
  url = {https://link.aps.org/doi/10.1103/PhysRevResearch.3.043099}
}

@article{yokoyama2013,
  author = {Yokoyama ,Hisatoshi and Ogata ,Masao and Tanaka ,Yukio and Kobayashi ,Kenji and Tsuchiura ,Hiroki},
  title = {Crossover between BCS Superconductor and Doped Mott Insulator of $d$-Wave Pairing State in Two-Dimensional Hubbard Model},
  journal = {J. Phys. Soc. Jpn.},
  volume = {82},
  number = {1},
  pages = {014707},
  year = {2013},
  doi = {10.7566/JPSJ.82.014707},
  URL = { https://doi.org/10.7566/JPSJ.82.014707}
}

@article{imada2019,
  author = {Imada ,Masatoshi and Suzuki ,Takafumi J.},
  title = {Excitons and Dark Fermions as Origins of Mott Gap, Pseudogap and Superconductivity in Cuprate Superconductors -- General Concept and Basic Formalism Based on Gap Physics},
  journal = {J. Phys. Soc. Jpn.},
  volume = {88},
  number = {2},
  pages = {024701},
  year = {2019},
  doi = {10.7566/JPSJ.88.024701},
  URL = { https://doi.org/10.7566/JPSJ.88.024701}
}

@article{imada2021,
  author = {Imada ,Masatoshi},
  title = {Charge Order and Superconductivity as Competing Brothers in Cuprate High-Tc Superconductors},
  journal = {J. Phys. Soc. Jpn.},
  volume = {90},
  number = {11},
  pages = {111009},
  year = {2021},
  doi = {10.7566/JPSJ.90.111009},
  URL = {https://doi.org/10.7566/JPSJ.90.111009}
}

@article{nilsson2019,
  title = {Dynamically screened Coulomb interaction in the parent compounds of hole-doped cuprates: Trends and exceptions},
  author = {Nilsson, F. and Karlsson, K. and Aryasetiawan, F.},
  journal = {Phys. Rev. B},
  volume = {99},
  issue = {7},
  pages = {075135},
  numpages = {9},
  year = {2019},
  month = {Feb},
  publisher = {American Physical Society},
  doi = {10.1103/PhysRevB.99.075135},
  url = {https://link.aps.org/doi/10.1103/PhysRevB.99.075135}
}

@article{zhao2017,
  title = {Variational Monte Carlo method for fermionic models combined with tensor networks and applications to the hole-doped two-dimensional Hubbard model},
  author = {Zhao, Hui-Hai and Ido, Kota and Morita, Satoshi and Imada, Masatoshi},
  journal = {Phys. Rev. B},
  volume = {96},
  issue = {8},
  pages = {085103},
  numpages = {16},
  year = {2017},
  month = {Aug},
  publisher = {American Physical Society},
  doi = {10.1103/PhysRevB.96.085103},
  url = {https://link.aps.org/doi/10.1103/PhysRevB.96.085103}
}

@article{yokoyama1987,
  author = {Yokoyama, Hisatoshi and Shiba, Hiroyuki},
  title = {Variational Monte-Carlo Studies of Hubbard Model. {I}},
  journal = {J. Phys. Soc. Jpn.},
  volume = {56},
  number = {4},
  pages = {1490--1506},
  year = {1987},
  doi = {10.1143/JPSJ.56.1490},
  URL = {https://doi.org/10.1143/JPSJ.56.1490}
}

@article{gros1987,
  title = {Antiferromagnetic correlations in almost-localized Fermi liquids},
  author = {Gros, C. and Joynt, R. and Rice, T. M.},
  journal = {Phys. Rev. B},
  volume = {36},
  issue = {1},
  pages = {381--393},
  numpages = {0},
  year = {1987},
  month = {Jul},
  publisher = {American Physical Society},
  doi = {10.1103/PhysRevB.36.381},
  url = {https://link.aps.org/doi/10.1103/PhysRevB.36.381}
}

@article{gros1988,
  title = {Superconductivity in correlated wave functions},
  author = {Gros, Claudius},
  journal = {Phys. Rev. B},
  volume = {38},
  issue = {1},
  pages = {931--934},
  numpages = {0},
  year = {1988},
  month = {Jul},
  publisher = {American Physical Society},
  doi = {10.1103/PhysRevB.38.931},
  url = {https://link.aps.org/doi/10.1103/PhysRevB.38.931}
}

@article{capriotti2001,
  title = {Resonating Valence Bond Wave Functions for Strongly Frustrated Spin Systems},
  author = {Capriotti, Luca and Becca, Federico and Parola, Alberto and Sorella, Sandro},
  journal = {Phys. Rev. Lett.},
  volume = {87},
  issue = {9},
  pages = {097201},
  numpages = {4},
  year = {2001},
  month = {Aug},
  publisher = {American Physical Society},
  doi = {10.1103/PhysRevLett.87.097201},
  url = {https://link.aps.org/doi/10.1103/PhysRevLett.87.097201}
}

@article{misawa2014a,
  title = {Origin of high-${T}_{c}$ superconductivity in doped Hubbard models and their extensions: Roles of uniform charge fluctuations},
  author = {Misawa, Takahiro and Imada, Masatoshi},
  journal = {Phys. Rev. B},
  volume = {90},
  issue = {11},
  pages = {115137},
  numpages = {16},
  year = {2014},
  month = {Sep},
  publisher = {American Physical Society},
  doi = {10.1103/PhysRevB.90.115137},
  url = {https://link.aps.org/doi/10.1103/PhysRevB.90.115137}
}

@article{wu2024,
  author = {Dian Wu  and Riccardo Rossi  and Filippo Vicentini  and Nikita Astrakhantsev  and Federico Becca  and Xiaodong Cao  and Juan Carrasquilla  and Francesco Ferrari  and Antoine Georges  and Mohamed Hibat-Allah  and Masatoshi Imada  and Andreas M. L\"{a}uchli  and Guglielmo Mazzola  and Antonio Mezzacapo  and Andrew Millis  and Javier Robledo Moreno  and Titus Neupert  and Yusuke Nomura  and Jannes Nys  and Olivier Parcollet  and Rico Pohle  and Imelda Romero  and Michael Schmid  and J. Maxwell Silvester  and Sandro Sorella  and Luca F. Tocchio  and Lei Wang  and Steven R. White  and Alexander Wietek  and Qi Yang  and Yiqi Yang  and Shiwei Zhang  and Giuseppe Carleo },
  title = {Variational benchmarks for quantum many-body problems},
  journal = {Science},
  volume = {386},
  number = {6719},
  pages = {296-301},
  year = {2024},
  doi = {10.1126/science.adg9774},
  URL = {https://www.science.org/doi/abs/10.1126/science.adg9774},
}

@article{cui2025,
  title={Ab initio quantum many-body description of superconducting trends in the cuprates},
  author={Cui, Zhi-Hao and Yang, Junjie and T{\"o}lle, Johannes and Ye, Hong-Zhou and Yuan, Shunyue and Zhai, Huanchen and Park, Gunhee and Kim, Raehyun and Zhang, Xing and Lin, Lin and Berkelbach, Timothy C.  and Chan, Garnet Kin-Lic},
  journal={Nat. Commun.},
  volume={16},
  number={1},
  pages={1845},
  year={2025},
  publisher={Nature Publishing Group UK London},
}

@article{kurokawa2023,
  title={Unveiling phase diagram of the lightly doped high-Tc cuprate superconductors with disorder removed},
  author={Kurokawa, Kifu and Isono, Shunsuke and Kohama, Yoshimitsu and Kunisada, So and Sakai, Shiro and Sekine, Ryotaro and Okubo, Makoto and Watson, Matthew D and Kim, Timur K and Cacho, Cephise and Shin, Shik and Tohyama, Takami and Tokiwa, Kazuyasu and Kondo, Takeshi},
  journal={Nat. Commun.},
  volume={14},
  number={1},
  pages={4064},
  year={2023},
  publisher={Nature Publishing Group UK London},
  doi={10.1038/s41467-023-39457-7}
}

@article{kotegawa2001,
  title = {Unusual magnetic and superconducting characteristics in multilayered high-${T}_{c}$ cuprates: ${}^{63}\mathrm{Cu}$ NMR study},
  author = {Kotegawa, H. and Tokunaga, Y. and Ishida, K. and Zheng, G.-q. and Kitaoka, Y. and Kito, H. and Iyo, A. and Tokiwa, K. and Watanabe, T. and Ihara, H.},
  journal = {Phys. Rev. B},
  volume = {64},
  issue = {6},
  pages = {064515},
  numpages = {5},
  year = {2001},
  month = {Jul},
  publisher = {American Physical Society},
  doi = {10.1103/PhysRevB.64.064515},
  url = {https://link.aps.org/doi/10.1103/PhysRevB.64.064515}
}

@article{mukuda2012,
  author = {Mukuda ,Hidekazu and Shimizu ,Sunao and Iyo ,Akira and Kitaoka ,Yoshio},
  title = {High-Tc Superconductivity and Antiferromagnetism in Multilayered Copper Oxides --A New Paradigm of Superconducting Mechanism--},
  journal={J. Phys. Soc. Jpn.},
  volume = {81},
  number = {1},
  pages = {011008},
  year = {2012},
  doi = {10.1143/JPSJ.81.011008},
}

@article{bacqlabreuil2025,
  title = {Toward an Ab Initio Theory of High-Temperature Superconductors: A Study of Multilayer Cuprates},
  author = {Bacq-Labreuil, Benjamin and Lacasse, Benjamin and Tremblay, A.-M. S. and S\'en\'echal, David and Haule, Kristjan},
  journal = {Phys. Rev. X},
  volume = {15},
  issue = {2},
  pages = {021071},
  numpages = {44},
  year = {2025},
  month = {May},
  publisher = {American Physical Society},
  doi = {10.1103/PhysRevX.15.021071},
  url = {https://link.aps.org/doi/10.1103/PhysRevX.15.021071}
}

@article{bouchaud1988,
  title={Pair wave functions for strongly correlated fermions and their determinantal representation},
  author={Bouchaud, JP and Georges, A and Lhuillier, C},
  journal={J. Phys. (Paris)},
  volume={49},
  number={4},
  pages={553--559},
  year={1988},
  publisher={Soci{\'e}t{\'e} fran{\c{c}}aise de physique},
  doi={10.1051/jphys:01988004904055300}
}

@article{bajdich2008,
  title = {Pfaffian pairing and backflow wavefunctions for electronic structure quantum Monte Carlo methods},
  author = {Bajdich, M. and Mitas, L. and Wagner, L. K. and Schmidt, K. E.},
  journal = {Phys. Rev. B},
  volume = {77},
  issue = {11},
  pages = {115112},
  numpages = {13},
  year = {2008},
  month = {Mar},
  publisher = {American Physical Society},
  doi = {10.1103/PhysRevB.77.115112},
  url = {https://link.aps.org/doi/10.1103/PhysRevB.77.115112}
}

@article{heeb1993,
  title={Systematic improvement of variational Monte Carlo using Lanczos iterations},
  author={Heeb, ES and Rice, TM},
  journal={Z. Phys. B: Condens. Matter},
  volume={90},
  number={1},
  pages={73--77},
  year={1993},
  publisher={Springer},
  doi={10.1007/BF01321035}
}

@article{tranquada1995,
  title={Evidence for stripe correlations of spins and holes in copper oxide superconductors},
  author={Tranquada, JM and Sternlieb, BJ and Axe, JD and Nakamura, Y and Uchida, Shin-ichi},
  journal={Nature},
  volume={375},
  number={6532},
  pages={561--563},
  year={1995},
  publisher={Nature Publishing Group UK London},
  doi={10.1038/375561a0}
}

@article{tranquada1997,
  title = {Coexistence of, and Competition between, Superconductivity and Charge-Stripe Order in ${\mathrm{La}}_{1.6\ensuremath{-}\mathit{x}}{\mathrm{Nd}}_{0.4}{\mathrm{Sr}}_{\mathit{x}}{\mathrm{CuO}}_{4}$},
  author = {Tranquada, J. M. and Axe, J. D. and Ichikawa, N. and Moodenbaugh, A. R. and Nakamura, Y. and Uchida, S.},
  journal = {Phys. Rev. Lett.},
  volume = {78},
  issue = {2},
  pages = {338--341},
  numpages = {0},
  year = {1997},
  month = {Jan},
  publisher = {American Physical Society},
  doi = {10.1103/PhysRevLett.78.338},
  url = {https://link.aps.org/doi/10.1103/PhysRevLett.78.338}
}

@article{putilin1993,
  title = {Superconductivity at 94 K in HgBa2Cu04+$\delta$},
  author = {Putilin, S. and Antipov, E. and Chmaissemt, O. and Mareziot, M.},
  journal = {Nature},
  volume = {362},
  issue = {},
  pages = {226},
  numpages = {3},
  year = {1993},
  month = {Mar},
  publisher = {Nature Publishing},
  doi = {10.1038/362226a0},
  url = {https://doi-org.utokyo.idm.oclc.org/10.1038/362226a0}
}

@article{schilling1993,
  title = {Superconductivity above 130 K in the Hg-Ba-Ca-Cu-O system},
  author = {Schilling, A. and Cantoni, M. and Guo, J. and Ott, H. R.},
  journal = {Nature},
  volume = {363},
  issue = {},
  pages = {56},
  numpages = {3},
  year = {1993},
  month = {May},
  publisher = {Nature Publishing},
  doi = {10.1038/363056a0},
  url = {https://doi-org/10.1038/363056a0}
}

@article{dai1995,
  title = {Synthesis and neutron powder diffraction study of the superconductor {HgBa$_2$Ca$_2$Cu$_3$O$_{8+\delta}$} by {Tl} substitution},
  journal = {Physica C},
  volume = {243},
  number = {3},
  pages = {201-206},
  year = {1995},
  issn = {0921-4534},
  doi = {https://doi.org/10.1016/0921-4534(94)02461-8},
  url = {https://www.sciencedirect.com/science/article/pii/0921453494024618},
  author = {Dai, P. and Chakoumakos, B. C. and Sun, G. F. and Wong, K.W. and Xin, Y. and Lu, D. F.}
}

@article{bednorz1986,
  title={Possible high T c superconductivity in the Ba- La- Cu- O system},
  author={Bednorz, J George and M{\"u}ller, K Alex},
  journal={Z. Phys. B: Condens. Matter},
  volume={64},
  number={2},
  pages={189--193},
  year={1986},
  publisher={Springer},
  doi={10.1007/BF01303701}
}

@article{nakamura2012,
  title = {Ab initio two-dimensional multiband low-energy models of EtMe${}_{3}$Sb[Pd(dmit)${}_{2}$]${}_{2}$ and $\ensuremath{\kappa}$-(BEDT-TTF)${}_{2}$Cu(NCS)${}_{2}$ with comparisons to single-band models},
  author = {Nakamura, Kazuma and Yoshimoto, Yoshihide and Imada, Masatoshi},
  journal = {Phys. Rev. B},
  volume = {86},
  issue = {20},
  pages = {205117},
  numpages = {9},
  year = {2012},
  month = {Nov},
  publisher = {American Physical Society},
  doi = {10.1103/PhysRevB.86.205117},
  url = {https://link.aps.org/doi/10.1103/PhysRevB.86.205117}
}

@article{iqbal2013,
  title = {Gapless spin-liquid phase in the kagome spin-$\frac{1}{2}$ Heisenberg antiferromagnet},
  author = {Iqbal, Yasir and Becca, Federico and Sorella, Sandro and Poilblanc, Didier},
  journal = {Phys. Rev. B},
  volume = {87},
  issue = {6},
  pages = {060405},
  numpages = {5},
  year = {2013},
  month = {Feb},
  publisher = {American Physical Society},
  doi = {10.1103/PhysRevB.87.060405},
  url = {https://link.aps.org/doi/10.1103/PhysRevB.87.060405}
}

@article{alldredge2008,
  author = {Alldredge, J. W. and Lee, J.  and Mcelroy, K. and Wang, M. and Fujita, K. and Kohsaka, Y. and Taylor, C. and
Eisaki, H. and Uchida, S. and Hirschfeld, P. J. and Davis, J. C.},
  journal = {Nat. Phys.},
  number = {},
  pages = {319-326},
  publisher = {Nature Publishing Group},
  title = {Evolution of the electronic excitation spectrum with strongly diminishing hole density in superconducting Bi$_2$Sr$_2$CaCu$_2$O$_{8+\delta}$},
  volume = {4},
  year = {2008},
  doi = {doi:10.1038/nphys917},
  URL = {https://doi.org/10.1038/nphys917}
}

@article{sakai2013,
  title = {Raman-Scattering Measurements and Theory of the Energy-Momentum Spectrum for Underdoped ${\mathrm{Bi}}_{2}{\mathrm{Sr}}_{2}{\mathrm{CaCuO}}_{8+\ensuremath{\delta}}$ Superconductors: Evidence of an $s$-Wave Structure for the Pseudogap},
  author = {Sakai, S. and Blanc, S. and Civelli, M. and Gallais, Y. and Cazayous, M. and M\'easson, M.-A. and Wen, J. S. and Xu, Z. J. and Gu, G. D. and Sangiovanni, G. and Motome, Y. and Held, K. and Sacuto, A. and Georges, A. and Imada, M.},
  journal = {Phys. Rev. Lett.},
  volume = {111},
  issue = {10},
  pages = {107001},
  numpages = {5},
  year = {2013},
  month = {Sep},
  publisher = {American Physical Society},
  doi = {10.1103/PhysRevLett.111.107001},
  url = {https://link.aps.org/doi/10.1103/PhysRevLett.111.107001}
}

@article{sakai2016,
  title = {Hidden Fermionic Excitation Boosting High-Temperature Superconductivity in Cuprates},
  author = {Sakai, Shiro and Civelli, Marcello and Imada, Masatoshi},
  journal = {Phys. Rev. Lett.},
  volume = {116},
  issue = {5},
  pages = {057003},
  numpages = {6},
  year = {2016},
  month = {Feb},
  publisher = {American Physical Society},
  doi = {10.1103/PhysRevLett.116.057003},
  url = {https://link.aps.org/doi/10.1103/PhysRevLett.116.057003}
}

@article{singh2022,
author={Singh, A.
and Huang, H. Y.
and Xie, J. D.
and Okamoto, J.
and Chen, C. T.
and Watanabe, T.
and Fujimori, A.
and Imada, M.
and Huang, D. J.},
title={Unconventional exciton evolution from the pseudogap to superconducting phases in cuprates},
journal={Nat. Commun.},
year={2022},
month={Dec},
day={23},
volume={13},
number={1},
pages={7906},
issn={2041-1723},
doi={10.1038/s41467-022-35210-8},
url={https://doi.org/10.1038/s41467-022-35210-8}
}

@article{sakai2025,
  title = {Unified Description of Cuprate Superconductors by Fractionalized Electrons Emerging from Integrated Analyses of Photoemission Spectra and Quasiparticle Interference},
  author = {Sakai, Shiro and Yamaji, Youhei and Imoto, Fumihiro and Tamegai, Tsuyoshi and Kaminski, Adam and Kondo, Takeshi and Kohsaka, Yuhki and Hanaguri, Tetsuo and Imada, Masatoshi},
  journal = {Phys. Rev. X},
  volume = {16},
  issue = {1},
  pages = {011018},
  numpages = {30},
  year = {2026},
  month = {Feb},
  publisher = {American Physical Society},
  doi = {10.1103/mww7-32gn},
  url = {https://link.aps.org/doi/10.1103/mww7-32gn}
}

@article{kamihara2008,
author = {Kamihara, Yoichi and Watanabe, Takumi and Hirano, Masahiro and Hosono, Hideo},
title = {Iron-Based Layered Superconductor La[O1-xFx]FeAs (x = 0.05--0.12) with Tc = 26 K},
journal = {J. Am. Chem. Soc.},
volume = {130},
number = {11},
pages = {3296-3297},
year = {2008},
doi = {10.1021/ja800073m},
URL = {https://doi.org/10.1021/ja800073m},
}

@article{li2019,
  title = {Superconductivity in an infinite-layer nickelate},
  author = {Li, Danfeng
and Lee, Kyuho
and Wang, Bai Yang
and Osada, Motoki
and Crossley, Samuel
and Lee, Hye Ryoung
and Cui, Yi
and Hikita, Yasuyuki
and Hwang, Harold Y.},
  journal = {Nature},
  volume = {572},
  issue = {},
  pages = {624-627},
  numpages = {4},
  year = {2019},
  month = {August},
  publisher = {Nature Publishing},
  doi = {10.1038/s41586-019-1496-5},
  url = {https://doi-org/10.1038/s41586-019-1496-5}
}

@article{aryasetiawan2009,
  title = {Downfolded Self-Energy of Many-Electron Systems},
  author = {Aryasetiawan, F. and Tomczak, J. M. and Miyake, T. and Sakuma, R.},
  journal = {Phys. Rev. Lett.},
  volume = {102},
  issue = {17},
  pages = {176402},
  numpages = {4},
  year = {2009},
  month = {Apr},
  publisher = {American Physical Society},
  doi = {10.1103/PhysRevLett.102.176402},
  url = {https://link.aps.org/doi/10.1103/PhysRevLett.102.176402}
}

@article{hirayama2013,
  title = {Derivation of static low-energy effective models by an ab initio downfolding method without double counting of Coulomb correlations: Application to SrVO${}_{3}$, FeSe, and FeTe},
  author = {Hirayama, Motoaki and Miyake, Takashi and Imada, Masatoshi},
  journal = {Phys. Rev. B},
  volume = {87},
  issue = {19},
  pages = {195144},
  numpages = {22},
  year = {2013},
  month = {May},
  publisher = {American Physical Society},
  doi = {10.1103/PhysRevB.87.195144},
  url = {https://link.aps.org/doi/10.1103/PhysRevB.87.195144}
}

@article{hirayama2017,
  title = {Low-energy effective Hamiltonians for correlated electron systems beyond density functional theory},
  author = {Hirayama, Motoaki and Miyake, Takashi and Imada, Masatoshi and Biermann, Silke},
  journal = {Phys. Rev. B},
  volume = {96},
  issue = {7},
  pages = {075102},
  numpages = {20},
  year = {2017},
  month = {Aug},
  publisher = {American Physical Society},
  doi = {10.1103/PhysRevB.96.075102},
  url = {https://link.aps.org/doi/10.1103/PhysRevB.96.075102}
}

@article{aryasetiawan2004,
  title = {Frequency-dependent local interactions and low-energy effective models from electronic structure calculations},
  author = {Aryasetiawan, F. and Imada, M. and Georges, A. and Kotliar, G. and Biermann, S. and Lichtenstein, A. I.},
  journal = {Phys. Rev. B},
  volume = {70},
  issue = {19},
  pages = {195104},
  numpages = {8},
  year = {2004},
  month = {Nov},
  publisher = {American Physical Society},
  doi = {10.1103/PhysRevB.70.195104},
  url = {https://link.aps.org/doi/10.1103/PhysRevB.70.195104}
}

@article{imada2010,
author = {Imada ,Masatoshi and Miyake ,Takashi},
title = {{Electronic Structure Calculation by First Principles for Strongly Correlated Electron Systems}},
journal={J. Phys. Soc. Jpn.},
volume = {79},
number = {11},
pages = {112001},
year = {2010},
doi = {10.1143/JPSJ.79.112001},
}

@article{ohta1991,
  title = {Apex oxygen and critical temperature in copper oxide superconductors: Universal correlation with the stability of local singlets},
  author = {Ohta, Y. and Tohyama, T. and Maekawa, S.},
  journal = {Phys. Rev. B},
  volume = {43},
  issue = {4},
  pages = {2968--2982},
  numpages = {0},
  year = {1991},
  month = {Feb},
  publisher = {American Physical Society},
  doi = {10.1103/PhysRevB.43.2968},
  url = {https://link.aps.org/doi/10.1103/PhysRevB.43.2968}
}

@article{al-Ruqaishi2026,
title = {Multi-variable empirical equation for critical temperature of cuprate superconductors},
journal = {Result. Phys.},
volume = {81},
pages = {108576},
year = {2026},
issn = {2211-3797},
doi = {https://doi.org/10.1016/j.rinp.2026.108576},
url = {https://www.sciencedirect.com/science/article/pii/S2211379726000021},
author = {Zakiya Al-Ruqaishi and C.H Raymond Ooi}
}

@unpublished{moree_unpublished,
title = {unpublished},
author = {Mor\'ee, Jean-Baptiste}
}

@misc{mVMC,
title ={https://www.pasums.issp.u-tokyo.ac.jp/mvmc/en/about
}}

@article{koblischka2020,
AUTHOR = {Koblischka, Michael Rudolf and Roth, Susanne and Koblischka-Veneva, Anjela and Karwoth, Thomas and Wiederhold, Alex and Zeng, Xian Lin and Fasoulas, Stefanos and Murakami, Masato},
TITLE = {Relation between Crystal Structure and Transition Temperature of Superconducting Metals and Alloys},
JOURNAL = {Metals},
VOLUME = {10},
YEAR = {2020},
NUMBER = {2},
PAGES = {158},
URL = {https://www.mdpi.com/2075-4701/10/2/158},
ISSN = {2075-4701},
DOI = {10.3390/met10020158}
}

@article{hobou2009,
  title = {Enhancement of the superconducting critical temperature in ${\text{Bi}}_{2}{\text{Sr}}_{2}{\text{CaCu}}_{2}{\text{O}}_{8+\ensuremath{\delta}}$ by controlling disorder outside ${\text{CuO}}_{2}$ planes},
  author = {Hobou, H. and Ishida, S. and Fujita, K. and Ishikado, M. and Kojima, K. M. and Eisaki, H. and Uchida, S.},
  journal = {Phys. Rev. B},
  volume = {79},
  issue = {6},
  pages = {064507},
  numpages = {6},
  year = {2009},
  month = {Feb},
  publisher = {American Physical Society},
  doi = {10.1103/PhysRevB.79.064507},
  url = {https://link.aps.org/doi/10.1103/PhysRevB.79.064507}
}

@article{attfield1998,
  title={Cation effects in doped La2CuO4 superconductors},
  author={Attfield, J. and Kharlanov, A. and McAllister, J.},
  journal={Nature},
  volume={394},
  number={},
  pages={157-159},
  year={1998},
  publisher={Nature Publishing Group},
  doi={. https://doi.org/10.1038/28120}
}

@article{liu2025,
  title = {Enhanced superconductivity via layer differentiation in the trilayer Hubbard model},
  author = {Liu, Xun and Jiang, Mi},
  journal = {Phys. Rev. B},
  volume = {112},
  issue = {20},
  pages = {L201103},
  numpages = {7},
  year = {2025},
  month = {Nov},
  publisher = {American Physical Society},
  doi = {10.1103/n28s-ggzc},
  url = {https://link.aps.org/doi/10.1103/n28s-ggzc}
}

@article{ideta2025,
  title = {Proximity-induced nodal metal in an extremely underdoped CuO2 plane in triple-layer cuprates
},
  author = "Ideta, Shin-ichiro
 and Adachi, Shintaro
 and Noji, Takashi
 and Yamaguchi, Shunpei
 and Sasaki, Nae
 and Ishida, Shigeyuki
 and Uchida, Shin-ichi
 and Fujii, Takenori
 and Watanabe, Takao
 and Wang, Wen O.
 and Moritz, Brian
 and Devereaux, Thomas P.
 and Arita, Masashi
 and Mou, Chung-Yu
 and Yoshida, Teppei
 and Tanaka, Kiyohisa
 and Lee, Ting-Kuo
 and Fujimori, Atsushi",
  journal = "Nat. Commun.",
  volume =  16,
  number =  1,
  pages = "9470",
  month =  Oct,
  year =  2025,
  doi = {10.1038/s41467-025-64492-x},
  url = {https://doi.org/10.1038/s41467-025-64492-x}
}

@article{BCS1957,
  title = {Theory of Superconductivity},
  author = {Bardeen, J. and Cooper, L. N. and Schrieffer, J. R.},
  journal = {Phys. Rev.},
  volume = {108},
  issue = {5},
  pages = {1175--1204},
  numpages = {0},
  year = {1957},
  month = {Dec},
  publisher = {American Physical Society},
  doi = {10.1103/PhysRev.108.1175},
  url = {https://link.aps.org/doi/10.1103/PhysRev.108.1175}
}

@article{uemura1991,
  title = {Basic similarities among cuprate, bismuthate, organic, Chevrel-phase, and heavy-fermion superconductors shown by penetration-depth measurements},
  author = {Uemura, Y. J. and Le, L. P. and Luke, G. M. and Sternlieb, B. J. and Wu, W. D. and Brewer, J. H. and Riseman, T. M. and Seaman, C. L. and Maple, M. B. and Ishikawa, M. and Hinks, D. G. and Jorgensen, J. D. and Saito, G. and Yamochi, H.},
  journal = {Phys. Rev. Lett.},
  volume = {66},
  issue = {20},
  pages = {2665--2668},
  numpages = {0},
  year = {1991},
  month = {May},
  publisher = {American Physical Society},
  doi = {10.1103/PhysRevLett.66.2665},
  url = {https://link.aps.org/doi/10.1103/PhysRevLett.66.2665}
}

@article{prokofev2001,
  title = {Critical Point of a Weakly Interacting Two-Dimensional Bose Gas},
  author = {Prokof'ev, Nikolay and Ruebenacker, Oliver and Svistunov, Boris},
  journal = {Phys. Rev. Lett.},
  volume = {87},
  issue = {27},
  pages = {270402},
  numpages = {4},
  year = {2001},
  month = {Dec},
  publisher = {American Physical Society},
  doi = {10.1103/PhysRevLett.87.270402},
  url = {https://link.aps.org/doi/10.1103/PhysRevLett.87.270402}
}

@article{luo2023,
  title={Electronic origin of high superconducting critical temperature in trilayer cuprates},
  author={Luo, X. and Chen, H. and Li, Y. and Gao, Qiang and Yin, Chaohui and yan, Hongtao and Miao, Taimin  and Luo, Hailan and Shu, Yingjie and Chen, Yiwen and Lin, Chengtian and Zhang, Shenjin and Wang, Zhimin and Zhang, Fengfeng and Yang, Feng and Peng, Qinjun and Liu, Guodong and Zhao, Lin and Xu, Zuyan and Xiang, Tao and Zhou, X. J. },
  journal={Nat. Phys.},
  volume={19},
  number={},
  pages={1841--1847},
  year={2023},
  publisher={Nature Publishing Group},
  doi={https://doi.org/10.1038/s41567-023-02206-0},
}

@article{fukuoka1997,
  title = {Dependence of ${\mathrm{T}}_{\mathrm{c}}$ and transport properties on the Cu valence in ${\mathrm{HgBa}}_{2}$${\mathrm{Ca}}_{\mathrm{n}\mathrm{\ensuremath{-}}1}$${\mathrm{Cu}}_{\mathrm{n}}$${\mathrm{O}}_{2(\mathrm{n}+1)+\mathrm{\ensuremath{\delta}}}$ (n=2,3) superconductors},
  author = {Fukuoka, A. and Tokiwa-Yamamoto, A. and Itoh, M. and Usami, R. and Adachi, S. and Tanabe, K.},
  journal = {Phys. Rev. B},
  volume = {55},
  issue = {10},
  pages = {6612--6620},
  numpages = {0},
  year = {1997},
  month = {Mar},
  publisher = {American Physical Society},
  doi = {10.1103/PhysRevB.55.6612},
  url = {https://link.aps.org/doi/10.1103/PhysRevB.55.6612}
}

@article{kitatani2023,
  title = {Optimizing Superconductivity: From Cuprates via Nickelates to Palladates},
  author = {Kitatani, Motoharu and Si, Liang and Worm, Paul and Tomczak, Jan M. and Arita, Ryotaro and Held, Karsten},
  journal = {Phys. Rev. Lett.},
  volume = {130},
  issue = {16},
  pages = {166002},
  numpages = {7},
  year = {2023},
  month = {Apr},
  publisher = {American Physical Society},
  doi = {10.1103/PhysRevLett.130.166002},
  url = {https://link.aps.org/doi/10.1103/PhysRevLett.130.166002}
}

@article{moree2024_2,
  title = {Universal chemical formula dependence of ab initio low-energy effective Hamiltonian in single-layer carrier-doped cuprate superconductors: Study using a hierarchical dependence extraction algorithm},
  author = {Mor\'ee, Jean-Baptiste and Arita, Ryotaro},
  journal = {Phys. Rev. B},
  volume = {110},
  issue = {1},
  pages = {014502},
  numpages = {29},
  year = {2024},
  month = {Jul},
  publisher = {American Physical Society},
  doi = {10.1103/PhysRevB.110.014502},
  url = {https://link.aps.org/doi/10.1103/PhysRevB.110.014502}
}

@article{sun2023,
   title={Signatures of superconductivity near 80K in a nickelate under high pressure},
   volume={621},
   ISSN={1476-4687},
   url={http://dx.doi.org/10.1038/s41586-023-06408-7},
   DOI={10.1038/s41586-023-06408-7},
   number={7979},
   journal={Nature},
   publisher={Springer Science and Business Media LLC},
   author={Sun, Hualei and Huo, Mengwu and Hu, Xunwu and Li, Jingyuan and Liu, Zengjia and Han, Yifeng and Tang, Lingyun and Mao, Zhongquan and Yang, Pengtao and Wang, Bosen and Cheng, Jinguang and Yao, Dao-Xin and Zhang, Guang-Ming and Wang, Meng},
   year={2023},
   month=jul, pages={493} }

@article{chen2010,
  title = {Enhancement of superconductivity by pressure-driven competition in electronic order},
  volume = {466},
  ISSN = {1476-4687},
  url = {http://dx.doi.org/10.1038/nature09293},
  DOI = {10.1038/nature09293},
  number = {7309},
  journal = {Nature},
  publisher = {Springer Science and Business Media LLC},
  author = {Chen,  Xiao-Jia and Struzhkin,  Viktor V. and Yu,  Yong and Goncharov,  Alexander F. and Lin,  Cheng-Tian and Mao,  Ho-kwang and Hemley,  Russell J.},
  year = {2010},
  month = Aug,
  pages = {950-953}
}

@article{Scalapino_1986,
  title = {$d$-wave pairing near a spin-density-wave instability},
  author = {Scalapino, D. J. and Loh, E. and Hirsch, J. E.},
  journal = {Phys. Rev. B},
  volume = {34},
  issue = {11},
  pages = {8190(R)--8192(R)},
  numpages = {0},
  year = {1986},
  month = {Dec},
  publisher = {American Physical Society},
  doi = {10.1103/PhysRevB.34.8190},
  url = {https://link.aps.org/doi/10.1103/PhysRevB.34.8190}
}

\onecolumngrid

\end{document}